\documentclass[fleqn,usenatbib]{mnras}

\usepackage{newtxtext,newtxmath}

\usepackage{ulem}

\usepackage[T1]{fontenc}

\DeclareRobustCommand{\VAN}[3]{#2}
\let\VANthebibliography\thebibliography
\def\thebibliography{\DeclareRobustCommand{\VAN}[3]{##3}\VANthebibliography}

\usepackage{graphicx}
\usepackage{amsmath}
\usepackage{placeins}

\usepackage[usenames,dvipsnames]{color}

\newcommand{\code}[1]{\texttt{#1}}

\title[IGWs in RGB stars]{3D hydrodynamics simulations of internal gravity waves in red giant branch stars}

\author[S. Blouin et al.]{Simon Blouin$^{1,\dagger}$\thanks{E-mail: sblouin@uvic.ca},
Huaqing Mao$^{2,\dagger}$,
Falk Herwig$^{1,\dagger}$,
Pavel Denissenkov$^{1,\dagger}$,
Paul R. Woodward$^{2,\dagger}$ and
\newauthor
William R. Thompson$^{1}$
\\
$^{1}$Department of Physics and Astronomy, University of Victoria, Victoria, BC V8W 2Y2, Canada\\
$^{2}$LCSE and Department of Astronomy, University of Minnesota, Minneapolis, MN 55455, USA\\
$^{\dagger}$Joint Institute for Nuclear Astrophysics -- Center for the Evolution of the Elements (JINA--CEE)
}

\date{Submitted: 7 October 2022, Resubmitted: 13 March 2023}

\pubyear{2022}

\begin{document}
\label{firstpage}
\pagerange{\pageref{firstpage}--\pageref{lastpage}}
\maketitle
\normalem

\begin{abstract}
We present the first 3D hydrodynamics simulations of the excitation and propagation of internal gravity waves (IGWs) in the radiative interiors of low-mass stars on the red giant branch (RGB). We use the \code{PPMstar} explicit gas dynamics code to simulate a portion of the convective envelope and all the radiative zone down to the hydrogen-burning shell of a $1.2\,M_{\odot}$ upper RGB star. We perform simulations for different grid resolutions (768$^3$, 1536$^3$ and 2880$^3$), a range of driving luminosities, and two different stratifications (corresponding to the bump luminosity and the tip of the RGB). Our RGB tip simulations can be directly performed at the nominal luminosity, circumventing the need for extrapolations to lower luminosities. A rich, continuous spectrum of IGWs is observed, with a significant amount of total power contained at high wavenumbers. By following the time evolution of a passive dye in the stable layers, we find that IGW mixing in our simulations is weaker than predicted by a simple analytical prescription based on shear mixing and not efficient enough to explain the missing RGB extra mixing. However, we may be underestimating the efficiency of IGW mixing given that our simulations include a limited portion of the convective envelope. Quadrupling its radial extent compared to our fiducial setup increases convective velocities by up to a factor 2 and IGW velocities by up to a factor 4. We also report the formation of a $\sim 0.2\,H_P$ penetration zone and evidence that IGWs are excited by plumes that overshoot into the stable layers.
\end{abstract}


\section{Introduction}
\label{sec:intro}
When exiting the main sequence and entering the red giant branch (RGB), the outer layers of low-mass stars expand, cool down, and as a result become more opaque to radiation. This triggers convection in their envelopes, which brings up to the surface the products of H fusion. After this ``first dredge-up'', no further changes to the atmospheric abundances of RGB stars are predicted by canonical stellar evolution theory, since the outer convective envelope remains isolated from the H-burning shell by a radiative zone that blocks the transport of species from the H-burning region to the surface.

But spectroscopic observations tell a different story. After the H-burning shell has crossed the composition discontinuity left at the maximal extent of the convective envelope during the first dredge-up (the so-called bump luminosity), the surface composition of nearly all low-mass RGB stars starts changing again. The $^{13}$C and $^{14}$N abundances increase, while the $^{12}$C and $^7$Li abundances decrease \citep{gilroy1989,pilachowski1993,charbonnel1994,gratton2000,mikolaitis2010,valenti2011}. Those abundance changes are the signpost of H fusion and imply that species can be transported across the radiative zone that separates the H-burning shell from the convective envelope \citep[e.g.,][]{karakas2014}.

The search for the extra mixing process responsible for this transport has been the subject of intense theoretical efforts for decades \citep{sweigart1979,smith1992,wasserburg1995,denissenkov2000,chaname2005,palacios2006,busso2007,denissenkov2003a,denissenkov2009,charbonnel2010}. The most commonly invoked mechanism is thermohaline mixing (or fingering convection). This instability is triggered in RGB stars due to $^3$He burning in the outskirts of the H-burning shell, which produces a depression in the mean molecular weight profile $\mu$ \citep{eggleton2006,charbonnel2007,cantiello2010}. The efficiency of this mixing mechanism crucially depends on the aspect ratio of the ``fingers'' formed as a result of the $\mu$ profile inversion. The aspect ratios needed to produce enough mixing far exceed those predicted by numerical simulations, where more blob-like structures are observed \citep{denissenkov2010,denissenkov2011,traxler2011,brown2013}. This suggests that thermohaline mixing alone is not sufficient to explain the observed extra mixing and the quest for additional mixing mechanisms is still ongoing.

Internal gravity waves (IGWs) are suspected to be an efficient species transport mechanism in the radiative zones of stellar interiors \citep{press1981,garcia1991,schatzman1996,denissenkov2003b,talon2005,schwab2020}. IGWs are caused by density perturbations in a stably stratified fluid and have buoyancy as a restoring force. In stellar interiors, they are stochastically excited by convective motions at the convective--radiative interface. Therefore, one can expect IGWs to be generated at the lower boundary of the convective envelope of RGB stars and to propagate inside the radiative zone that separates the H-burning shell from the convective envelope. This could provide an additional mixing mechanism for RGB stars. 

In addition to transporting species, IGWs may also transport angular momentum \citep{ringot1998,kumar1999,talon2002,rogers2013,pincon2017}. Asteroseismological observations have revealed that the cores of RGB stars spin much faster than their envelopes \citep{beck2012,beck2014,deheuvels2012,deheuvels2014}, but also much slower than if there was no additional angular momentum transport mechanism coupling the core to the envelope \citep{eggenberger2012,marques2013,cantiello2014}. Current evolutionary models fail to predict this (relatively) slow rotation: an additional process that extracts angular momentum from the core is needed. IGWs \citep[and mixed modes,][]{belkacem2015} have been considered as a solution to this problem \citep{fuller2014}.

Due to the complex interplay between convective motions and the excitation and propagation of IGWs, detailed hydrodynamics simulations are required to accurately determine the properties of IGWs in stellar interiors and ultimately assess their impact on stellar evolution. Analytical approaches are also possible \citep{kumar1999,montalban2000,lecoanet2013}, but they inevitably rely on a number of assumptions (e.g., the wave excitation mechanism, the shape of the power spectrum). Multi-dimensional hydrodynamics simulations of IGWs excitation and propagation have been performed for main-sequence stars \citep{rogers2005,dintrans2005,rogers2013,brun2011,alvan2014,alvan2015,edelmann2019,horst2020,ratnasingam2020,saux2022,herwig2023,thompson2023}, but to our knowledge no results currently exist for RGB stars. In this work, we present the first hydrodynamics simulations of IGW excitation and propagation in RGB stars. We use the \code{PPMstar} gas dynamics code to perform three-dimensional, full-sphere, high-resolution simulations of two different phases in the evolution of a 1.2\,$M_{\odot}$ star on the RGB. Based on these simulations, we present a first estimate of the mixing enabled by IGWs in RGB stars.

In Section~\ref{sec:methods}, we describe the \code{MESA} models that we use as base states for our hydrodynamics simulations and explain how the latter are set up in \code{PPMstar}. We then discuss the overall properties of our simulations in Section~\ref{sec:morpho}, where we present high-resolution renderings of our simulations as well as radial profiles and power spectra. Mixing by IGWs is investigated in Section~\ref{sec:IGWmixing} using two different approaches. Section~\ref{sec:envelope} is devoted to the influence of the size of the convective envelope on our results and Section~\ref{sec:boundary} to the properties of the convective boundary. Finally, our conclusions are stated in Section~\ref{sec:conclu}.

\section{Methods}
\label{sec:methods}
\subsection{\code{MESA} models}
\label{sec:mesa}
\begin{figure}
  \includegraphics[width=\columnwidth]{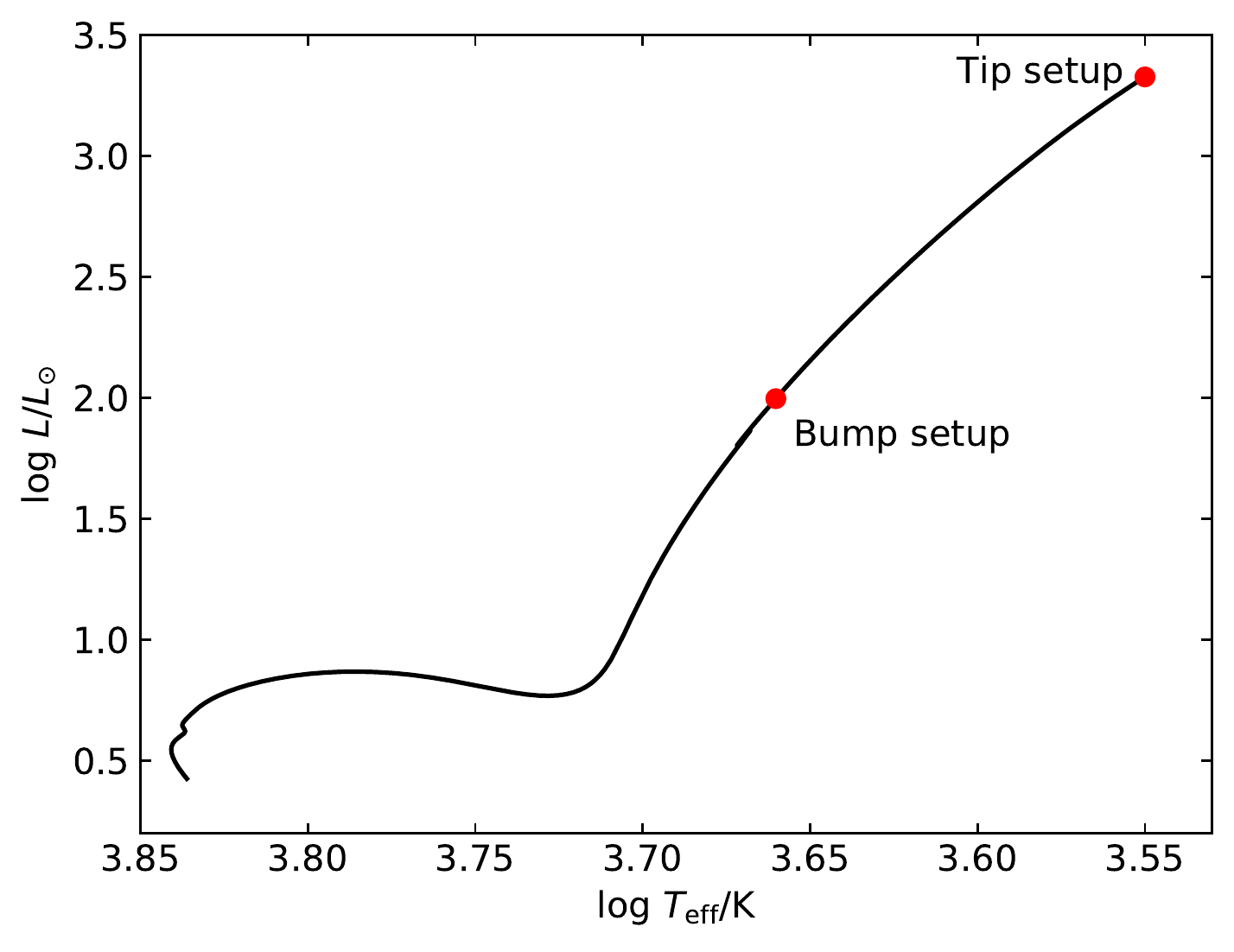}
    \caption{Hertzsprung--Russell diagram of our $1.2\,M_{\odot}$ \code{MESA} model from the zero-age main sequence to the tip of the RGB. The two phases simulated in this work are marked with red circles.}
    \label{fig:HR}
\end{figure}

We use \code{MESA} version 7624 \citep{paxton2011,paxton2013,paxton2015} to generate the initial states of our hydrodynamics simulations. We calculated the evolution of a $1.2\,M_{\odot}$ star from the pre-main sequence to the tip of the RGB. We assume an initial metallicity of $[{\rm Fe/H}]=-0.3$, and the mixing length theory (MLT) is used, with mixing length $\ell_{\rm MLT} = 2 H_P$ (where $H_P$ is the pressure scale height). Figure~\ref{fig:HR} displays the evolution of our model from the zero-age main sequence to the tip of the RGB in the theoretical Hertzsprung--Russell diagram. Two circles identify the two phases simulated in 3D in this work. The first one is just after the bump luminosity at $\log L/L_{\odot}=2.00$ (this is our ``bump setup''), and the second one is very close to the tip of the RGB at $\log L/L_{\odot}=3.33$ (this is our ``tip setup'').

\begin{figure}
  \includegraphics[width=\columnwidth]{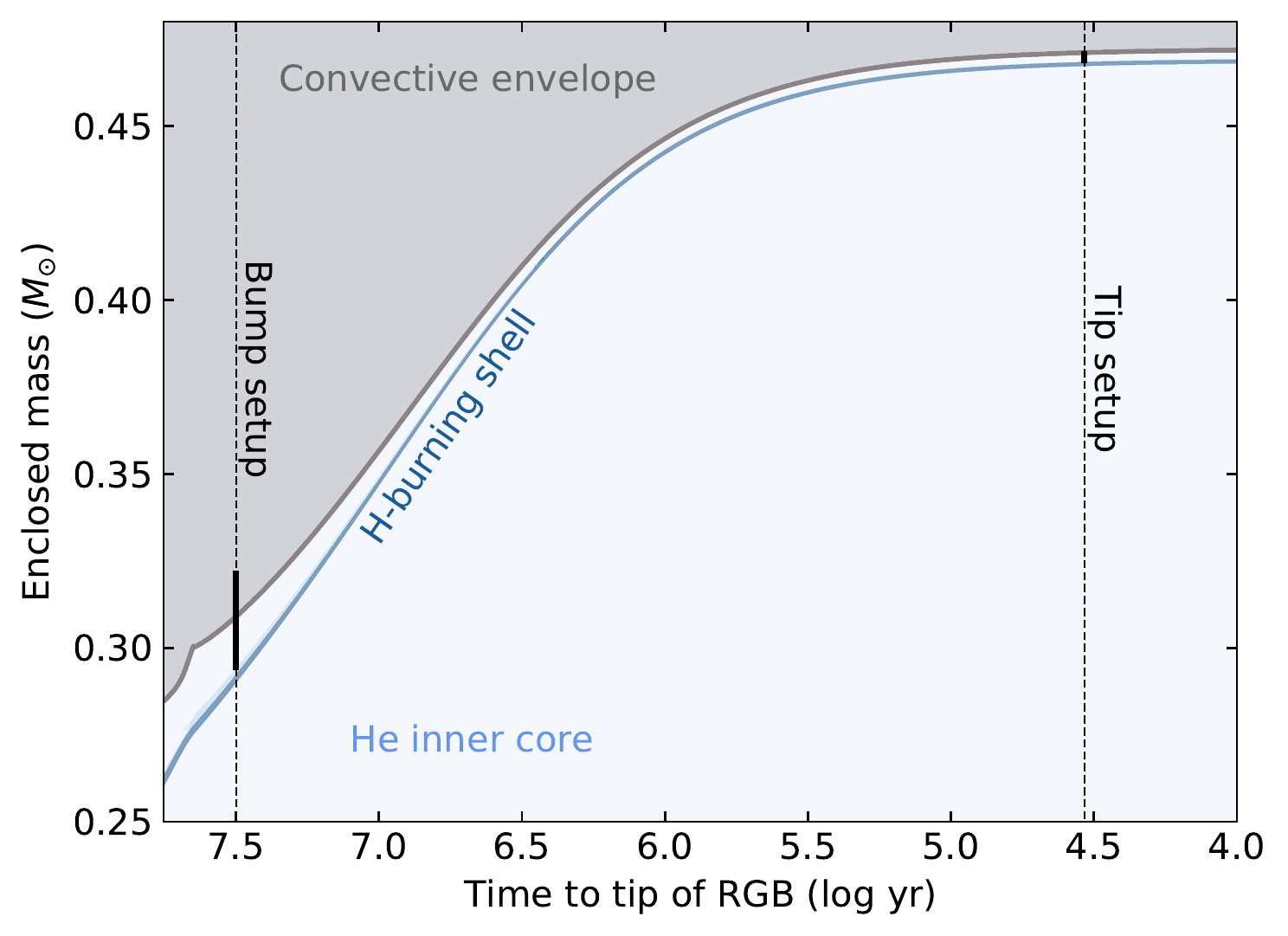}
    \caption{Kippenhahn diagram of our $1.2\,M_{\odot}$ \code{MESA} model on the RGB. H burning takes place in a shell represented by the thin blue line. The narrow region above this H-burning shell is the radiative zone where IGWs are excited by convective motions in the convective envelope above (in grey). The two RGB phases examined in this work are marked by the vertical dashed lines. In both cases, a small solid line indicates the actual mass included in our 3D simulations.}
    \label{fig:kip}
\end{figure}

Figure~\ref{fig:kip} shows the Kippenhahn diagram of our \code{MESA} model along the RGB. The convective envelope is in grey and the H-burning shell is displayed as a thin blue line; the narrow region in between is the radiative zone where IGWs propagate. Two dashed vertical lines indicate the models that correspond to our bump and tip setups and the thick solid lines show the portion of those models that we actually simulate in 3D. It would be prohibitively expensive to simulate the whole star and choices have to be made regarding which regions to include. For both setups, we omit the inner $R < 40\,$Mm, which corresponds to truncating our setups just above the H-burning shell. The equation of state currently included in \code{PPMstar} does not account for the degeneracy pressure that becomes prominent in the dense inner core, and in any case this region is not strictly needed to study IGWs in the radiative zone above. Furthermore, recent 2D hydrodynamics simulations of a Sun-like star show that the location of the inner simulation boundary has a negligible effect on the IGWs \citep{vlaykov2022}; a similar behaviour can be expected for the RGB. The mass contained inside this inner 40\,Mm is taken into account when computing the gravitational acceleration in our 3D simulations. For the upper boundary, we adopt a maximum radius of $R_{\rm max}=900$\,Mm for our fiducial simulations, which represents only 8\% of the stellar radius for the bump setup and 1\% for the tip setup. This is enough to include all the radiative zone and a portion of the convective envelope. However, we recognize that artificially blocking the flow at 900\,Mm (\code{PPMstar} uses reflective boundary conditions) alters the convective motions in the envelope by impeding the development of large-scale convective modes. But extending our simulation sphere further out would degrade the grid resolution in the radiative zone where the IGWs propagate, and we thus settled on $R\leq 900$\,Mm as a compromise between those two effects. We will explore how including a larger envelope affects our results in Section~\ref{sec:envelope}.

\begin{figure}
  \includegraphics[width=\columnwidth]{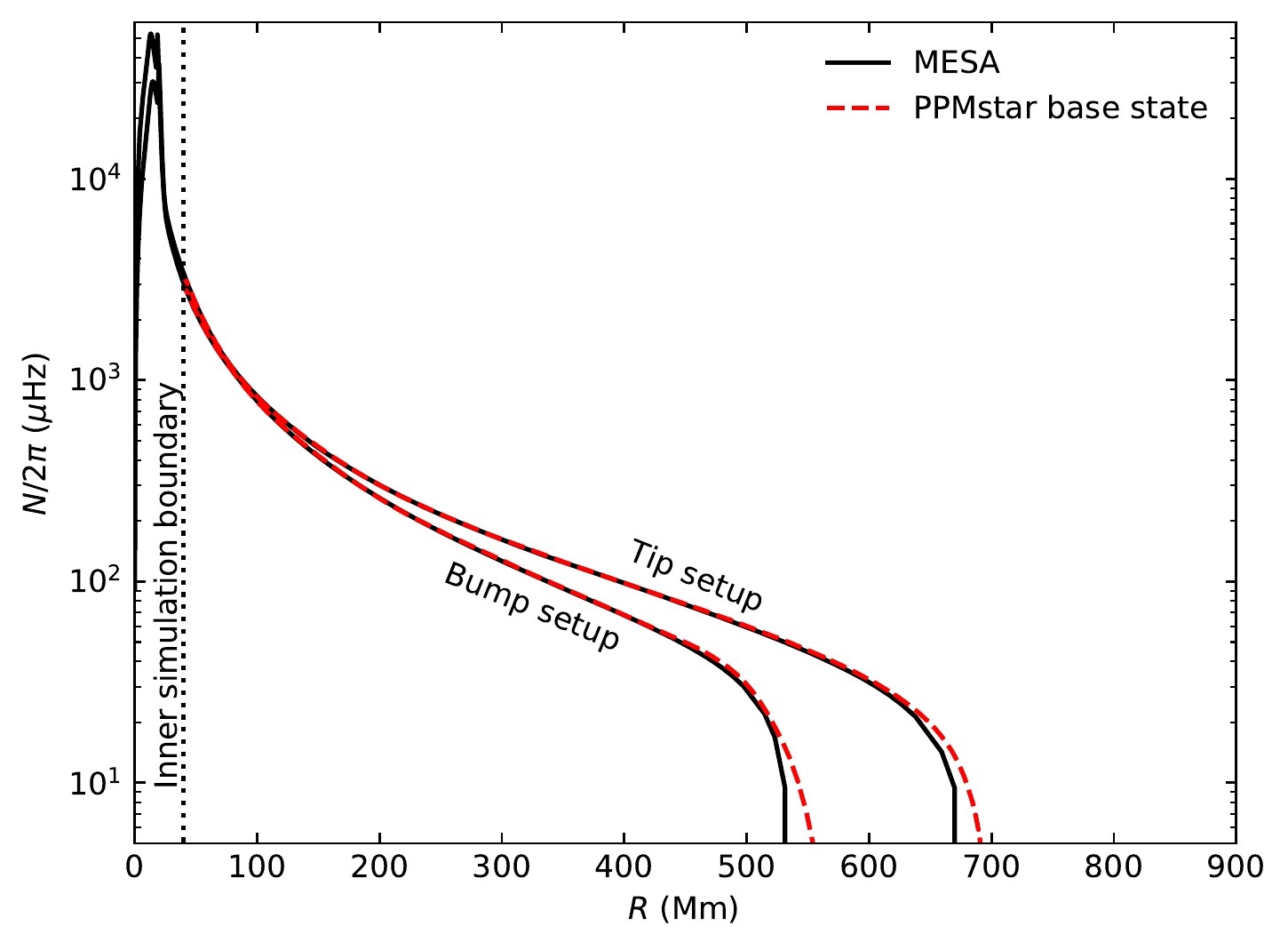}
    \caption{Brunt--V\"ais\"al\"a frequency $N$ as a function of radius for our two RGB setups. The \code{MESA} profiles (solid black lines) are compared to the $N$ profiles used as base states of our \code{PPMstar} simulations (red dashed lines). The vertical dotted line indicates the location of the inner simulation boundary.}
    \label{fig:Nbase}
\end{figure}

To initialize the 3D hydrodynamics simulations, we only require (1) the mass contained within the inner simulation radius (i.e., within $R<40\,$Mm), (2) the pressure at the inner simulation radius, and (3) the entropy profile up to the outer simulation radius. From those quantities, the initial pressure ($P$), density ($\rho$), temperature ($T$), and mass ($M_r$) stratifications can be recovered by integrating the hydrostatic equilibrium equation from the inner boundary. Note that there is no composition gradient in our RGB setups, including across the convective boundary, meaning that only one fluid with the appropriate $\mu$ is needed in our 3D simulations. As in \cite{herwig2023}, we smooth the \code{MESA} entropy profile to remove small-scale noise and we force a constant entropy in the convective envelope. Figure~\ref{fig:Nbase} compares the Brunt--V\"ais\"al\"a frequency $N$ obtained by this procedure to the original \code{MESA} profile. Only at the convective boundary (when $N \rightarrow 0$) is there a small discrepancy between both profiles due to our smoothing procedure.

\subsection{\code{PPMstar} simulations}
\label{sec:ppmstar}
We use the \code{PPMstar} explicit gas dynamics code \citep{woodward2015,jones2017,andrassy2020,herwig2023,mao2023,woodward2023} to perform our 3D hydrodynamics simulations. As described in \cite{mao2023}, a more realistic equation of state that includes both the ideal gas pressure and the radiation pressure is now implemented in \code{PPMstar}. Contributions from electron degeneracy pressure and Coulomb interactions remain negligible in the regions we simulate ($R \geq 40\,{\rm Mm}$). In addition, radiation diffusion is now also included \citep{mao2023}. The Rosseland mean opacity is calculated using polynomial fits that are generated before each run by fitting OPAL opacity tables \citep{iglesias1996} within the restricted $\rho$--$T$ domain relevant to each setup. Here, those fits depend only on $\rho$ and $T$ since the composition is uniform throughout the simulated region. We opted for this approach instead of a direct interpolation of the OPAL tables in the interest of code execution speed.

Convection is driven by cooling down (i.e., removing heat) from the uppermost 50\,Mm. The rate at which heat is removed corresponds to the luminosity that we are simulating (which is not necessarily the same as the nominal stellar luminosity as discussed in the next paragraph). In addition, we inject heat at the same rate in the first 20\,Mm above the inner boundary of our simulations. This is to compensate the loss of heat at the upper boundary with the aim of keeping the thermal content of the star constant over the simulation. This driving strategy closely imitates energy transport in the real star. Heat is produced close to the centre in the H-burning shell, which is what our heating term at the inner simulation boundary mimics. Similarly, convection carries heat all the way to the surface in a real star, which is what our cooling term at the outer simulation boundary simulates.

The explicit scheme used by \code{PPMstar} sets a lower limit on the Mach numbers that can be simulated. A very low Mach number flow would demand prohibitively small grid cells. Consequently, for our bump setup, we cannot perform the simulations at the nominal luminosity $L_\star$. Our fiducial bump simulations use $L=1000\,L_\star$ to drive the convection zone. To extrapolate our results to nominal heating, we perform a series of simulations with different luminosity boost factors as indicated in Table~\ref{tab:runs}. Note that the radiative diffusivity $K$ is also increased by the same factor so that the energy transported by radiation scales as the driving luminosity (in order to conserve energy).

\begin{table}
  \centering
  \caption{Summary of simulations used in this paper.}
  \label{tab:runs}
  \begin{tabular}{lcccccc} 
    \hline
    ID & setup & $L/L_\star$ & grid & $R_{\rm max}$ (Mm) & $t$ (h) & \# dumps \\
    \hline
    X17 & bump & $10^4$ & $768^3$ & 900 & 797 & 308\\
    X18 & bump & $10^{3.5}$ & $768^3$ & 900 & 789 & 305 \\
    X14 & bump & $10^3$ & $768^3$ & 900 & 1058 & 419 \\
    X22 & bump & $10^3$ & $1536^3$ & 900 & 831 & 480 \\
    X21 & bump & $10^{2.5}$ & $768^3$ & 900 & 1552 & 600 \\
    X15 & bump & $10^2$ & $768^3$ & 900 & 935 & 723\\
    X16 & bump & 10 & $768^3$ & 900 & 1497 & 579 \\
    X24 & tip  & 1 & $768^3$ & 900 & 1695 & 1280\\
    X25 & tip  & 1 & $1536^3$ & 1800 & 1352 & 510\\
    X26 & tip  & 1 & $1536^3$ & 900 & 682 & 515\\
    X30 & tip  & 1 & $2880^3$ & 1100 & 552 & 700\\
    X32$^\dagger$ & tip  & 1 & $768^3$ & 900 & 850 & 642\\
    X33$^\dagger$ & tip  & 1 & $1536^3$ & 900 & 633 & 478 \\
        \hline
        \end{tabular}
{\raggedright $^\dagger$ No radiation diffusion ($K=0$) \par}
\end{table}
 
The tip setup is different. Thanks to its higher luminosity and different stratification, the Mach numbers in its convection zone (${\rm Ma} \simeq 0.010-0.015$) are high enough that the star can be simulated at nominal luminosity. This is also true for the IGWs in the stable layers. This gives our RGB tip simulations an exceptional degree of fidelity, providing solutions of the full conservation equations in 3D and over the $4\pi$ sphere, at the actual stellar luminosity, and including radiation diffusion with realistic opacities. To our knowledge, these are the first stellar interior hydrodynamic simulations of this kind. The one approximation we are making is that only a small portion of the envelope is included. This can admittedly have a large impact both on the convective and IGW motions \citep[e.g.,][]{vlaykov2022}. We return to this question in Section~\ref{sec:envelope}.

\begin{figure}
  \includegraphics[width=\columnwidth]{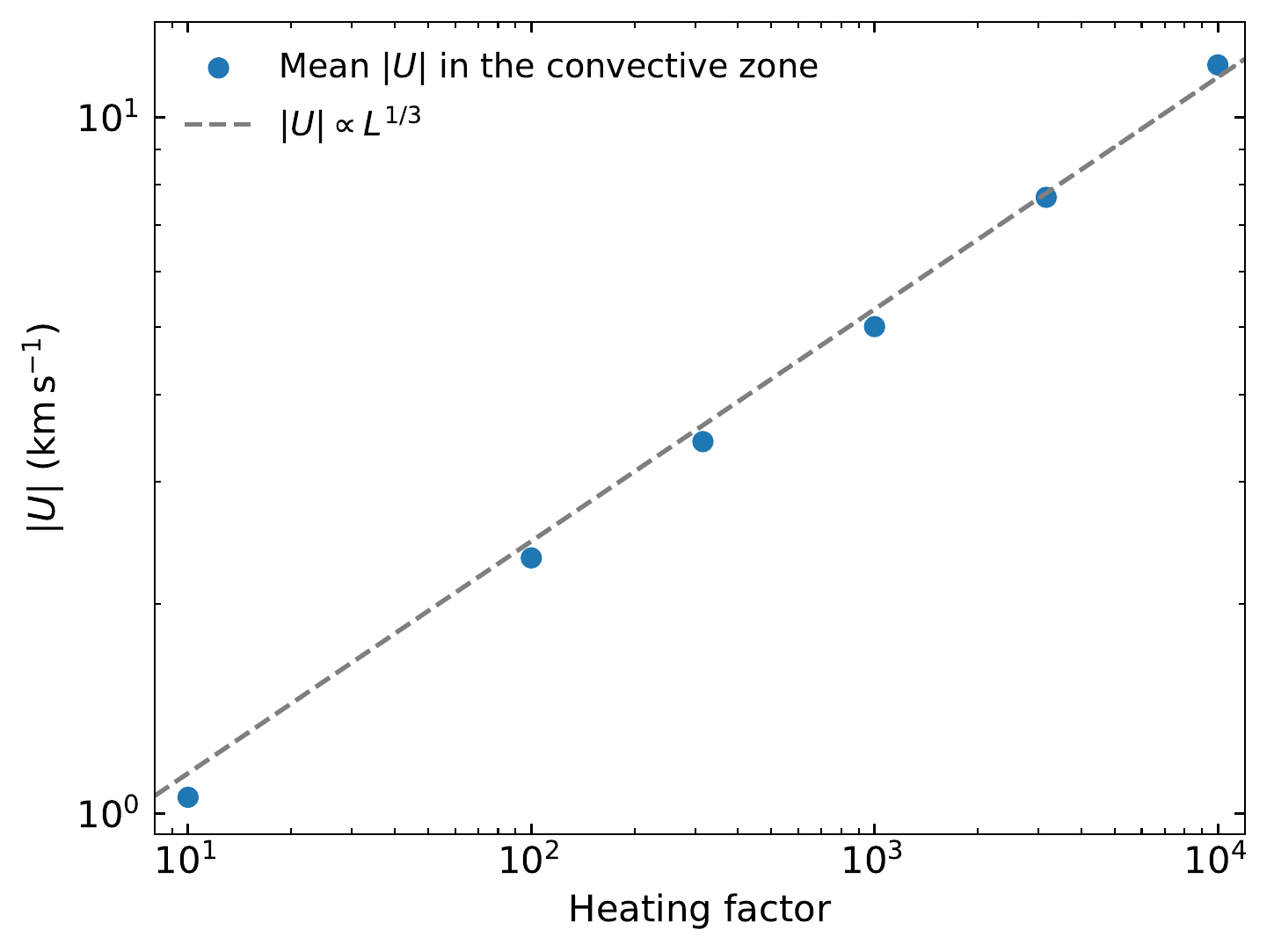}
  \includegraphics[width=\columnwidth]{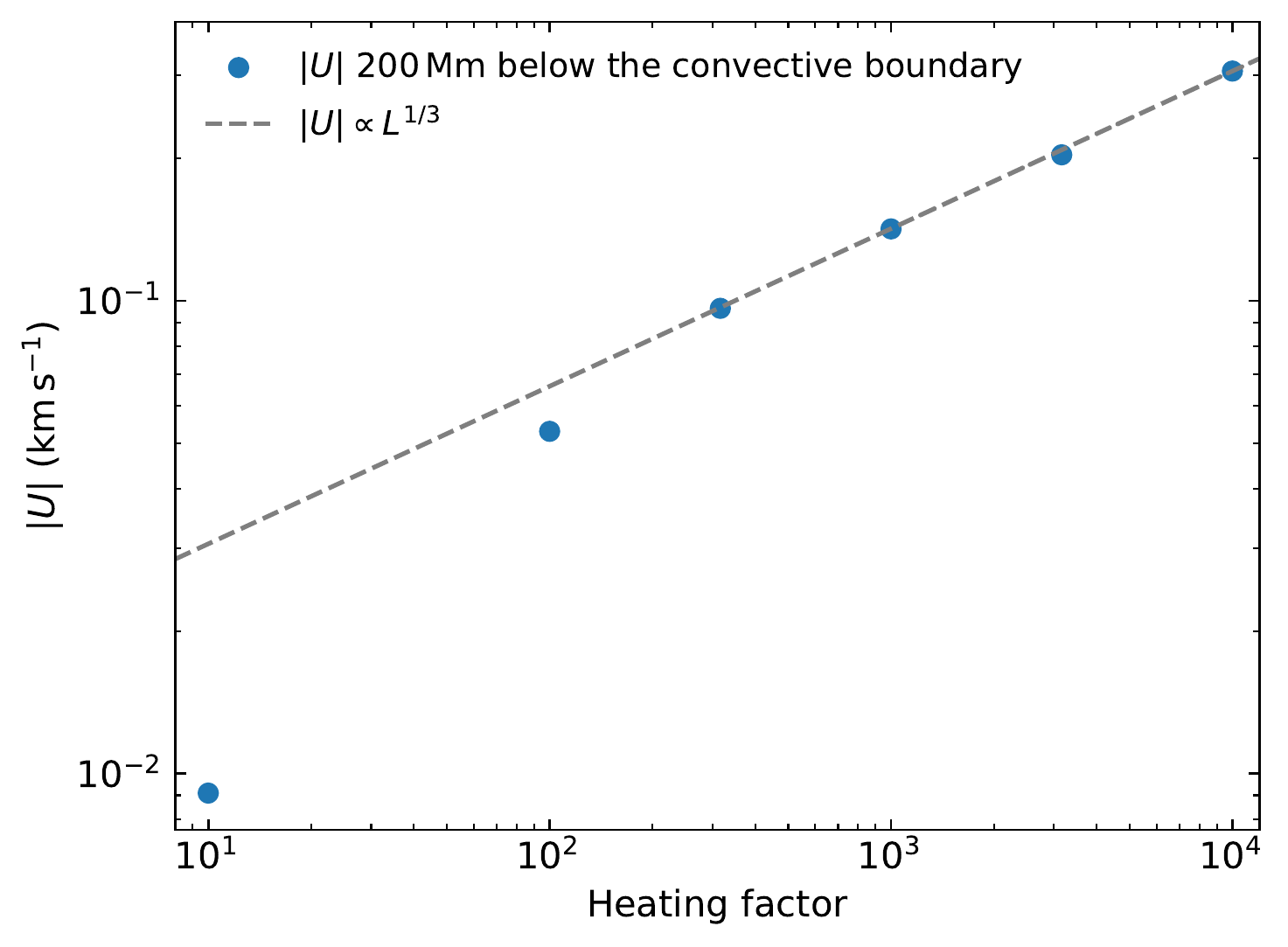}
    \caption{Root mean square velocity in the convection zone (top panel) and in the stable layers $200\,{\rm Mm}\simeq 1\,H_P$ below the convective boundary (bottom panel) as a function of the boost factor applied to the luminosity. In each case except X17, the vorticity is averaged over the last 100 dumps. For X17, earlier dumps ($100-120$, $t=260$ to 310\,h) are used to prevent selecting dumps where the convective boundary has moved below $R=400$\,Mm and the flow velocity is tainted by the numerical artifacts in the innermost regions of the simulation. Those runs are all for the RGB bump setup and a 768$^3$ grid.}
    \label{fig:scaling_U}
\end{figure}

To demonstrate that the tip setup can reliably be simulated at nominal heating, we show in Figure~\ref{fig:scaling_U} how, for the bump setup, the rms velocity in the convective envelope and in the stable layers scales as a function of the boost factor applied to the heating luminosity. The convective velocities follow a well-defined $L^{1/3}$ scaling relation, as established in previous 3D hydrodynamics simulations \citep{porter2000,muller2016,jones2017,baraffe2021,herwig2023}. In the stable layers, we observe the same $L^{1/3}$ dependence down to at least $L=10^{2.5}\,L_{\star}$ (this differs from the results of \citealt{herwig2023}, a point to which we return in Section~\ref{sec:heating_series}). This suggests that there are no numerical convergence issues at those luminosities. The smallest velocity that still adheres to the scaling law is $|U| \simeq 0.1\,{\rm km/s}$ ($L=10^{2.5}\,L_{\star}$ in the bottom panel of Figure~\ref{fig:scaling_U}), which corresponds to a Mach number of $\simeq 0.0003$. As shown in Figure~\ref{fig:velcomp_mach}, the Mach numbers of our RGB tip simulations are above that threshold for $R \gtrsim 300\,{\rm Mm}$. This indicates that our nominal-heating RGB tip simulations can be trusted everywhere except in the innermost portion of the radiative interior. Finally, note that this analysis (Figure~\ref{fig:scaling_U}) was performed with the lowest-resolution grid ($768^3$) used in this work. This is therefore a pessimistic assessment of convergence errors as the scaling laws are expected to hold down to lower luminosities and smaller velocities when the grid resolution is increased (e.g., Figure~32 of \citealt{herwig2023}).

\begin{figure}
  \includegraphics[width=\columnwidth]{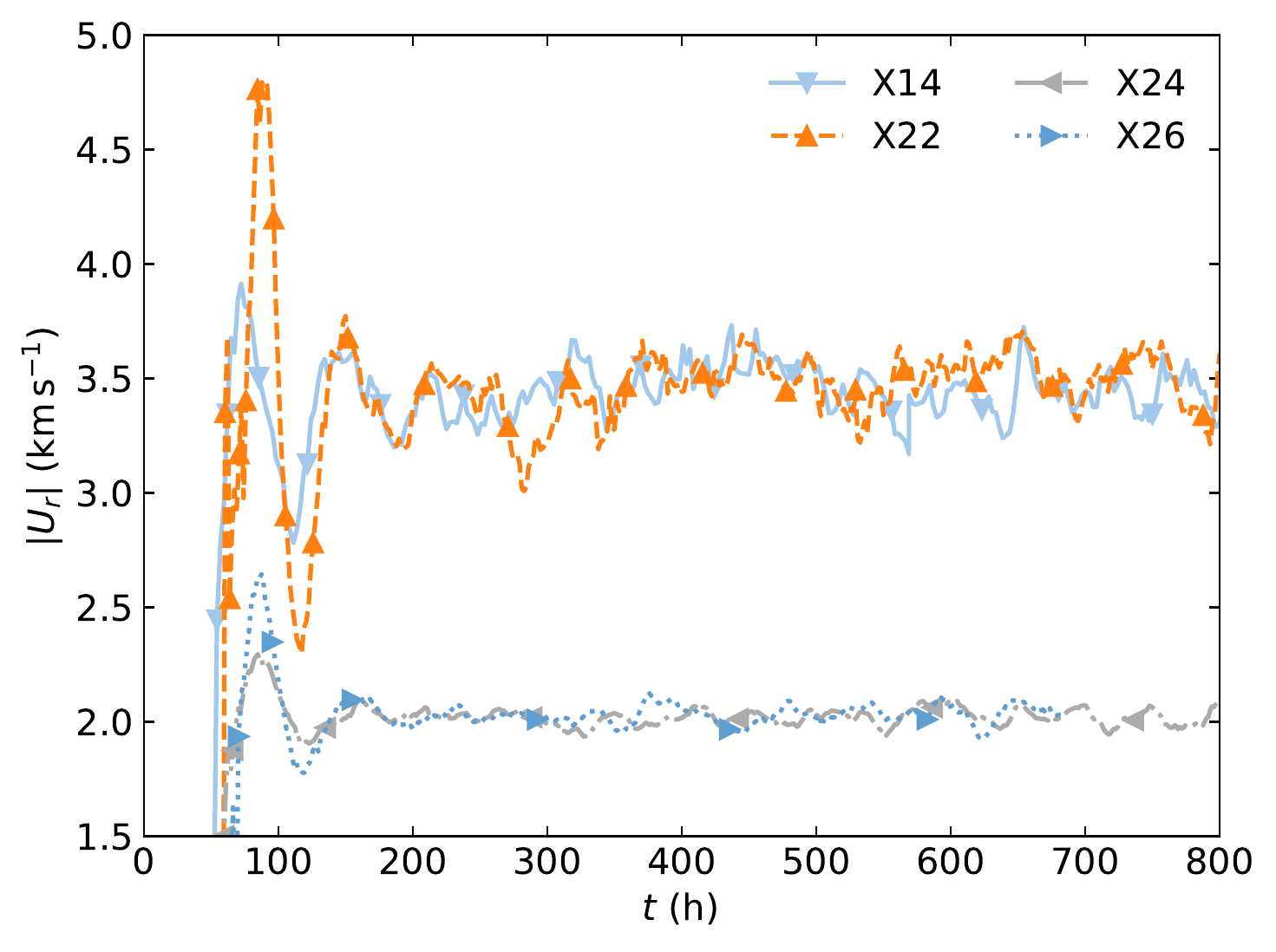}
  \includegraphics[width=\columnwidth]{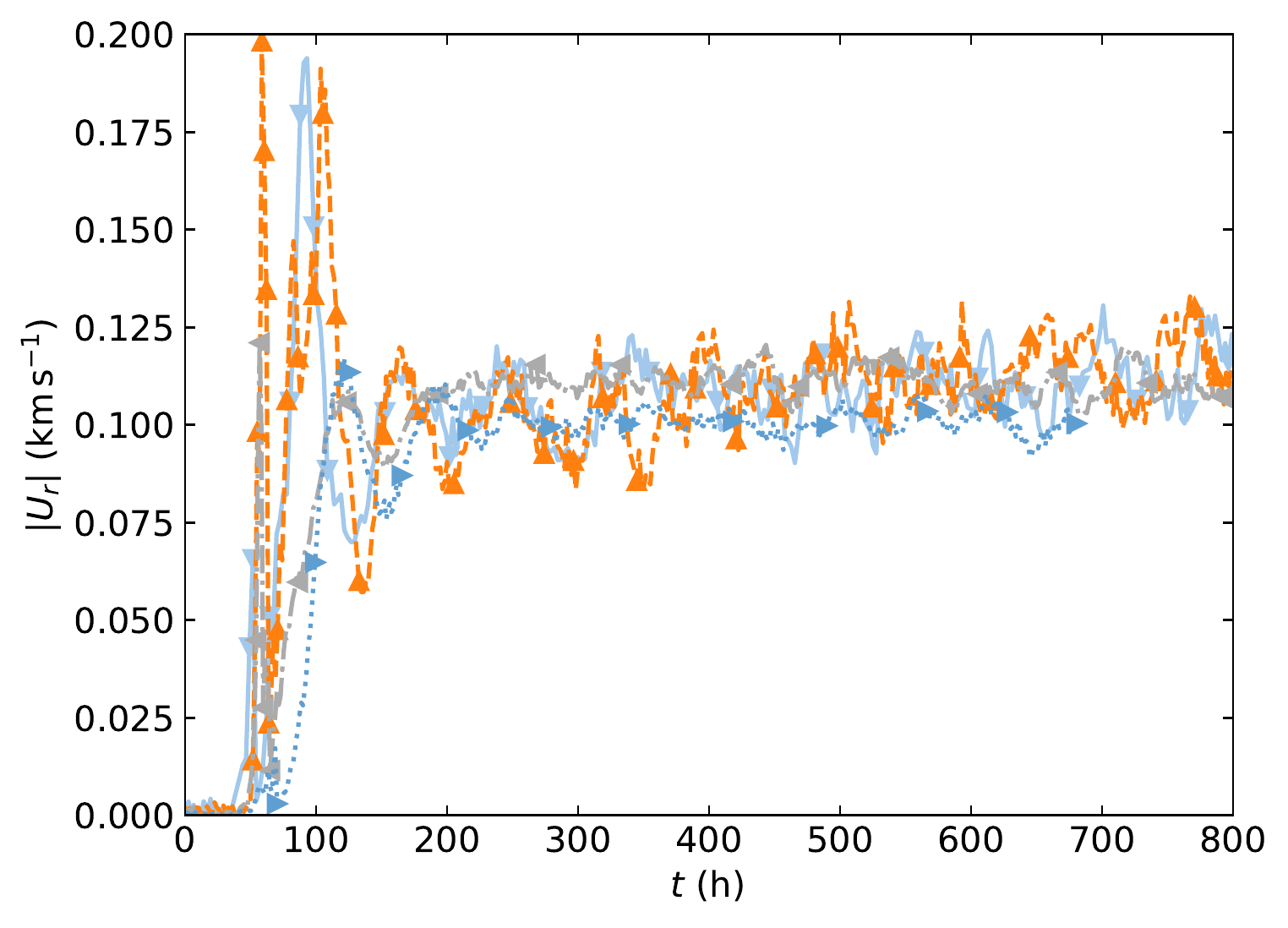}
    \caption{Time evolution of the spherically averaged radial velocity amplitude for runs X14, X22, X24, and X26 (see Table~\ref{tab:runs}). The top panel shows $|U_r|$ $0.5\,H_P$ above the convective boundary and the bottom panel shows the same quantity $0.5\,H_P$ below the convective boundary ($H_P \simeq 200\,{\rm Mm}$ at the convective boundary).}
    \label{fig:convergence}
\end{figure}

Our simulations are performed on Cartesian grids of $768^3$, $1536^3$ or $2880^3$ and run on average for $t \simeq 1000\,$h of star time, as indicated in Table~\ref{tab:runs}. With a characteristic convective turnover timescale of 1~day, the simulations are long enough to determine the flow properties after eliminating the initial transient phase of a few hundred hours (Figure~\ref{fig:convergence}). We note however that our simulations are not long enough to achieve a thermal equilibrium state. Since the gas dynamics in the 3D simulations is different than what is assumed in the initial 1D MLT-based \code{MESA} stratifications, our setups are inevitably out of thermal equilibrium. This results in the gradual migration of the convective boundary. Furthermore, in all our simulations except X30, our choice of external boundary condition also leads to a spurious migration of the Schwarzschild boundary (see Section~\ref{sec:gaussians}). For those reasons, the convective boundary does not remain at its original location shown in Figure~\ref{fig:Nbase}. Finding configurations that are in equilibrium when simulated with 3D hydrodynamics is an interesting endeavour in itself, but one that is outside the scope of this work.

The time steps are adjusted as to maintain a Courant number of 0.85. In the case of our $768^3$-grid simulations, this corresponds to $\Delta t=3.2\,$s for the bump setup and 2.6\,s for the tip setup ($\Delta t$ is half that when using a $1536^3$ grid with the same $R_{\rm max}$). A detailed output of the simulations (or dump) is produced every 2.6\,h for the bump setup and every 1.3\,h for the fiducial tip setup. The content of those dumps is detailed in \cite{herwig2023}.

\code{PPMstar}’s use of Cartesian coordinates optimizes numerical accuracy for a general fluid flow problem. It gives rise to a simple and highly effective design in which the computation proceeds in symmetrized sequences of 1D passes in the three coordinate directions (i.e., directional operator splitting). One consequence of our coordinate choice is that the application of boundary conditions becomes more difficult. Boundary conditions are currently implemented at specific radii (here, an inner boundary at 40\,Mm and an outer boundary at $R_{\rm max}$), placed well away from the main region of interest to minimize potential numerical artifacts. We approximate these bounding spheres by the nearest set of cubical grid cell faces, which implies that these spheres are ragged at the scale of the grid. We impose a reflecting boundary condition at the bounding spheres using ghost cells that mirror the cells across the bounding surfaces. This is done in each 1D pass, and in each such pass the bounding surface is perpendicular to the direction of the pass, but it is not perpendicular to the gravitational acceleration vector. For convenience, we therefore smoothly turn off gravity beginning a few grid cell widths in radius before the bounding sphere is reached, allowing us to implement a simple boundary condition in each 1D pass. The downside of this approach is the introduction of a very thin layer next to the boundary where gravitational acceleration smoothly drops to zero. This approach has so far caused no noticeable problems \citep[e.g.,][]{woodward2015,jones2017,andrassy2020,herwig2023}, but the current context is trickier because the reflection of IGWs at the inner boundary could matter. We find that over the course of the computation, motions are set going and grow in the thin layer next to the inner boundary. However, these do not come near to the observed IGW motions that are set in motion by the convective envelope (as we will see in Section~\ref{sec:damping}, radiative damping dissipates the IGWs well before they reach the inner boundary), and they should not affect the computed results in the regions of interest. 

\section{Main properties of the flow}
\label{sec:morpho}
\subsection{Centre-plane slice renderings}
\label{sec:bobs}
\begin{figure*}
    \centering
    \includegraphics[width=\columnwidth]{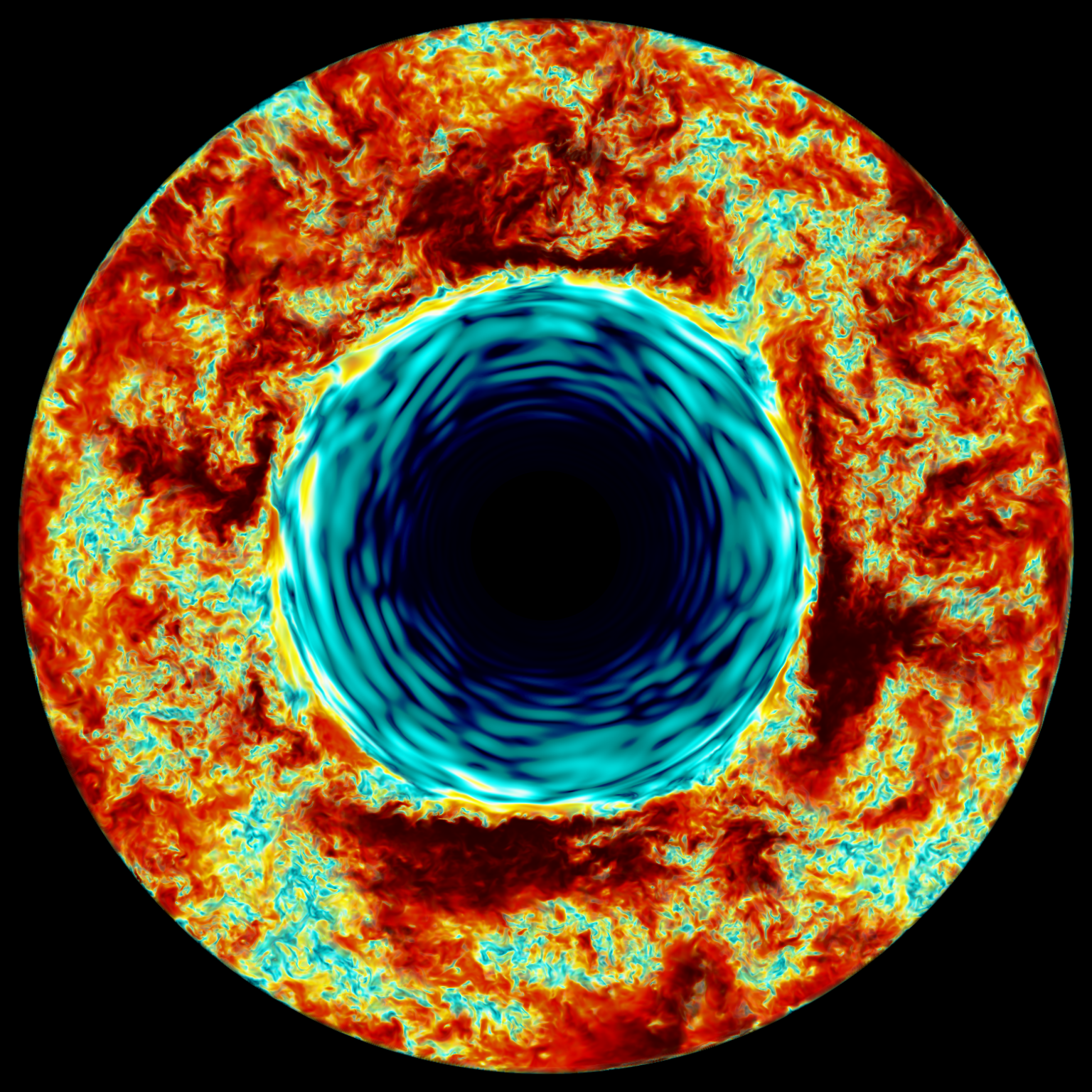}
    \includegraphics[width=\columnwidth]{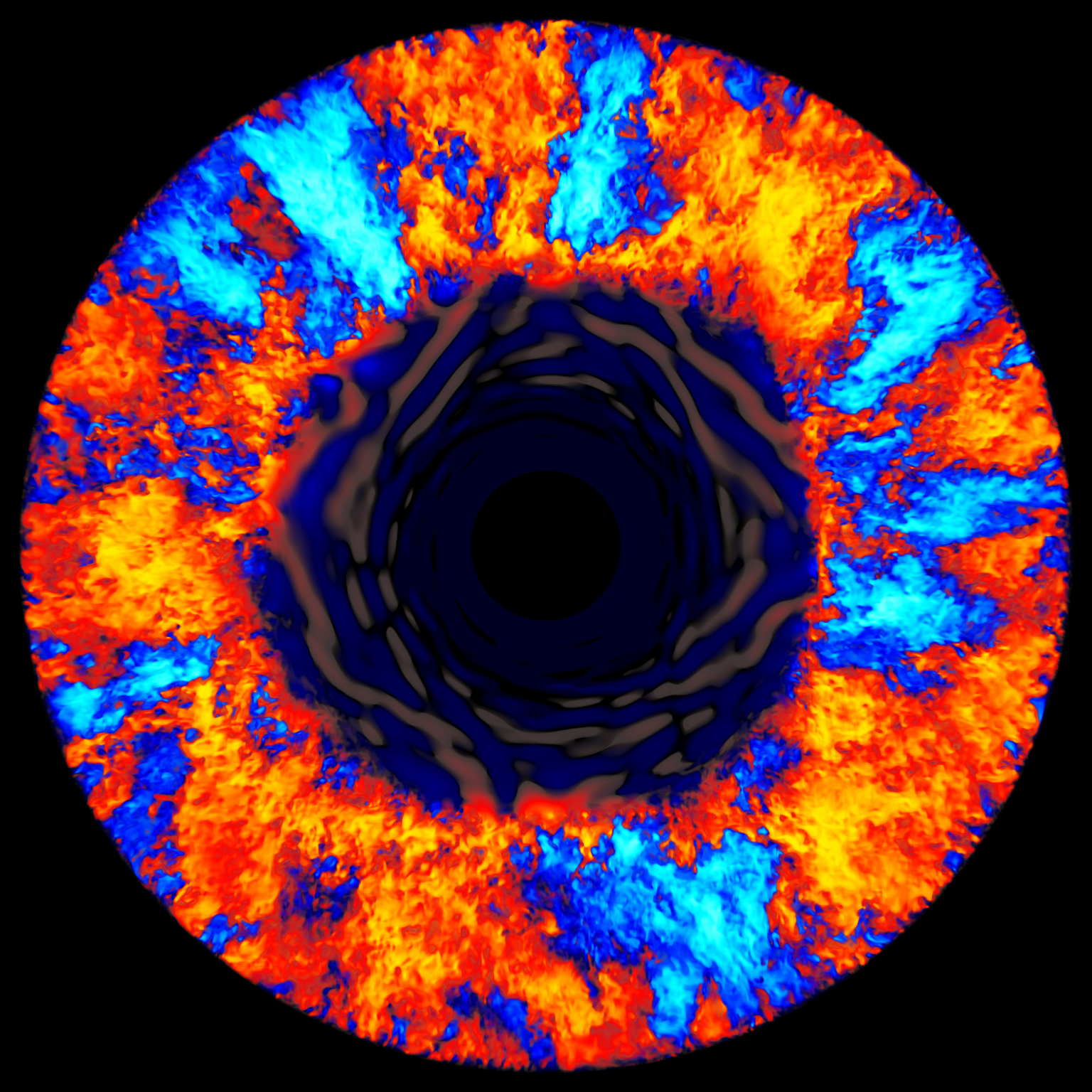}
    \caption{Centre-plane slice rendering of run X22 (bump setup, 1536$^3$ grid) at dump~390 ($t=672\,$h). {\it Left}: magnitude of the tangential velocity component $|U_t|$ (i.e., perpendicular to the radial direction), with dark blue, turquoise, yellow, red, and dark red representing a sequence of increasing velocities. {\it Right}: radial velocity $U_r$, with blue colours representing inward-moving flows and red-orange colours outward-moving flows. Those renderings were generated to qualitatively visualize the important features of our simulations. The inner 120\,Mm were masked to remove the artifacts introduced by the inner simulation boundary. High-resolution movies are available at \url{https://www.ppmstar.org}.}
    \label{fig:bobs}
\end{figure*}

\begin{figure}
    \centering
    \includegraphics[width=\columnwidth]{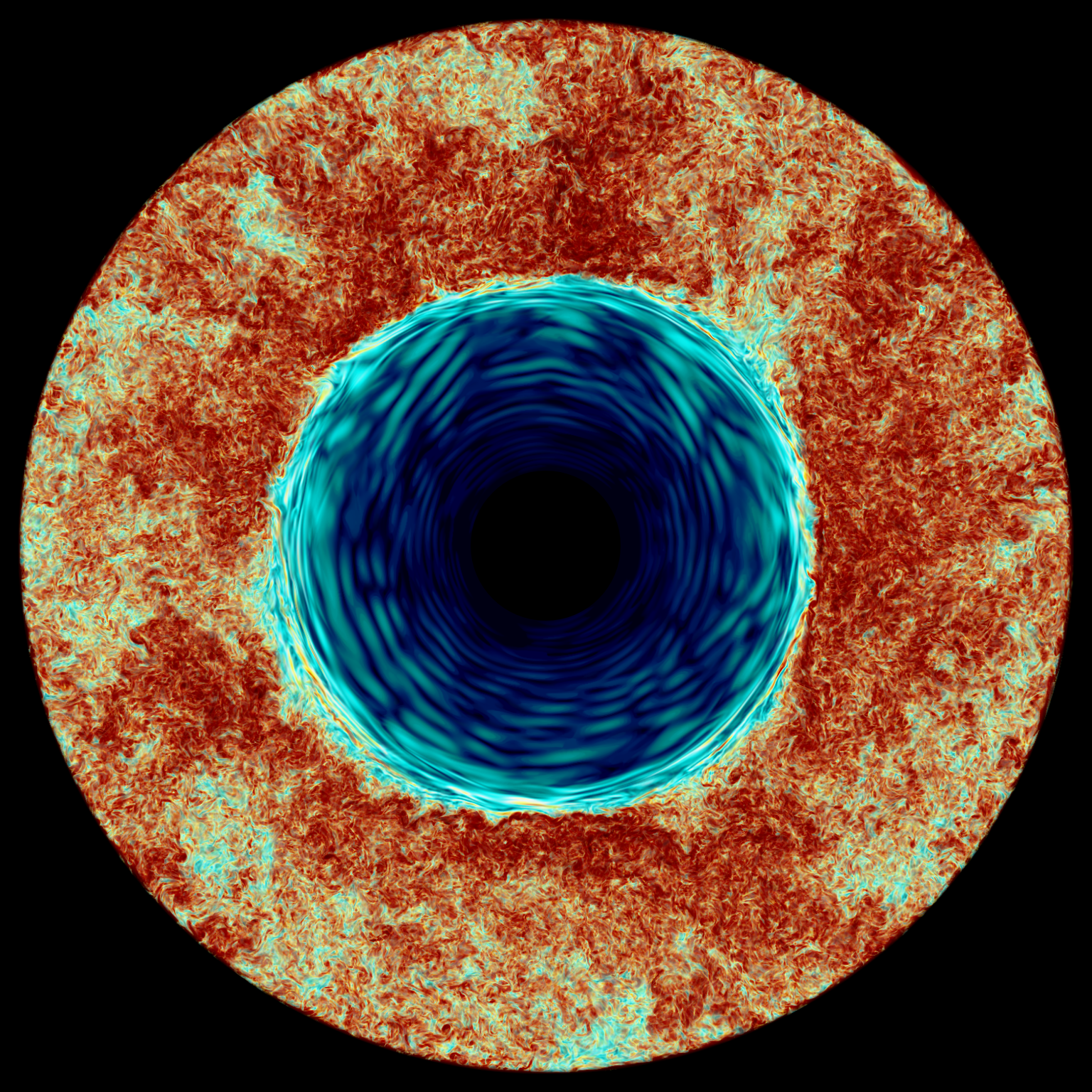}
    \caption{Centre-plane slice rendering of the vorticity magnitude (same colour sequence as for $|U_t|$ in Figure~\ref{fig:bobs}) for run X22 at dump~390 ($t=672\,$h). The inner 120\,Mm were masked to remove the artifacts introduced by the inner simulation boundary.}
    \label{fig:bobs2}
\end{figure}

To visualize the important features of our simulations, Figures~\ref{fig:bobs} and~\ref{fig:bobs2} show renderings of the tangential velocity magnitude $|U_t|$, radial velocity $U_r$, and vorticity magnitude $| \nabla \times U|$ for dump~390 ($t=672\,$h) of X22 ($L=1000 L_{\star}$, $1536^3$ grid). Those renderings display a centre-plane slice of the full 3D simulation sphere. In all three renderings, the convective envelope is easily distinguished from the radiative zone. The former is characterized by large velocities and is highly turbulent as revealed by the fine-scale structures in the vorticity rendering. In contrast, the radiative zone shows slower, more organized, wave-like flows. As we will see in Section~\ref{sec:spectra}, the almost circular structures visible in the $|U_t|$ and vorticity renderings correspond to the superposition of several IGWs with different spatial and temporal frequencies.

While the behaviour of our simulations is most easily visualized in the movies available at \url{https://www.ppmstar.org}, those static renderings nevertheless offer important insights. The rendering of the radial velocity component reveals that the convective envelope hosts several convective cells, with alternating downdrafts and updrafts (shown as an alternation of blue and red-orange colours) as we rotate around the sphere. This is reminiscent of the flow patterns observed in \code{PPMstar} simulations of He-shell flash convection in rapidly accreting white dwarfs \citep{stephens2021} and of O-burning shells in massive stars \citep{jones2017,andrassy2020}. However, this is very different from the behaviour found in our recent simulations of core convection in non-rotating massive main sequence stars, where a single dipole mode dominates the convective flow \citep{herwig2023}. This also differs from the results of \cite{brun2009}, whose anelastic 3D simulations of the envelope of slowly rotating RGB stars are also dominated by a large dipole mode, as previously established by \cite{porter2000b}. We attribute this difference to the artificial boundary imposed on the flow at 900\,Mm. The largest-scale mode that develops in the convection zone is limited by the vertical extent of the convection zone, and therefore a large dipole mode is prevented from forming in our simulations. We investigate this question in Section~\ref{sec:envelope}.

The tangential velocity rendering displays a few high-$|U_t|$ (dark red) structures inside the convection zone that follow the contour of the convective boundary. Those structures are caused by the inward moving flows that collide with the convective--radiative interface. Unable to continue inward, those flows are forced to turn and continue in a perpendicular direction, thereby creating high-$|U_t|$ structures. For example, the downdraft seen just a few degrees East from North in the $U_r$ rendering of Figure~\ref{fig:bobs} creates a high-$|U_t|$ double wedge structure where it hits the convective boundary. This behaviour is entirely analogous to the one described in \cite{herwig2023} for core convection in massive main sequence stars. In Section~\ref{sec:boundary}, we will see that it is where those flows impact (or overshoot past) the convective boundary that waves are excited in the radiative zone. The $|U_t|$, $U_r$ and $| \nabla \times U |$ renderings of the RGB tip simulations are qualitatively similar to those shown for the bump setup in Figures~\ref{fig:bobs} and~\ref{fig:bobs2}. We omit them for conciseness, but they are available, along with high-resolution movies, at \url{https://www.ppmstar.org}.

\subsection{Radial profiles}
\label{sec:prfs}
\begin{figure}
    \centering
    \includegraphics[width=\columnwidth]{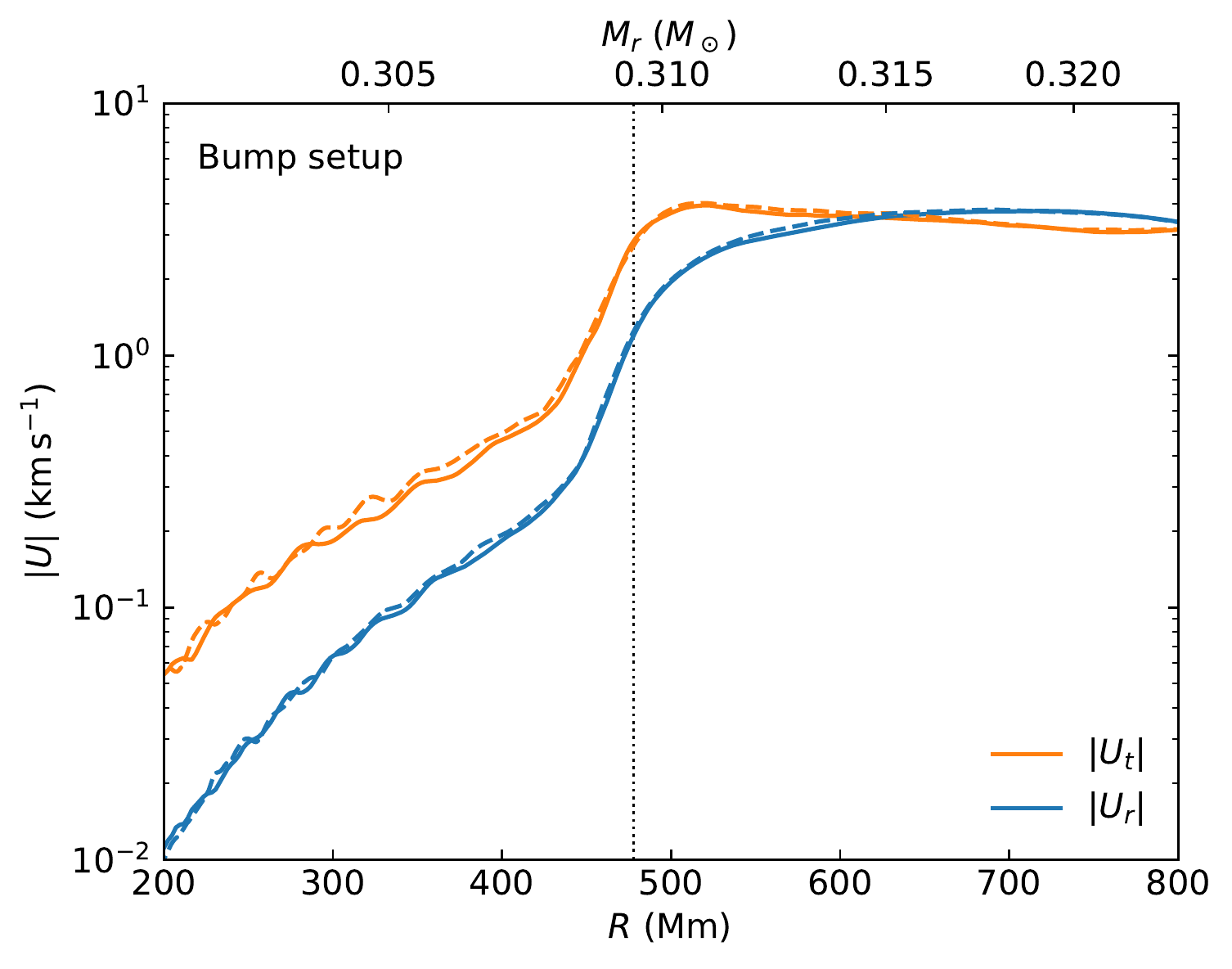}
    \includegraphics[width=\columnwidth]{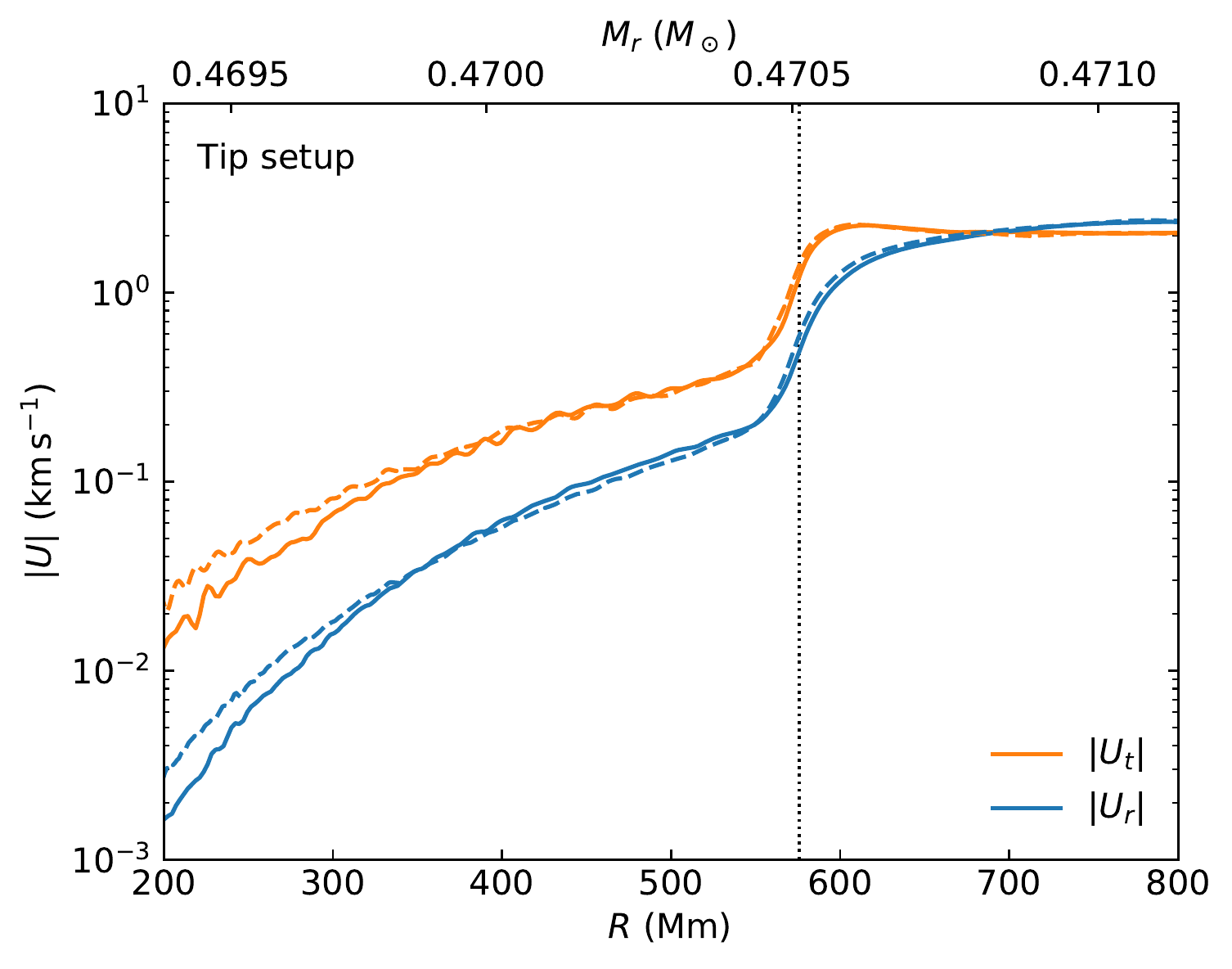}
    \caption{Rms radial and tangential velocity for the bump (top panel, X14 and X22) and tip setups (bottom panel, X24 and X26) at $t=672\,$h (same time as in Figure~\ref{fig:bobs}). Profiles calculated from runs on a 768$^3$ grid are shown as solid lines; those from 1536$^3$ grids are shown as dashed lines. The vertical dotted lines indicate the location of the convective boundaries, determined by finding the location of the maximum $U_t$ gradient as in \protect\cite{jones2017}. Note that the X14 and X22 simulations shown on the top panel were driven with $1000 \times$ the nominal luminosity (see Table~\ref{tab:runs}).}
    \label{fig:velcomp}
\end{figure}

Now that the overall morphology of the flow has been established, we investigate its properties more quantitatively. Figure~\ref{fig:velcomp} displays radial profiles of $|U_t|$ and $|U_r|$ for both the bump and tip setups (see also Figure~\ref{fig:velcomp_mach}, which displays the same quantities but this time in terms of Mach numbers). We omit the $40\,{\rm Mm} < R < 200\,{\rm Mm}$ region in this and subsequent radial profile figures as the behaviour of the flow in this region is tainted by artifacts introduced by the inner boundary conditions of our simulations (Section~\ref{sec:ppmstar}). The results of both the 768$^3$ and 1536$^3$ simulations are shown and the agreement between both grid resolutions is excellent. Above 300\,Mm, the difference never exceeds 20\%. Larger discrepancies are apparent at $R<300$\,Mm for the tip setup. This can be at least partially explained by the fact that a smaller grid cell size is required to satisfactorily resolve the slower flow at small radii (consistent with our discussion of Figure~\ref{fig:scaling_U}), but as we will see numerical heat diffusion also plays a role.

In the convective envelope, the radial and tangential velocity components have similar magnitudes, except near the convective boundary where $|U_t|>|U_r|$. This is the result of the turning of the flow when downdrafts collide against the convective boundary, as described in the previous section and illustrated in Figure~\ref{fig:bobs}. Inside the radiative zone, the tangential component of the velocity remains 2 to 8 times larger than the radial component. This qualitatively matches the expected behaviour for a flow dominated by IGWs, since these waves have higher amplitudes in the tangential direction.

\begin{figure}
    \centering
    \includegraphics[width=\columnwidth]{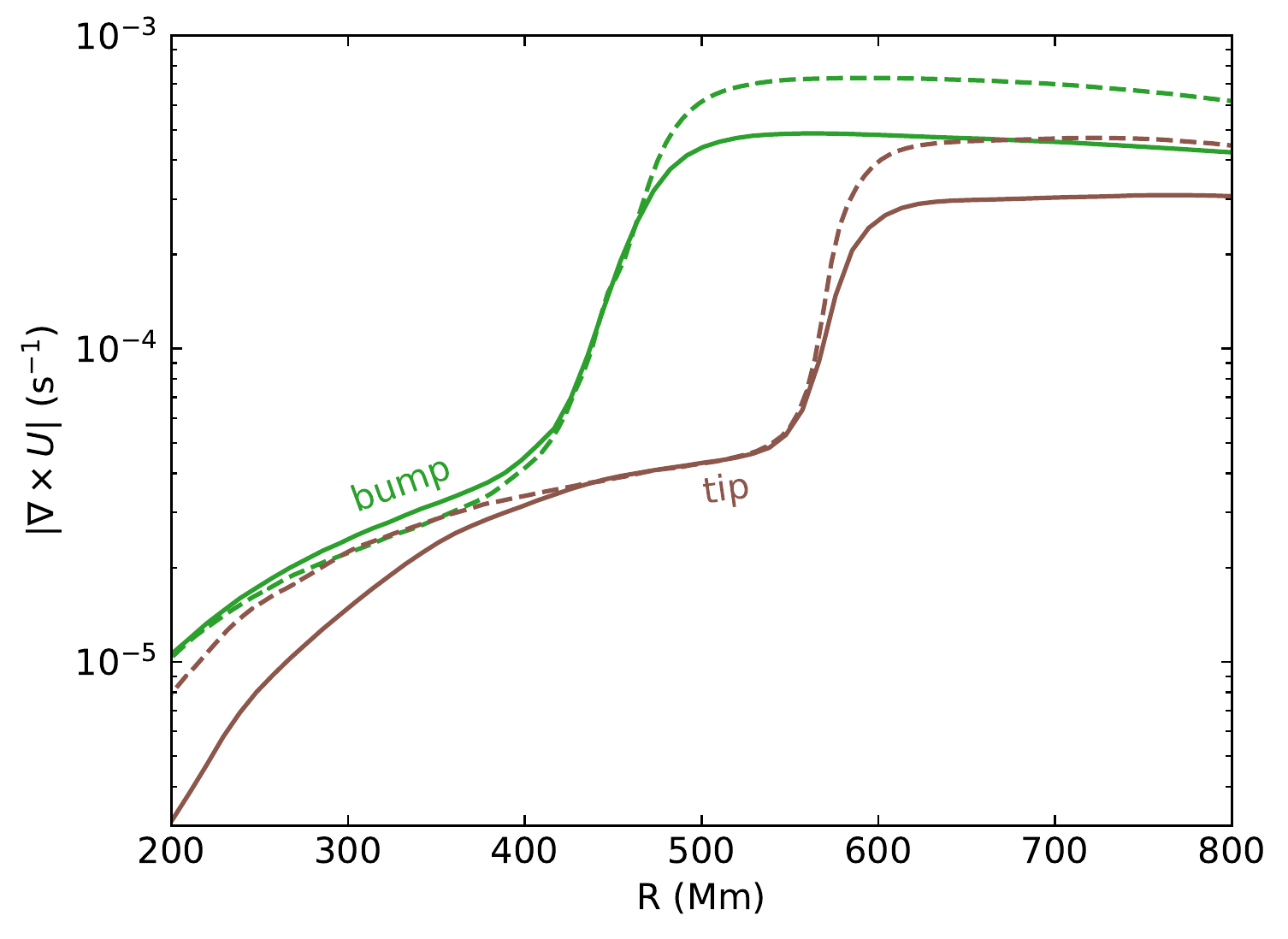}
    \caption{Magnitude of the vorticity for the bump (green) and tip (brown) setups and for runs on 768$^3$ (solid lines) and 1536$^3$ (dashed lines) grids. The vorticity profiles are averaged overs dumps 400 to 420. The simulations shown here are X14, X22, X24 and X26. Note that the bump simulations displayed here were driven with $1000 \times$ the nominal luminosity.}
    \label{fig:vort}
\end{figure}

Figure~\ref{fig:vort} shows radial profiles of the vorticity magnitude, $| \nabla \times U |$, for both setups and two grid resolutions. As expected, the vorticity is much higher in the turbulent convective envelope than in the stable layers dominated by wave-like flows. In the convective envelope, doubling the resolution results in a $\simeq 50\%$ increase of the vorticity. A similar increase was also observed in the convective cores of our recent \code{PPMstar} simulations of massive main sequence stars \citep[Figure~29 of][]{herwig2023}. This is due to the fact that the turbulent cascade in the convective envelope extends to the smallest scales resolved by the simulation grid. As will be shown in Figure~\ref{fig:spectra_rescomp} (see Section~\ref{sec:spectra}), doubling the resolution allows this cascade to extend to smaller spatial scales and therefore increases the vorticity. This behaviour is to be contrasted with what is observed in the stable layers. For the bump setup, the vorticity profiles obtained at 768$^3$ and 1536$^3$ are in excellent agreement. This indicates good convergence with respect to the grid resolution: there are no smaller-scale motions to be resolved. The tip setup behaves somewhat differently as we find progressively larger discrepancies between both grid resolutions below 400\,Mm.

\subsection{Power spectra}
\label{sec:spectra}
\begin{figure}
    \centering
    \includegraphics[width=\columnwidth]{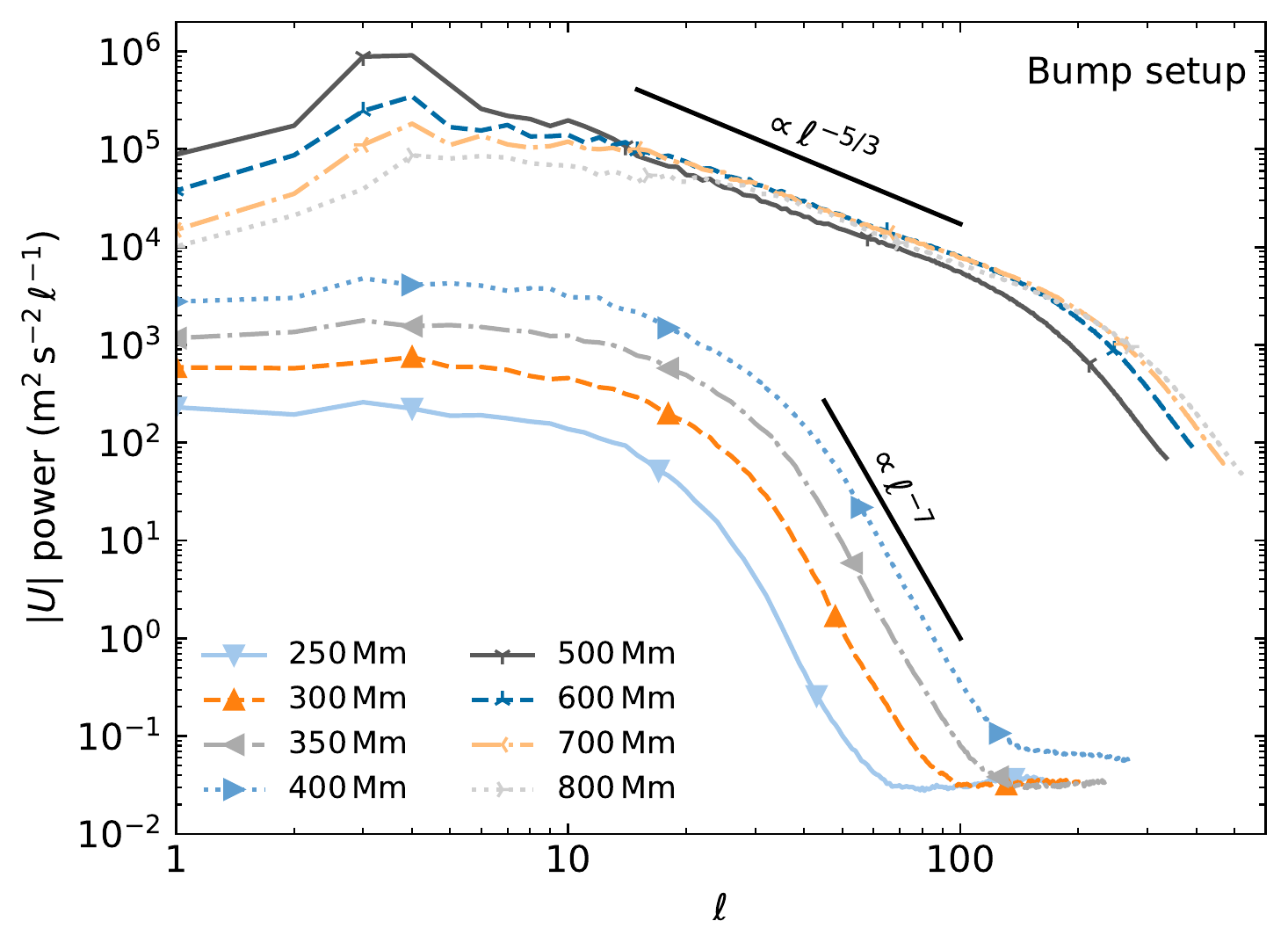}
    \includegraphics[width=\columnwidth]{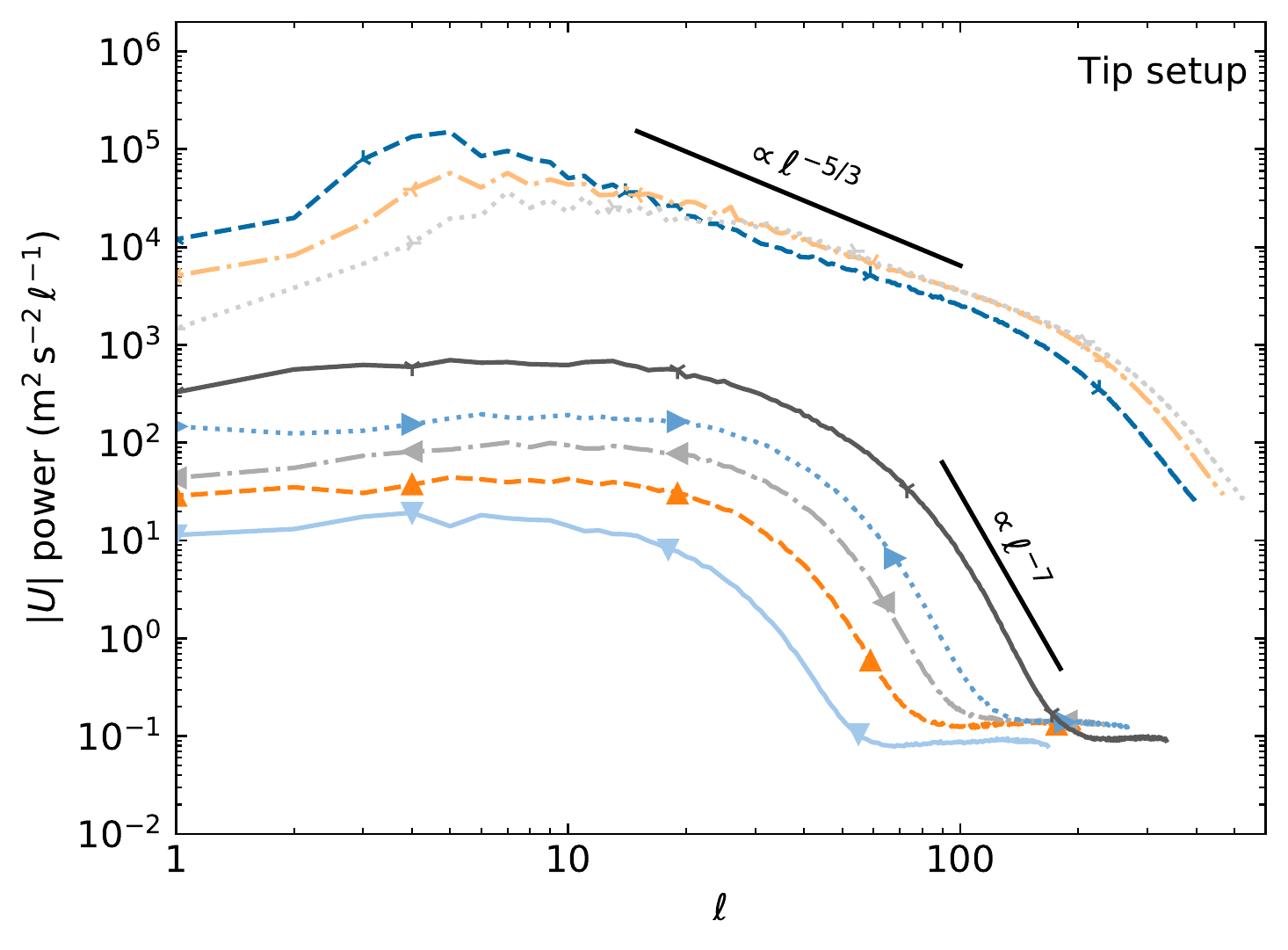}
    \caption{Power spectrum of $|U|$ at different radii for X22 (bump setup, 1536$^3$ grid) and X26 (tip setup, 1536$^3$ grid). The power is binned as a function of the spherical harmonics angular degree $\ell$. The four largest radii are in the convective envelope and the four smallest are in the radiative zone. The spectra were computed by averaging over the last 100~dumps of the simulations. Note that the sole purpose of the symbols on the lines is to help distinguish the curves from one another.}
    \label{fig:power}
\end{figure}

To characterize the spatial structure of the flow, we show in Figure~\ref{fig:power} the power spectra of $|U|$ at different radii for runs X22 and X26 (1536$^3$ grid, bump and tip setups, respectively). In those figures, the power is decomposed into spherical harmonics, each represented by their angular degree (or spherical wavenumber) $\ell$.\footnote{The angular degree $\ell$ is related to the horizontal wavenumber $k_h$ by $k_h = \sqrt{\ell (\ell +1 )}/R$.} The three largest radii are in the convection zone, the four smallest are in the stable layers, and $R=500\,$Mm is in the convection zone for X22 and in the radiative zone for X26 (the convective boundary is at $R_c \simeq 470\,$Mm for X22 and at $R_c \simeq\,580\,$Mm for X26). The maximum $\ell$ value shown in those spectra is set by the highest-degree spherical harmonics that can be resolved given the angular resolution of the Cartesian simulation grid when projected on a sphere at a given radius. Note that the spectra are calculated from the filtered briquette data output \citep{stephens2021}, for which the grid size in each direction is four times smaller than the computational grid. In the convection zone, we recover the expected Kolmogorov spectrum at small spatial scales. The spectra depart from the Kolmogorov $\ell^{-5/3}$ scaling at $\ell \lesssim 4$, which reflects the fact that the $\ell=1$ and $2$ modes are not able to fully develop in the relatively small envelope included in our simulations \citep[but would presumably be dominating the power in the larger convective envelope of a real RGB star,][]{porter2000b,brun2009}. Note that we find the same results if we use a coarser grid resolution, as shown in Figure~\ref{fig:spectra_rescomp}. The only difference is that in the high-$\ell$ limit the spectrum departs from the $\ell^{-5/3}$ scaling at lower $\ell$ since smaller-scale turbulence is not resolved, consistent with our discussion of Figure~\ref{fig:vort} in Section~\ref{sec:prfs}.

In this Section, it will become clear from the properties of the flow, and in particular from the frequencies containing most of the power, that the stable layers are dominated by IGW motions. This IGW-dominated region has a very different spectrum compared to the unstable layers. First, the total power decreases rapidly as we move away from the convective boundary. This signals a strong damping of the IGWs (also visible in Figures~\ref{fig:velcomp}--\ref{fig:vort}), which we attribute to radiative diffusion in Section~\ref{sec:damping}. Secondly, for both setups, the spectrum resembles a broken power law with a flat portion up to $\ell \simeq 10-40$, followed by a very sharp ($\sim \ell^{-7}$) decline at higher wavenumbers. As with the convective layers, the general shape of the spectrum is insensitive to the grid resolution (Figure~\ref{fig:spectra_rescomp}). From this observation, we can infer that the shift in the $\ell$ value where the spectrum goes from flat to rapidly decreasing is not simply due to a change in the effective angular grid resolution with $R$. As they travel toward the centre of the star, the high-$\ell$ waves are more readily damped than their low-$\ell$ counterparts. This IGW spectrum has both important similarities and differences with previous hydrodynamics simulations. The steep power law at large $\ell$ is reminiscent of the \cite{alvan2014} simulations of IGWs in the Sun (where the resonant cavity is similar to that considered here, with IGWs propagating in a radiative zone surrounded by a convective envelope) and of the massive main-sequence star simulations of \cite{rogers2013}. The former find a steep $\sim -5$ to $-7$ power law (see their Figure~15) and the latter find a similarly steep $\sim -4$ to $-6$ power law at large $\ell$ (see their Figure~6). However, a major difference is that both \cite{alvan2014} and \cite{rogers2013} find a spectrum that is monotonically decreasing with respect to $\ell$, whereas we obtain a flat spectrum at low $\ell$.

We note that the comparison of the IGW spectra of Figure~\ref{fig:power} to those presented in \cite{rogers2013} and \cite{alvan2014} is not rigorous as in those works the IGW spectrum is separated into its spatial and temporal dependencies,
\begin{equation}
    E(\ell, \nu) = f(\ell) g(\nu).
    \label{eq:decomposition}
\end{equation}
In Figure~\ref{fig:power}, the dependence on the temporal frequency $\nu$ was ignored but nevertheless affects the power spectra since the spherical wavenumber $\ell$ is coupled to $\nu$ through a dispersion relation. To independently study the dependencies on $\ell$ and $\nu$, \cite{rogers2013} and \cite{alvan2014} decompose their spectra into the form given by Equation~\eqref{eq:decomposition} using singular value decomposition. Naturally, the spectrum cannot be entirely separated as in Equation~\eqref{eq:decomposition}, and in our case we found that the singular value decomposition led to a poor representation of the full spectrum. We therefore opted not to use singular value decomposition to present our IGW spectra. The same considerations apply to the temporal frequency power spectra presented below.

\begin{figure}
    \centering
    \includegraphics[width=\columnwidth]{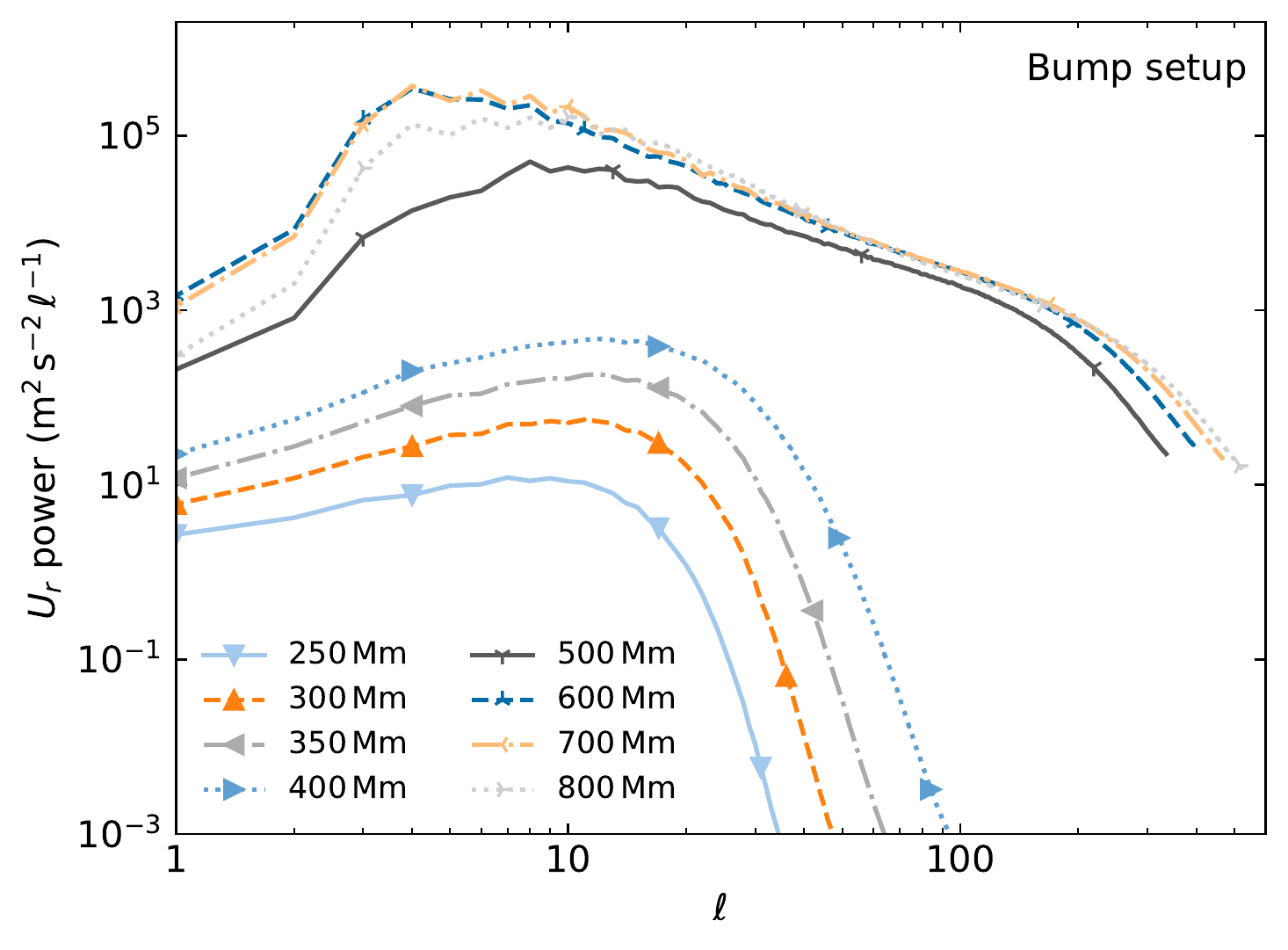}
    \includegraphics[width=\columnwidth]{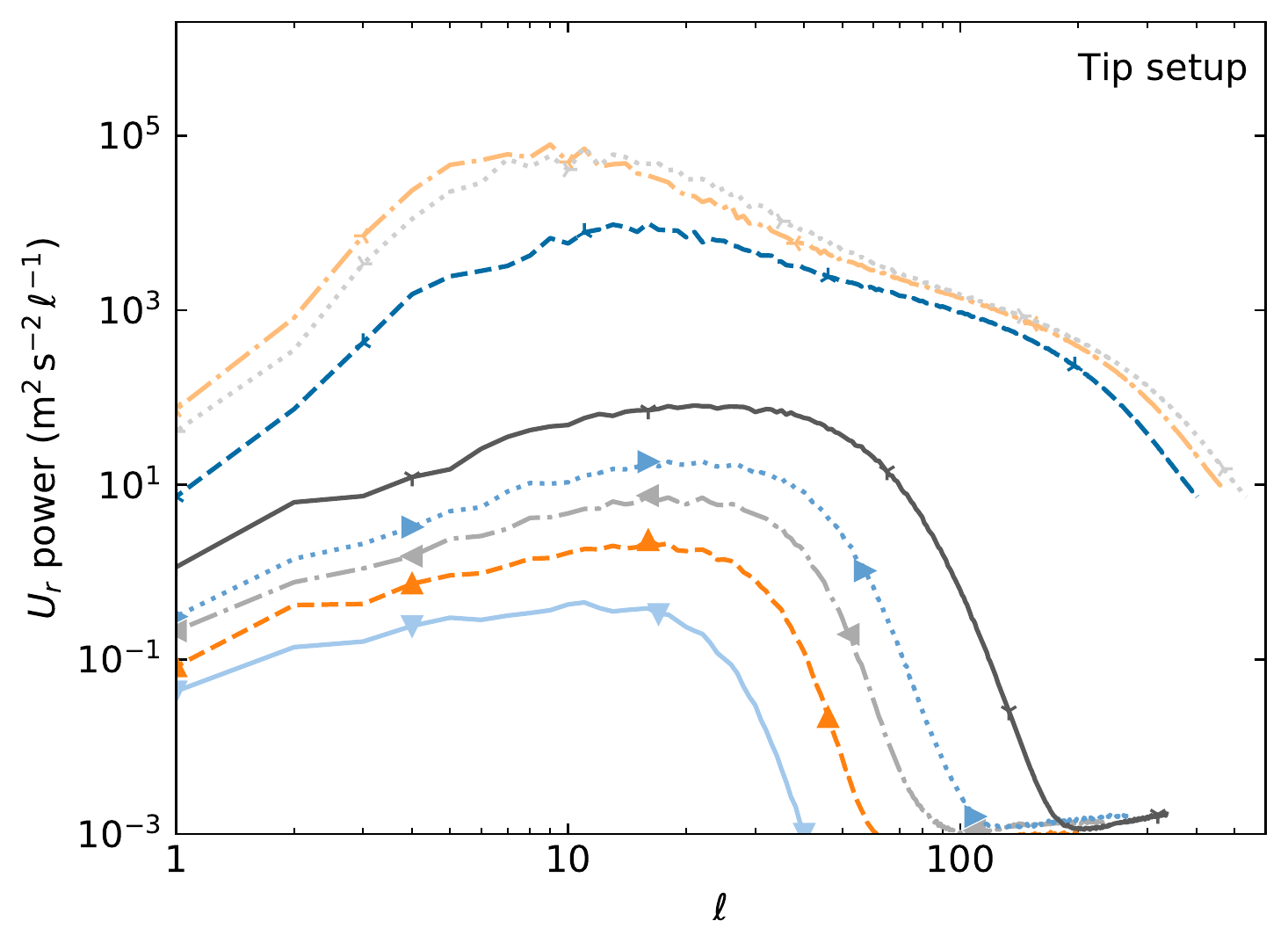}
    \caption{Identical to Figure~\ref{fig:power}, but this time only the radial velocity component is considered in the calculation of the power spectra.}
    \label{fig:Ur_power}
\end{figure}

In Figure~\ref{fig:power}, we showed the spectra of the total power, that is, due to displacements both in the radial and horizontal directions. It is instructive to compare those spectra to spectra computed by considering only the radial velocity component (Figure~\ref{fig:Ur_power}). The most important difference is that instead of being flat at low $\ell$, the $U_r$ IGW spectra instead increase up to $\ell \simeq 10-40$. This behaviour is expected. For IGWs, the ratio $U_r/U_t$ increases with frequency (see \citealt{herwig2023}) as the waves approach the Brunt--V\"ais\"al\"a frequency and the vertical motions become more important compared to the horizontal motions. This can explain the positive slope observed in Figure~\ref{fig:Ur_power}. A similar trend is also observed in the IGW spectra of our massive main sequence stars simulations \citep[Figure~19 of][]{herwig2023} and in recent 3D simulations of late-type F stars \citep[][Figure~8]{breton2022}.

\begin{figure}
    \centering
    \includegraphics[width=\columnwidth]{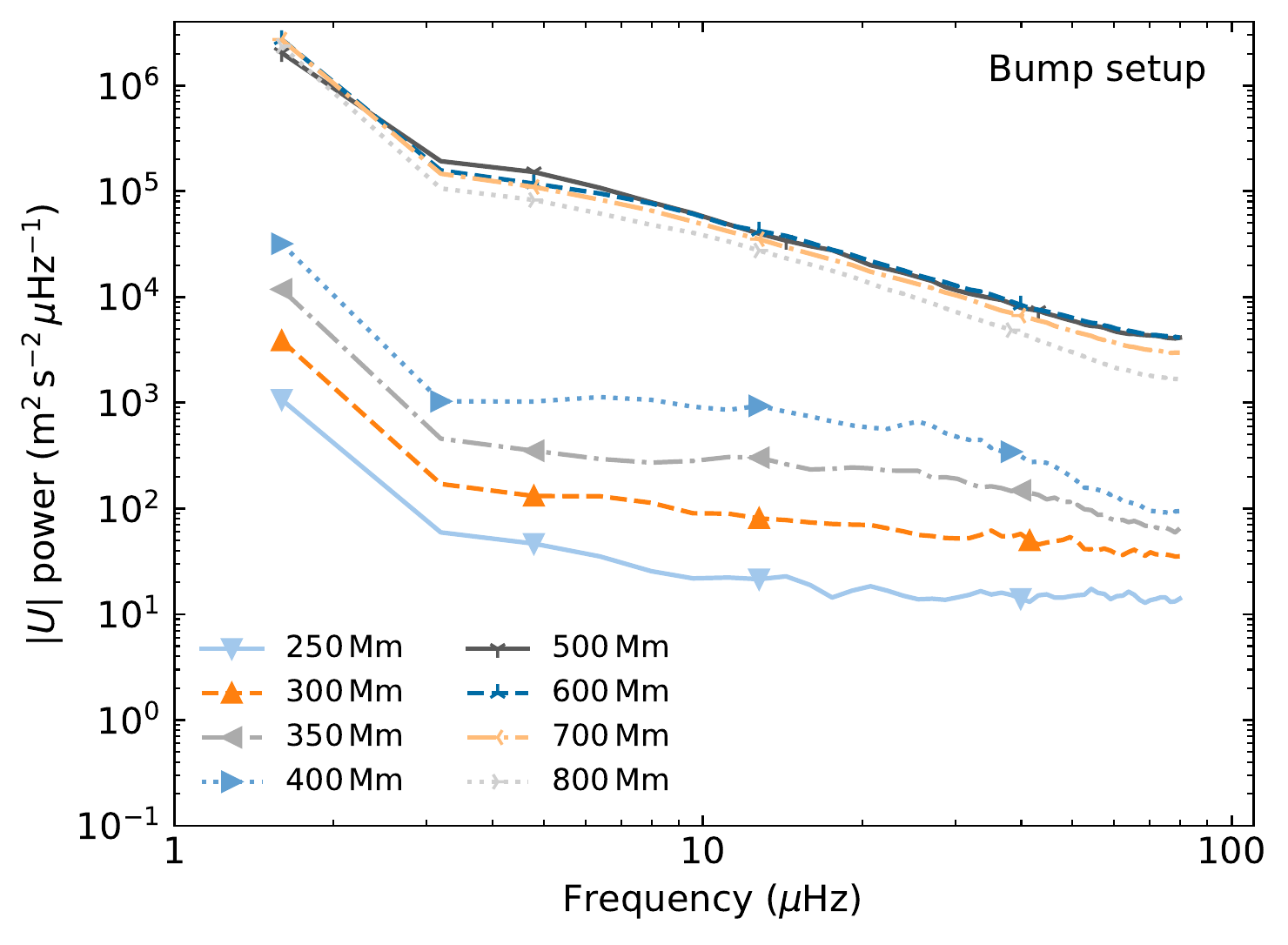}
    \includegraphics[width=\columnwidth]{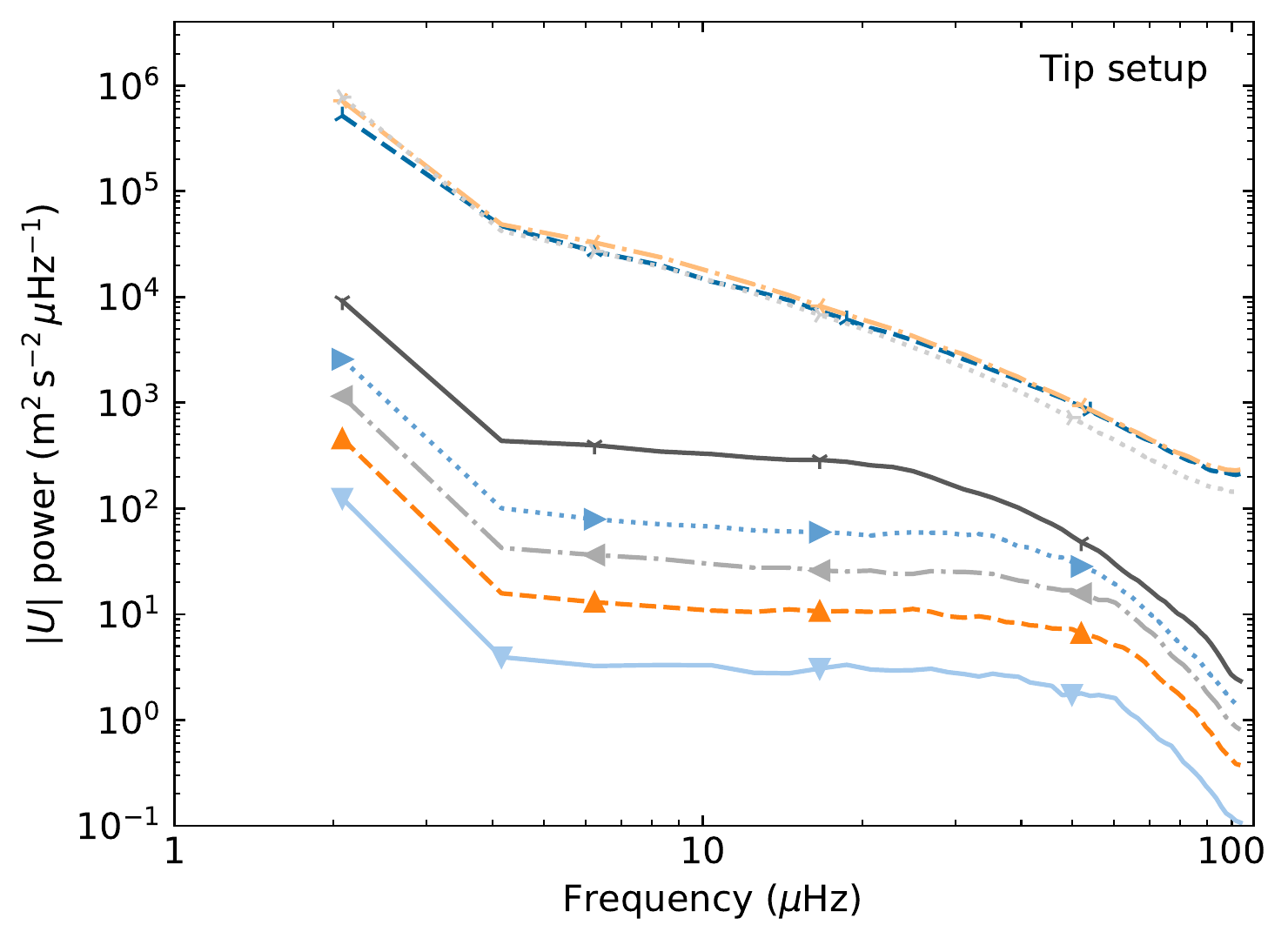}
    \caption{Frequency power spectrum at different radii for X22 (bump setup, 1536$^3$ grid) and X26 (tip setup, 1536$^3$ grid). The spectra were computed by considering the last 100~dumps of the simulations.}
    \label{fig:Ur_power_freq}
\end{figure}

In Figure~\ref{fig:Ur_power_freq} we now show the power spectra for the same radii as in Figures~\ref{fig:power} and~\ref{fig:Ur_power}, but this time in the temporal frequency space. The spectra are truncated at $\nu = 80\,\mu$Hz (bump setup) and $\nu = 104\,\mu$Hz (tip setup), which correspond to the Nyquist cut-off frequencies given the time interval that separates each dump in our simulations (i.e., higher frequency modes cannot be resolved). To be clear, higher frequencies are resolved in the simulations themselves, but the detailed outputs that allow us to reconstruct the power spectra are written to disk only every $\sim 2000$ time steps. In the stable layers, the spectra are relatively flat, although for the tip setup there is a noticeable decrease of the power at high frequencies. This high-frequency quenching is more pronounced close to the convective boundary. This is consistent with these waves being IGWs. IGWs can only propagate when their frequencies are smaller than the local Brunt--V\"ais\"al\"a frequency. Here, the relative amount of power at high frequencies grows as we move inward and $N$ increases (Figure~\ref{fig:Nbase}).

It is instructive to compare the frequencies that contain most of the IGW power to the convective frequency $\nu_c$ at the bottom of the envelope. The convective turnover timescale is not a precisely defined quantity, but as the convective cells occupy the maximum space available to them in the envelope, we can estimate it by taking the thickness of the envelope ($\simeq 400\,$Mm for X22 and 300\,Mm for X26) and dividing it by the average convective velocity ($|U|\simeq 5\,{\rm km}\,{\rm s}^{-1}$ for X22 and 3\,km\,s$^{-1}$ for X26). This yields a convective turnover timescale of 22\,h ($\nu_c \simeq 13\,\mu$Hz) for the bump setup and 28\,h ($\nu_c \simeq 10\,\mu$Hz) for the tip setup. Figure~\ref{fig:Ur_power_freq} shows that a large fraction of the IGW power is contained at frequencies that exceed $\nu_c$. This result is consistent with the fact that the power spectra in the convective envelope are flat (Figure~\ref{fig:Ur_power_freq}). If there is no correlation between the convective turnover timescale and the convective spectrum, then we also expect to observe no correlation between the convective turnover timescale and the IGW spectrum.

Comparing our wave spectra to asteroseismological observations of RGB stars would be interesting, but this exercise is complicated by the fact that the IGWs propagating in the radiative interior are coupled with pressure modes in the envelope \citep[those are known as ``mixed'' modes,][]{aerts2010,hekker2017}. Because of this and given the truncation of the envelope in our simulations, we cannot make a direct comparison between the frequencies of the mixed modes detected in upper RGB stars and the frequency spectra of Figure~\ref{fig:Ur_power_freq}. Nonetheless, we note that the frequency at maximum oscillation power for mixed modes in a $1.2\,M_{\odot}$ RGB star at the bump luminosity, $\nu_{\rm max} \simeq 40\,\mu$Hz \citep[Figure~1 of][]{khan2018}, is within the range of frequencies where IGWs are excited in our simulations (Figure~\ref{fig:Ur_power_freq}).

\begin{figure}
    \centering
\includegraphics[width=\columnwidth]{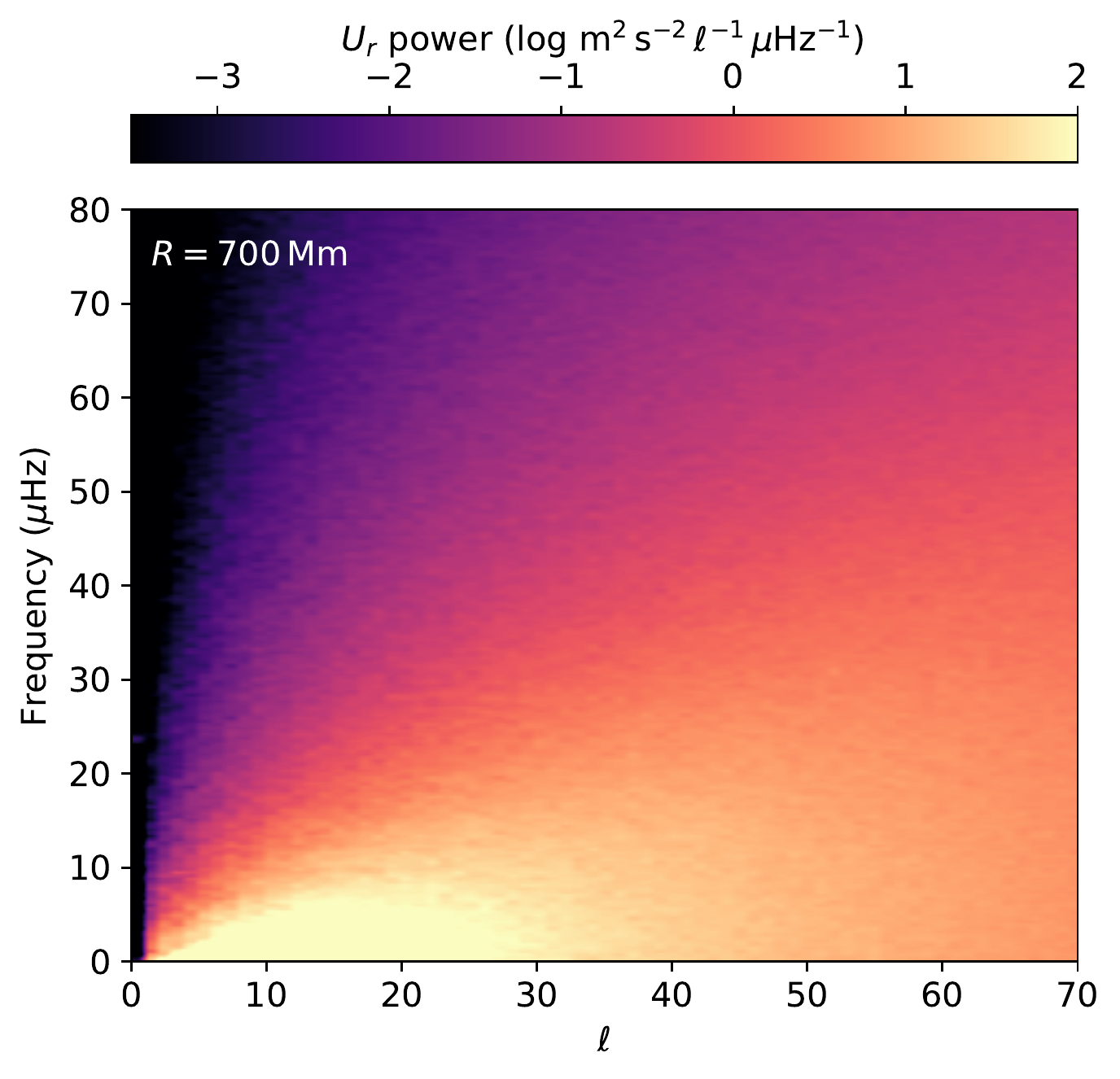}
    \includegraphics[width=\columnwidth]{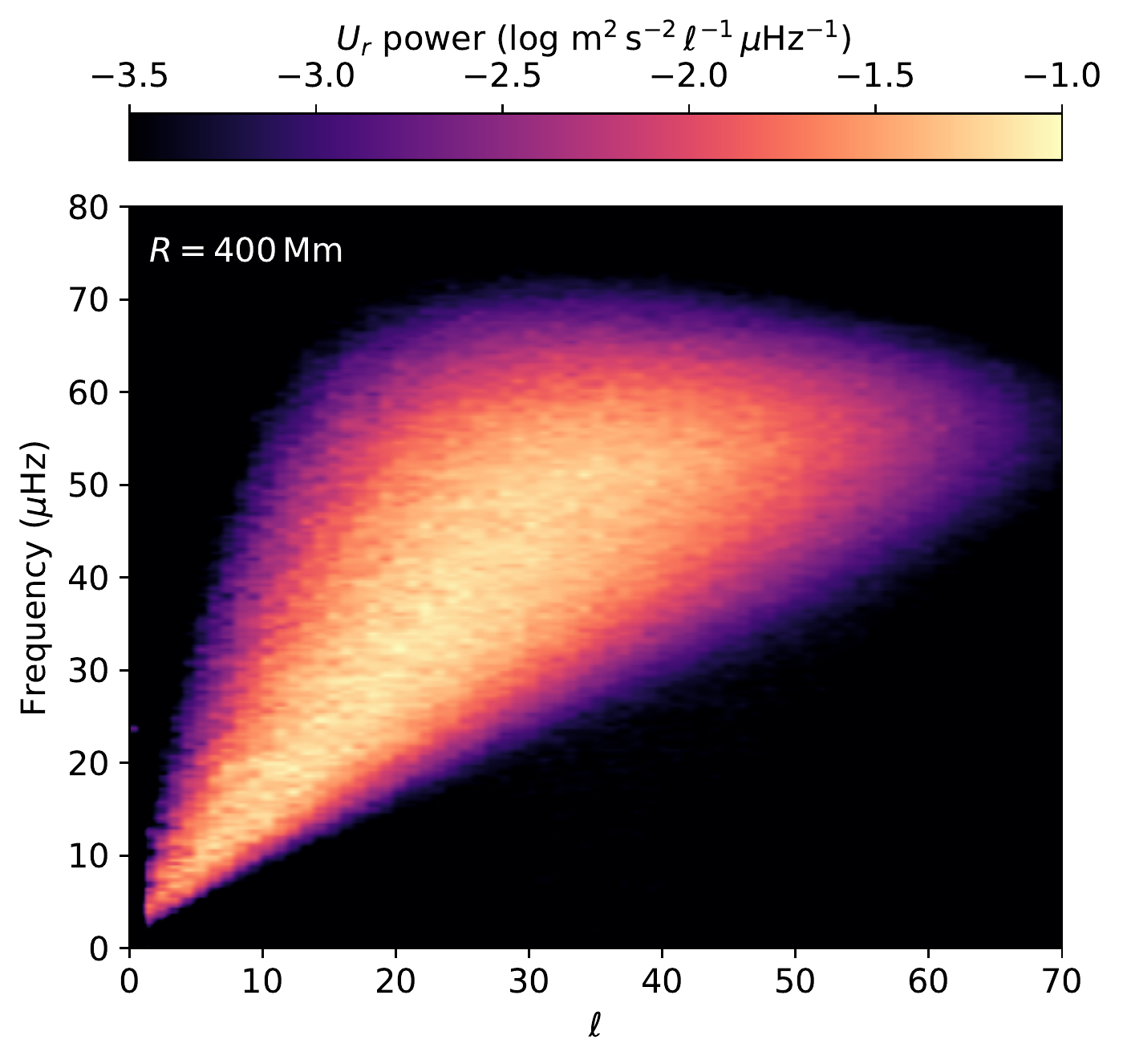}
    \caption{Power spectra of $U_r$ at $R=700\,$Mm (top panel) and $R=400\,$Mm (bottom panel) as a function of the angular degree $\ell$ and the temporal frequency for X24 (tip setup, 768$^3$ grid). The spectra were obtained by considering the last 600~dumps of the simulation. Note the different colour scale between the two panels.}
    \label{fig:kw}
\end{figure}

To conclude our analysis of the wave spectra, we show in Figure~\ref{fig:kw} power spectra of the radial velocity component as a function of both the angular degree and the frequency for an RGB tip simulation (X24). We use our longest simulation for this analysis in order to attain a finer frequency sampling in the Fourier decomposition. The power spectrum in the convective envelope (top panel) differs greatly from the spectrum obtained in the stable layers $\simeq 0.8\,H_P$ below the convective boundary (bottom panel). In the convection zone, the power spectrum is very smooth and has no specific features, as expected for turbulence. In the IGW-dominated region, we see a more distinctive power distribution, with no power at high frequencies (due to the quenching of IGWs with frequencies exceeding $N$) and in the low-frequency, high-$\ell$ region of the diagram. The most striking aspect of this wave spectrum is it blurriness. In contrast, hydrodynamics simulations of IGWs in stellar radiative interiors usually yield spectra where the power is predominantly contained in a set of discrete, well-defined ridges in the $\ell-\nu$ space \citep{alvan2014,alvan2015,rogers2013,horst2020,thompson2023}, with each ridge corresponding to a specific radial order of standing IGW modes (or $g$ modes). The absence of such ridges in Figure~\ref{fig:kw} suggests that standing modes are not formed in our simulations or, in other words, that the mode lifetime is very short.

For standing waves to form, two progressive (or travelling) IGWs have to constructively interfere with each other. Here, this would require that IGWs excited at the convective boundary and travelling inward undergo a reflection somewhere close to the centre of the star. This reflection would generate outward travelling waves that could constructively interfere with the inward travelling waves to create standing modes. The absence of standing modes in Figure~\ref{fig:kw} therefore suggests that only inward moving progressive waves exist in our simulations. This situation is analogous to that described in \cite{alvan2015} for a solar-like star, where low-frequency IGWs are damped before reaching their reflection point near the centre. In the solar case, the reflection point is located where the Brunt--V\"ais\"al\"a frequency $N$ becomes equal to the IGW frequency. Here, $N$ remains larger than the IGW frequency all the way to the inner simulation radius, effectively meaning that the waves can only be reflected at that radius (remember that we use reflective boundary conditions). But Figures~\ref{fig:power} to~\ref{fig:Ur_power_freq} show that the power contained in the IGWs is strongly damped as the waves propagate inward, which most likely explains the absence of reflected waves (and, by extension, of standing modes).

\subsection{Internal gravity wave damping}
\label{sec:damping}
\begin{figure}
    \centering
    \includegraphics[width=\columnwidth]{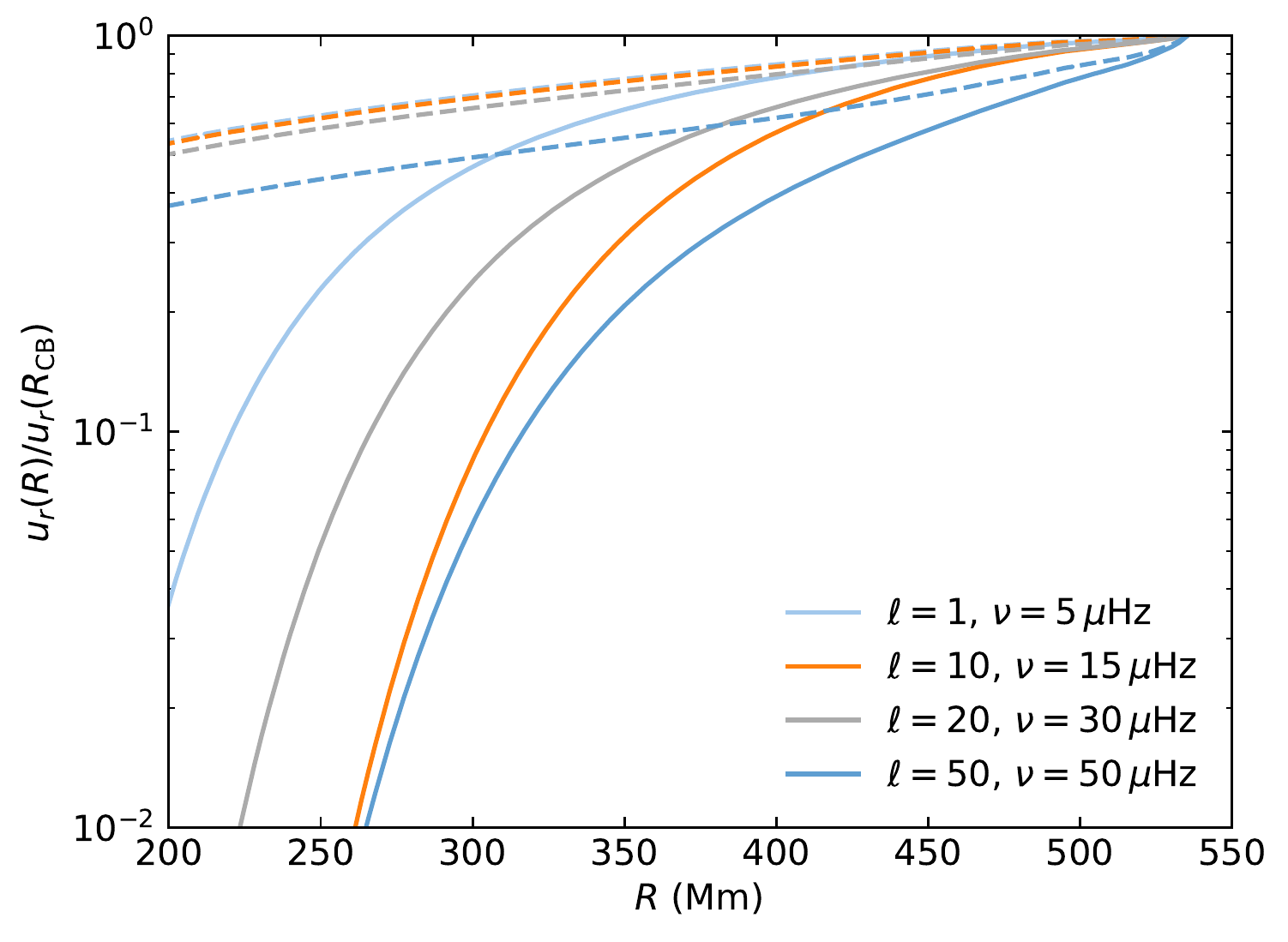}
    \caption{Predicted decrease of the velocity of individual IGWs as they travel inward from the convective boundary in the tip setup. The dashed lines correspond to the case without radiative diffusion (constant wave luminosity, Equation~\ref{eq:L2}), while the solid lines show the case with radiation diffusion. The $(\ell,\nu)$ pairs shown here are typical values for the IGWs observed in our simulations (see Figure~\ref{fig:kw}).}
    \label{fig:damping}
\end{figure}

In the previous sections, we have seen how the amplitude of IGWs drops as they propagate inward (Figures~\ref{fig:velcomp}--\ref{fig:Ur_power_freq}). What is the cause of this decrease? In a simulation without any heat diffusion and without any nonlinear interactions between waves, we expect the luminosity of each wave (i.e., the kinetic energy transported by wave packets moving radially at the group velocity $U_{{\rm gr},r}$),
\begin{equation}
L^{\rm IGW}_{\ell,\nu} = 4 \pi R^2 \rho (U_r^2 + U_h^2) U_{{\rm gr},r},
\label{eq:L1}
\end{equation}
to remain constant as a function of $R$. Using the IGW dispersion relation \citep[e.g.,][]{press1981}, Equation~\eqref{eq:L1} can be expressed more conviently as
\begin{equation}
L^{\rm IGW}_{\ell,\nu} = 4 \pi R^3 \rho U_r^2 \frac{N}{\sqrt{\ell ( \ell +1 )}} \sqrt{ 1 - \omega^2/N^2},
\label{eq:L2}
\end{equation}
where $\omega= 2 \pi \nu$. In Figure~\ref{fig:damping}, the dashed lines show the decrease in $U_r$ predicted using this equation for four different $(\ell,\nu)$ pairs representative of the IGWs present in our simulations. The IGW velocities are predicted to only decrease by a factor $\sim 2$ between the convective boundary and $R=200\,{\rm Mm}$, a much more modest effect than the two orders of magnitude drop observed in Figure~\ref{fig:velcomp}. A nonadiabatic effect must therefore be invoked to explain the wave amplitude damping observed in our simulations.

If radiative diffusion is considered, the wave velocities are damped by an additional attenuation factor $e^{-\tau}$, where $\tau$ is analogous to an optical depth and is given by \citep{zahn1997}
\begin{equation}
\tau = \frac{\left[ \ell (\ell+1) \right]^{3/2}}{2} \int_{R_{\rm CB}}^R K \frac{N N_T^2}{\omega^4} \sqrt{ \frac{N^2}{N^2-\omega^2}} \frac{ |dr|}{r^3},
\end{equation}
where $K$ is the thermal diffusivity,
\begin{equation}
K = \frac{4 a c T^3}{3 \kappa c_{\scriptscriptstyle P} \rho^2},
\end{equation}
with $a$ the radiation constant, $c$ the speed of light, $\kappa$ the Rosseland mean opacity and $c_{\scriptscriptstyle P}$ the specific heat at constant pressure, and $N_T$ is the thermal component of $N$,
\begin{equation}
N_T^2 = -\frac{g}{H_P} \frac{\partial \ln \rho}{\partial \ln T} \left( \nabla_{\rm ad} - \nabla \right),
\end{equation}
where the temperature gradients have their usual meanings. The solid lines of Figure~\ref{fig:damping} show what happens if we include this radiative damping in our predictions of the IGW velocities. A sharp $\sim 100 \times$ damping is predicted between the convective boundary and $R \simeq 200\,{\rm Mm}$, which agrees very well with our simulations (compare to Figure~\ref{fig:velcomp}).

\begin{figure}
    \centering
    \includegraphics[width=\columnwidth]{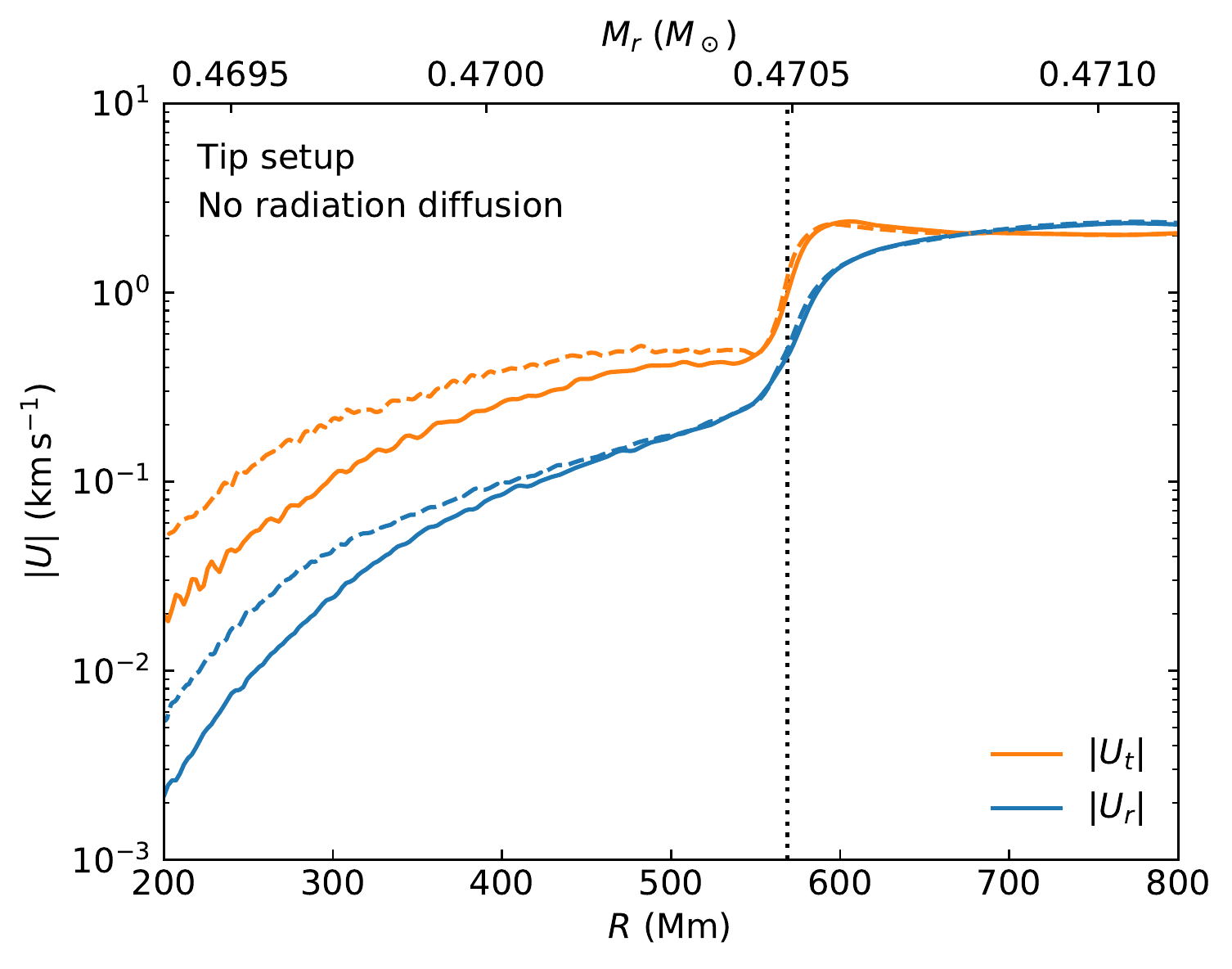}
    \caption{Rms radial and tangential velocity for the tip setup simulations without radiative diffusion (X32 and X33) at $t=629\,$h. Solid lines are for X32 (768$^3$ grid) and dashed lines are for X33 (1536$^3$ grid). The vertical dotted lines indicate the location of the convective boundaries.}
    \label{fig:velcomp_norad}
\end{figure}

To verify that radiative damping is indeed the mechanism leading to the IGW velocity decrease, we have performed two additional nominal-luminosity RGB tip simulations without radiative diffusion (X32 and X33, respectively on $768^3$ and $1536^3$ grids). Heating at the inner boundary is turned off for these runs, since with $K=0$ this extra heat would be spuriously trapped in the radiative cavity. Figure~\ref{fig:velcomp_norad} compares the spherically averaged velocity profiles of X32 and X33. It is entirely analogous to the bottom panel of Figure~\ref{fig:velcomp}: the same grid resolutions are considered and the same simulation setup is used. The only difference is the omission of radiative diffusion. Clearly, the IGW velocities still undergo a considerable decrease between the convective boundary and $R=200\,{\rm Mm}$ even if radiative diffusion is omitted. Remember that only a factor $\sim 2$ decrease is expected in the purely adiabatic case (dashed lines in Figure~\ref{fig:damping}). The stronger decline must be due to a nonadiabatic effect that we attribute here to the numerical diffusion of entropy, which mimics radiative diffusion. This interpretation is supported by the fact that the velocities decline faster when the grid resolution is lower (compare the dashed and solid lines in Figure~\ref{fig:velcomp_norad}), pointing to a numerical effect. Note that we have also observed numerical heat diffusion in our recent core-convection simulations \citep{herwig2023}. 

We stress that the observed velocity damping is not due to numerical viscosity. If it were the case, we would expect to also see the same significant difference between both grid resolutions for simulations where radiative diffusion is included, as this effect operates independently from radiative diffusion. But this is not the case: the difference between the $768^3$ and $1536^3$ grid resolutions is much smaller when the simulations include radiative diffusion (Figure~\ref{fig:velcomp}). This observation is naturally explained by numerical heat diffusion. Numerical heat diffusion is hardly noticeable in Figure~\ref{fig:velcomp} because radiative diffusion has a considerably larger effect. In contrast, numerical heat diffusion is very important in the case where $K=0$, since it becomes the only mechanism available for entropy diffusion. 

Three important conclusions follow from the analysis presented in this section. First, numerical heat diffusion operates in our simulations, but is much weaker than radiative diffusion and can therefore be ignored. Secondly, radiative damping can naturally explain most of the decrease of the IGW velocities observed in our simulations. Thirdly, the spurious behaviour of the flow at the inner simulation boundary (Section~\ref{sec:ppmstar}) can be ignored, since radiative damping strongly suppresses the IGWs before they reach that radius.

\section{Mixing by internal gravity waves}
\label{sec:IGWmixing}
Now that we have established the main properties of the flow, we turn to the problem of estimating the mixing enabled by IGWs in the radiative zone. To do so, we first perform an estimate of the diffusion coefficient $D$ based on the vorticities and a simple analytical prescription (Section~\ref{sec:Dzahn}), and we then attempt to measure $D$ more directly by studying the time evolution of a passive fluid added to our simulations (Section~\ref{sec:gaussians}). As we will see, the second, more robust method disagrees with the first.

\subsection{Estimating the mixing from the vorticity}
\label{sec:Dzahn}
\subsubsection{Theoretical framework}
IGWs can cause vertical mixing in the stable layers of stellar interiors if their horizontal velocity shear is high enough to overcome the tendency of the fluid to remain stratified. In the adiabatic case (no heat diffusion), this is expected to take place only if the Richardson number drops below $1/4$,
\begin{equation}
    {\rm Ri} = \frac{N^2}{\left( d U_t / d R \right)^2} < \frac{1}{4}.
\label{eq:Ri}
\end{equation}
In our RGB stars, ${\rm Ri} > 1/4$ throughout the radiative zone (except within $\simeq 0.1\,H_p$ from the convective boundary) and no vertical mixing is therefore expected based on this simple criterion. But in real stars (and in our simulations), radiative diffusion modifies Equation~\eqref{eq:Ri} and may allow mixing to take place at larger Ri \citep{zahn1974}. The diffusion coefficient due to IGW shear proposed by \cite{zahn1992} is then given by
\begin{equation}
    D \simeq \eta \frac{K}{{\rm Ri}},
\end{equation}
where $\eta$ is a dimensionless parameter of order 0.1 \citep{prat2013,prat2016,garaud2016}. In the present context, this equation can be written as \citep[Section~2.4 of][]{herwig2023}
\begin{equation}
    D \simeq 0.1 \frac{K \left( \nabla \times U \right)^2}{N^2},
    \label{eq:zahn}
\end{equation}
where we have also assumed that $\eta = 0.1$.

\subsubsection{IGW vorticity scaling for the bump setup}
\label{sec:heating_series}
In order to estimate $D$ using Equation~\eqref{eq:zahn}, we need the vorticity profiles $|\nabla \times U|$ from our simulations. We have already shown those quantities in Figure~\ref{fig:vort}, but only the vorticity profile for the tip setup is directly usable. As the RGB bump simulations are driven with a luminosity that is much higher than the nominal value (see Table~\ref{tab:runs}), the flow velocities and vorticities are necessarily overestimated.

\begin{figure}
  \includegraphics[width=\columnwidth]{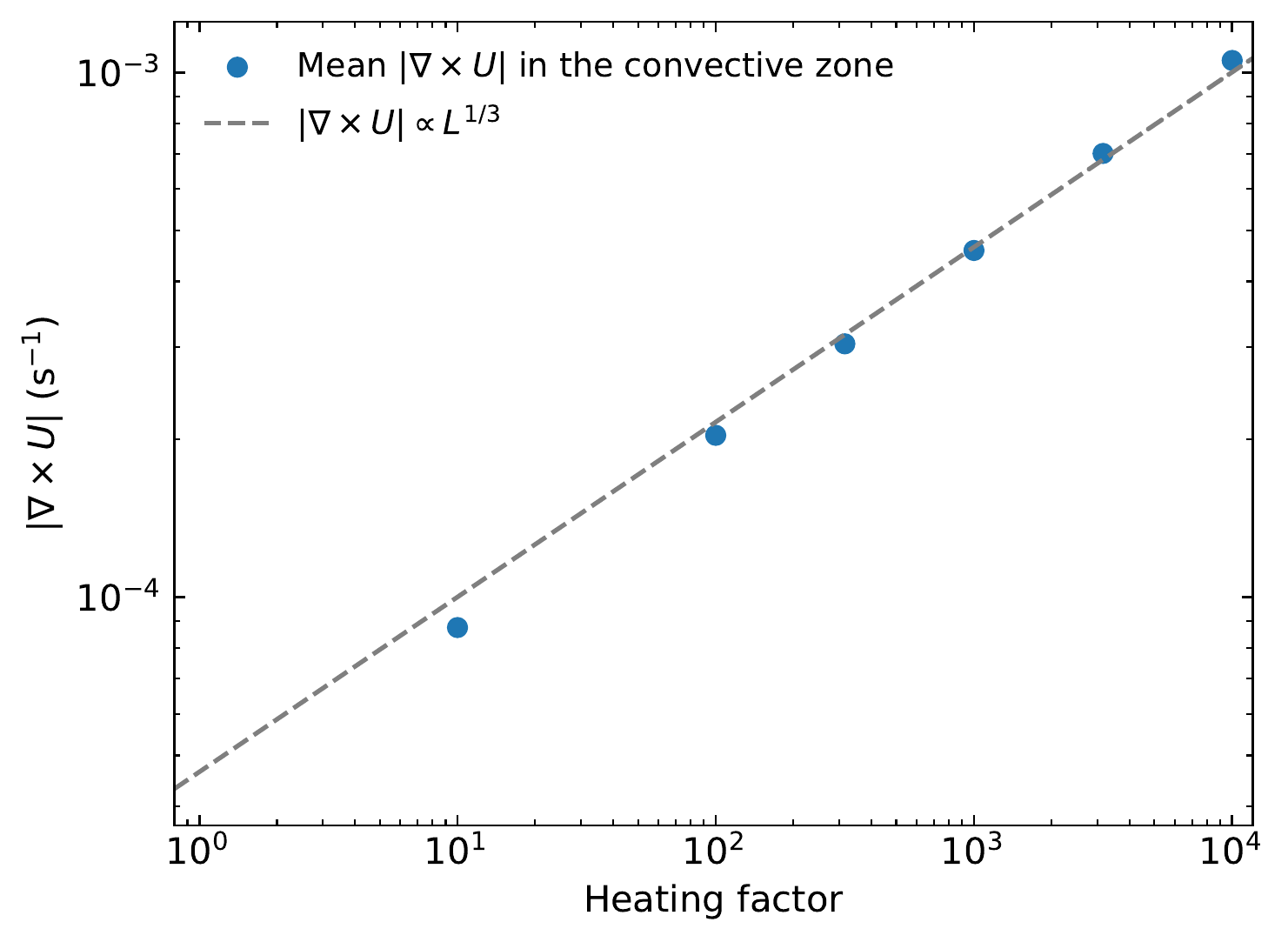}
  \includegraphics[width=\columnwidth]{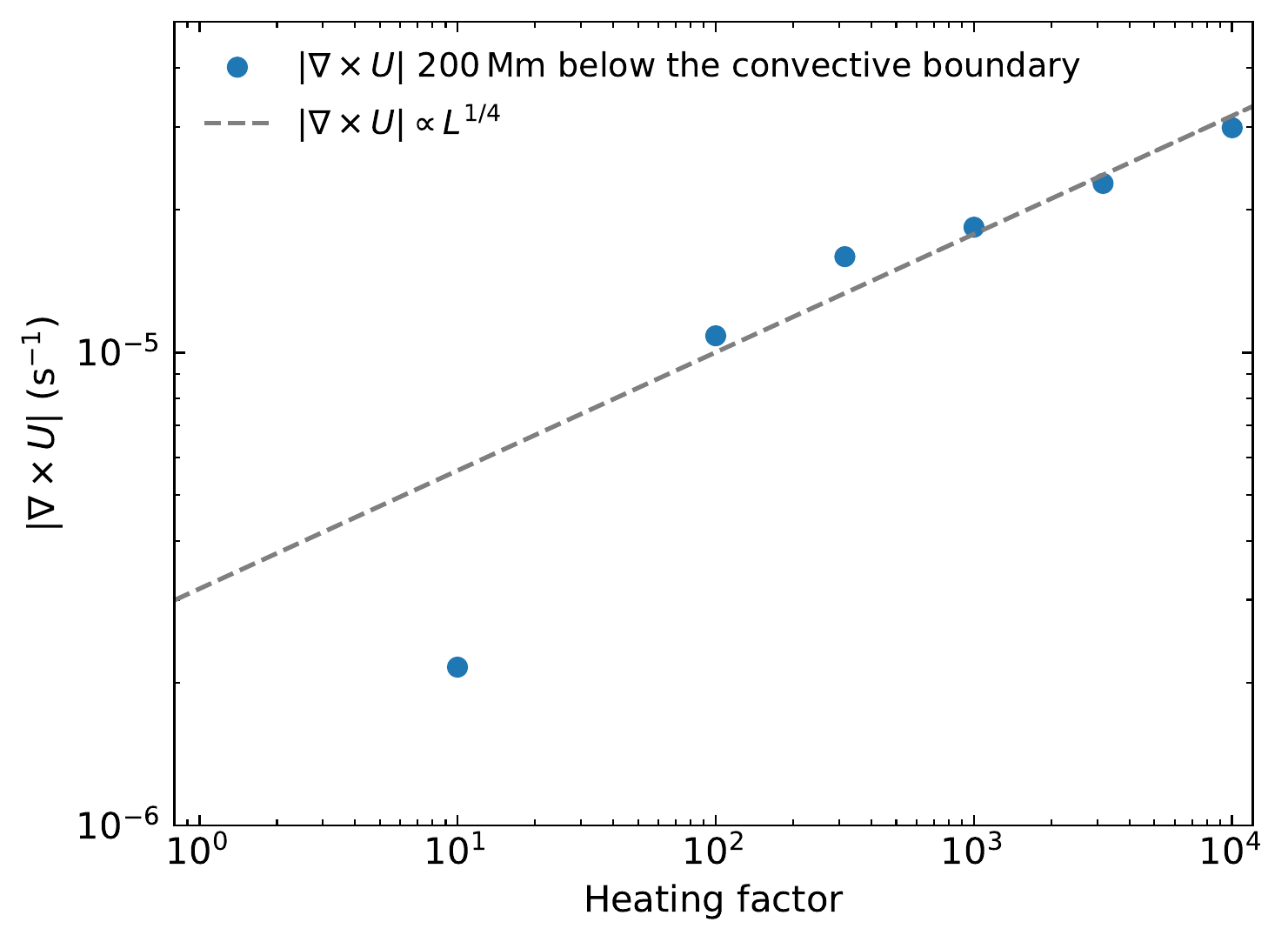}
    \caption{Mean vorticity in the convection zone (top panel) and in the stable layers 200\,Mm ($\simeq 1\,H_P$) below the convective boundary (bottom panel) as a function of the boost factor applied to the luminosity. The vorticities are averaged over the same dumps as in Figure~\ref{fig:scaling_U}. These simulations are all for the RGB bump setup and a 768$^3$ grid.}
    \label{fig:scaling_vort}
\end{figure}

To extrapolate our RGB bump results to nominal luminosity, we use our series of 768$^3$ simulations with different heating rates. The mean vorticity measured in the convection zone is shown in the top panel of Figure~\ref{fig:scaling_vort}, where it can be seen that it follows a $L^{1/3}$ scaling law as previously established in the case of core convection in massive main sequence stars \citep{herwig2023}. This is consistent with Figure~\ref{fig:scaling_U}, which shows that the velocities also scale with $L^{1/3}$, as observed in previous 3D hydrodynamics simulations (see Section~\ref{sec:ppmstar}). As argued by \cite{herwig2023}, the fact that both the velocity and the vorticity scale with $L^{1/3}$ requires that the spatial spectrum in the convection zone is independent of the heating factor.

We find a shallower $\sim L^{1/4}$ scaling when we measure the vorticity $200\,{\rm Mm} \simeq 1\,H_P$ below the convective boundary (bottom panel of Figure~\ref{fig:scaling_vort}). Note that the vorticities for the lowest heating factors should be interpreted with caution as they correspond to very low velocities (${\rm Ma} \lesssim 10^{-4}$ for $L \leq 100 L_{\star}$), a regime where the validity of our simulations is questionable. This explains why the $L=10 L_{\star}$ point does not line up with the trend established at higher luminosities (see also Figure~\ref{fig:scaling_U}). Note also that this $L^{1/4}$ scaling law is not well defined (e.g., $ L^{1/5}$ would give a similar fit to the data), but we assume $ L^{1/4}$ in what follows. This result strongly differs from the $L^{2/3}$ scaling found in \cite{herwig2023}. As shown in Section~\ref{sec:ppmstar}, we also observe a different scaling of the IGW velocities ($L^{1/3}$ here compared to $L^{2/3}$ in the core convection case) and in this respect it is not surprising to also find a different scaling of the vorticities.

However, unlike \cite{herwig2023}, we do not observe the same scaling for the velocities and for the vorticities. This implies that the spatial IGW spectrum changes as a function of heating rate. Figure~\ref{fig:spectra_heating} shows the $|U|$ power spectra measured 200\,Mm below the convective boundary (as in Figures~\ref{fig:scaling_U} and~\ref{fig:scaling_vort}) for different heating rates. It reveals that there is indeed a strong dependence of the power spectra on the heating rate, with more power at low $\ell$ relative to high $\ell$ when the luminosity is increased. This is qualitatively similar to the findings of \citet[][Figure~5]{saux2022}, but this comparison is misleading for at least two reasons. First, the IGW spectra measured in our simulations (Figure~\ref{fig:kw}) are very different from those reported in \cite{saux2022}. Their spectra show well-defined IGW standing modes, but also contain a significant amount of power at low frequencies and over all wavenumbers. Secondly, the spurious migration of the convective boundary in our simulations (Section~\ref{sec:ppmstar}) operates more rapidly when the heating rate is increased, meaning that the spectra shown in Figure~\ref{fig:spectra_heating} are not all determined at the same radius since they are measured at a fixed distance from the convective boundary. Moreover, this implies that the IGWs are not excited at the same location in the star. Ideally, we should compare the wave spectra and the IGW vorticities at the same radius and for simulations where the convective boundary is at the same location. This cannot be accomplished with our current simulations. Using earlier dumps for high-$L$ runs where the boundary moves rapidly is not possible because the boundary migrates before the dynamics has the time to reach a steady state. We are therefore forced to conclude that the scaling laws that we have determined for the IGW velocities and vorticities in the RGB bump setup are most likely incorrect. In the absence of any suitable alternative, we will nevertheless use them in what follows. Fortunately, this problem does not affect our RGB tip simulations, where no extrapolation to low luminosities is required, and it does not impact any of the conclusions of this work.

\subsubsection{Diffusivity estimates}
\begin{figure}
  \includegraphics[width=\columnwidth]{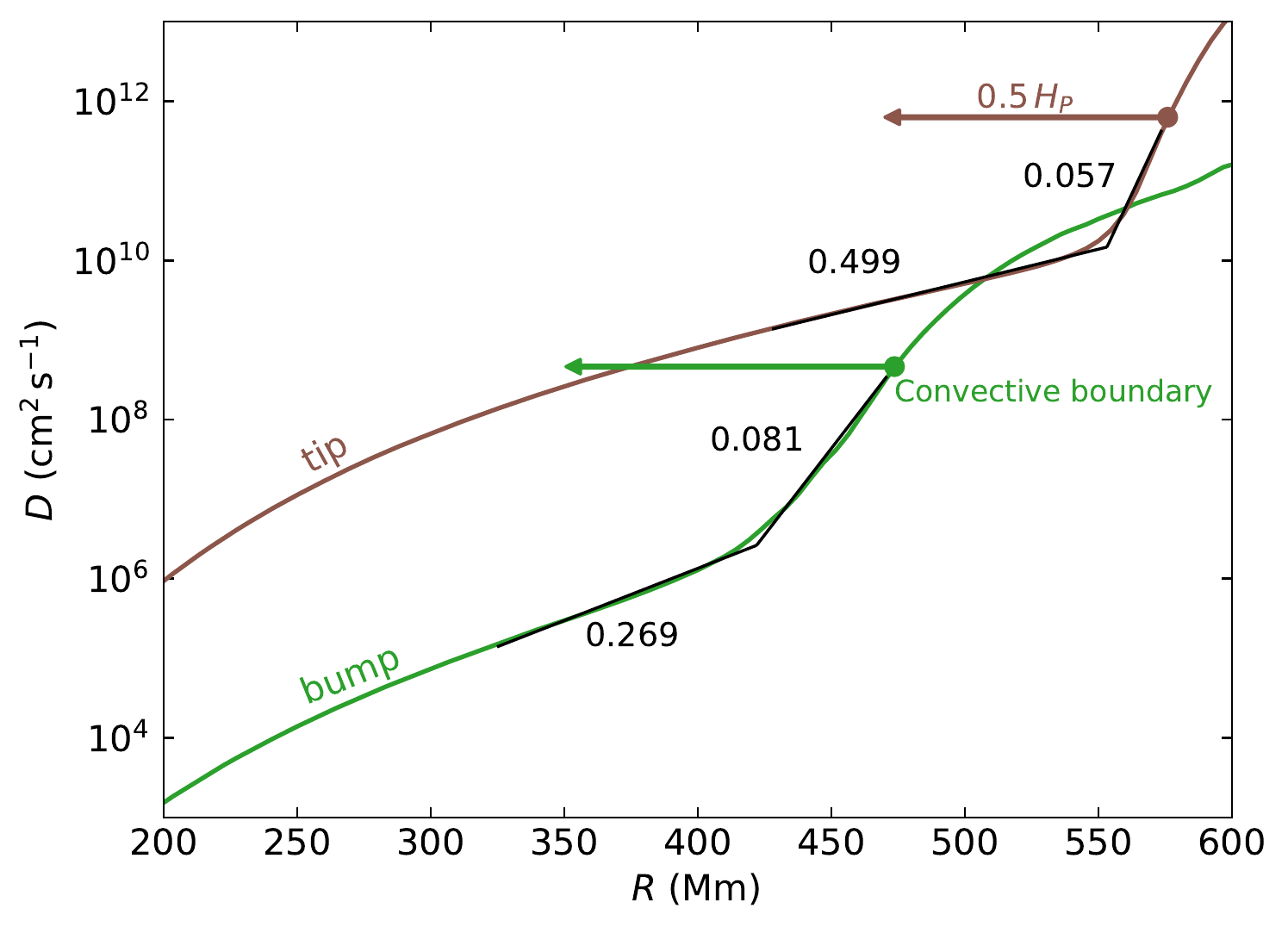}
    \caption{Estimate of the diffusion coefficient in the radiative zone due to IGW mixing based on the Zahn formula (Equation~\ref{eq:zahn}). Those results are based on the vorticities measured in X22 and X26. For the bump setup, the $L^{1/4}$ scaling law (Figure~\ref{fig:scaling_vort}) was used to extrapolate the vorticity to nominal luminosity. For each setup, a circles marks the location of the convective boundary and a horizontal arrow indicates the radial extent of half a pressure scale height. The thin black lines correspond to the best fit to $D$ assuming the double-exponential convective boundary mixing prescription of Equation~\eqref{eq:double-exp}. The values of the $f$ parameters are given next to each segment.}
    \label{fig:Dzahn}
\end{figure}

Equipped with our scaling relation for $|\nabla \times U|$ in the radiative zone of the bump setup, we can now estimate $D$ by virtue of Equation~\eqref{eq:zahn} for both setups. Figure~\ref{fig:Dzahn} shows the resulting diffusivity profiles. The magnitude of $D$ in the stable layers is quite significant and suggests that IGW mixing could alter the evolution of upper RGB stars. For reference, at the bump luminosity, a diffusion coefficient of $\sim 10^9\,{\rm cm}^2\,{\rm s}^{-1}$ at the base of the envelope is needed to explain the observed RGB extra mixing \citep{denissenkov2003a}.

The diffusion coefficient profiles of Figure~\ref{fig:Dzahn} follow a distinctive double-exponential decay near the convective boundary, a conclusion that does not depend on the uncertain RGB bump vorticity scaling law. $D$ initially decreases rapidly close to the radiative--convective interface but then exhibits a shallower decay further inside the radiative zone. This behaviour is reminiscent of the prescription used by \cite{battino2016}, and based on the stellar hydrodynamics simulations of \cite{Herwig:2007wh}, to model mixing below the convective envelopes of thermally pulsating asymptotic giant branch (AGB) stars during the third dredge-up. With this prescription, $D$ at a distance $z$ below the convective boundary is given by
\begin{equation}
    D(z) = \begin{cases}
          D_0 \exp \left[ -2z /(f_1 H_{P,0}) \right] & z \leq z_2\\
         D_2 \exp \left[ -2(z-z_2)/(f_2 H_{P,0}) \right] & z > z_2,
\end{cases}
\label{eq:double-exp}
\end{equation}
where 
\begin{equation}
    D_2 = D_0 \exp \left[ -2 z_2 /(f_1 H_{P,0}) \right],
\end{equation}
$D_0$ is the diffusion coefficient at the convective boundary, $H_{P,0}$ is the pressure scale height at the boundary, and $f_1 H_{P,0}$ and $f_2 H_{P,0}$ are the overshoot scale heights \citep{freytag1996,herwig2000}. We show in Figure~\ref{fig:Dzahn} the values of $f_1$ and $f_2$ obtained by fitting Equation~\eqref{eq:double-exp} to the $D$ profiles for the first 150\,Mm below the convective boundary. If the Zahn formula is correct, if $\eta$ remains constant from the bump luminosity to the tip of the RGB, and if $\eta$ is constant throughout the radiative zone (we return to this point in Section~\ref{sec:gaussians}), then Figure~\ref{fig:Dzahn} implies that the $f$ values change throughout the RGB evolution and that a single prescription cannot be used to describe all the upper RGB evolution.

\cite{battino2016} adjusted the free parameters of Equation~\eqref{eq:double-exp} in order to reproduce the IGW mixing calculations of \cite{denissenkov2003b}, based on the IGW mixing prescription of \cite{garcia1991}. The extra mixing generated with this simple model was shown to be able to generate a $^{13}$C pocket in the radiative zone of AGB stars that is large enough to obtain $s$-process yields that are compatible with observations. While promising, this mixing prescription has not yet been verified with multi-dimensional hydrodynamics simulations of the stable layers below a convective envelope. In this context, our results shown in Figure~\ref{fig:Dzahn} offer additional support for the double-exponential prescription used in AGB models. While the radial stratification of a thermally pulsating AGB star obviously differs from that of an upper RGB star, there are significant similarities (e.g., comparable luminosities, analogous geometries). It is therefore encouraging to recover a double-exponential profile in our simulations, especially since the $f$ values we find are approximately similar to those assumed by \citet[$f_1=0.014$ and $f_2=0.25$]{battino2016}.

\subsection{Constraints on the diffusivity from the time evolution of a tracer fluid}
\label{sec:gaussians}
The diffusivity estimates presented in the previous section are far from robust. Apart from concerns regarding the RGB bump vorticity scaling law, their validity depends on the correctness of the Zahn formula (Equation~\ref{eq:zahn}) and on the assumed $\eta$ value. Here, we attempt to measure the mixing more directly using a second passive fluid. \code{PPMstar} follows species advection using the high-order \code{PPB} scheme \citep{woodward2015}. In setups that include a composition gradient \citep[e.g.,][]{herwig2023}, a passive dye cannot be directly inserted in the simulations as \code{PPMstar} is currently a two-fluid code. Fortunately, our RGB setups have a uniform composition and a second fluid with the same mean molecular weight as the first fluid can be added to our base states. This is equivalent to adding a passive tracer. It has no effect on the flow but allows us to directly measure species mixing.

We initialize the concentration of this second fluid as a series of spherical shells with Gaussian radial profiles. The expectation is that shear-induced IGW mixing will spread those Gaussian profiles in the stable layers. The diffusion coefficient can then be recovered by measuring the rate at which the fractional volume of the second fluid at the peak of each Gaussian (FV$_{\rm max}$) decreases with time. By applying the diffusion equation to a Gaussian, one can show that the diffusion coefficient can be calculated as
\begin{equation}
    D = -\frac{d {\rm FV}_{\rm max}/dt}{{\rm FV}_{\rm{max},0}} \sigma_0^2,
    \label{eq:gaussian}
\end{equation}
where $d {\rm FV}_{\rm max}/dt$ is the rate at which the FV at the peak of a Gaussian decreases, ${\rm FV}_{{\rm max},0}$ is its initial value, and $\sigma_0$ is the initial standard deviation of the Gaussian. Note that Equation~\eqref{eq:gaussian} assumes that the width of the Gaussian remains constant, which is an excellent approximation for the relatively short time scales over which our simulations are performed.

Previous experience has taught us that the grid resolutions we have been using so far in this work ($768^3$ and $1536^3$) are insufficient to measure diffusion coefficients with this technique. In fact, in \cite{herwig2023b} we show that in massive main-sequence stars the measured $D$ steadily decreases with increasing grid resolution due to numerical entropy diffusion up to at least $2688^3$. We therefore jump directly to a very high $2880^3$ resolution for this analysis (run X30, see Table~\ref{tab:runs}), and any mixing measured at that resolution should be interpreted as an upper limit given the results of \cite{herwig2023b}. The setup for this new run is identical to that described above for the tip RGB X26 run, except that (1) FV Gaussians with $\sigma_0=8.3\,$Mm (corresponding to a full width at half maximum of $\simeq 25$ grid cells) are placed 100\,Mm apart in the stable layers, (2) the convective envelope now extends further out to $R_{\rm max}=1100\,$Mm (we can afford this extension given the larger 2880$^3$ grid resolution), (3) we have disabled heat conduction at the inner and outer boundaries. This last change was made after we realized that simultaneously cooling the top layers and allowing heat to escape through radiative diffusion at the outer boundary was effectively cooling the star by more than $1\,L_{\star}$. By omitting radiative diffusion at the boundary, we can now precisely set the luminosity to $1\,L_{\star}$ by cooling the upper layers at that exact rate. Due to the large computational cost of the $2880^3$ grid, X30 ran for a shorter total duration than other simulations presented so far (552\,h of star time, see Table~\ref{tab:runs}). Nevertheless, as we will show below, this is sufficient to establish useful constraints on $D$.

\begin{figure}
  \includegraphics[width=\columnwidth]{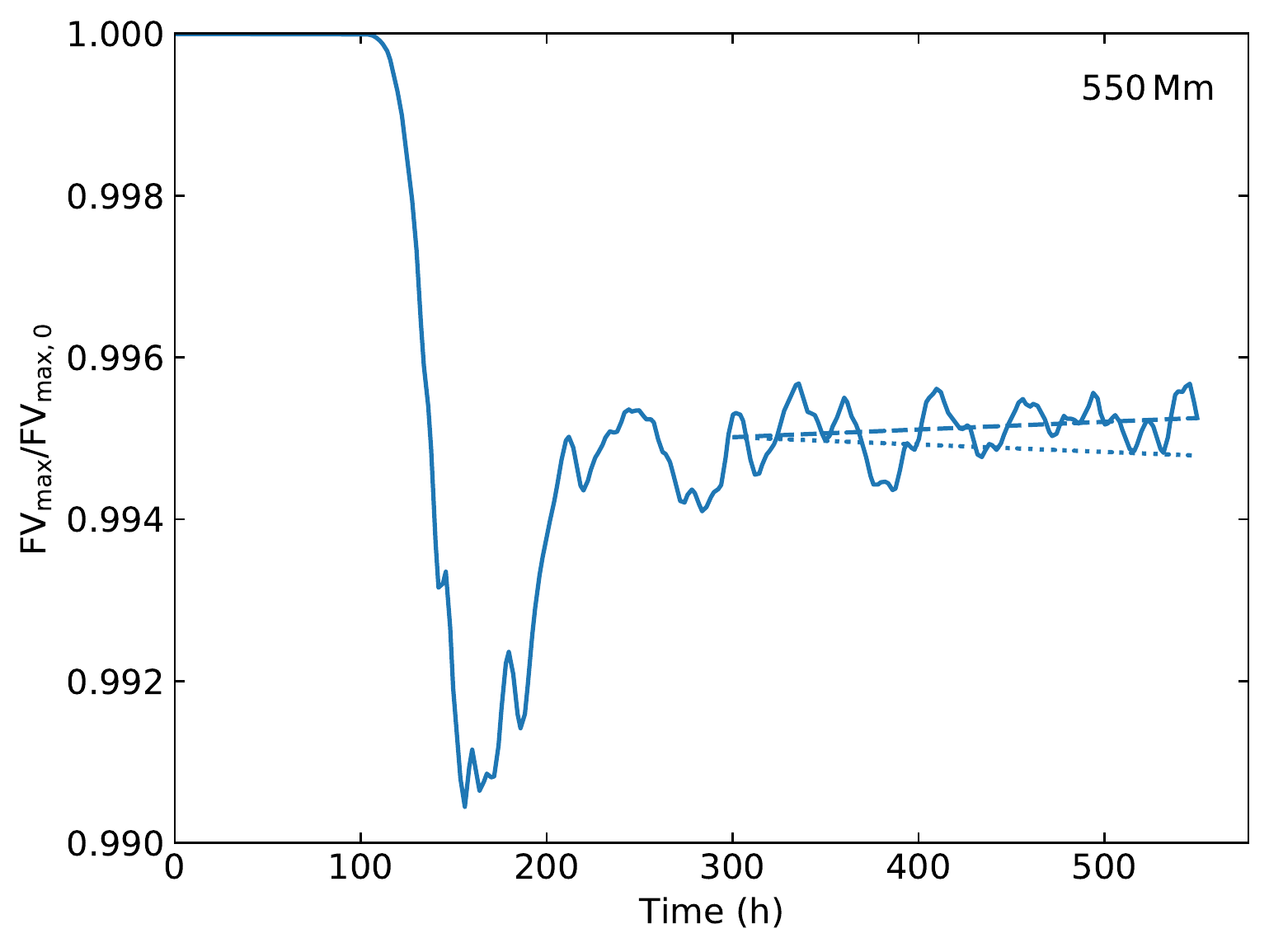}
  \includegraphics[width=\columnwidth]{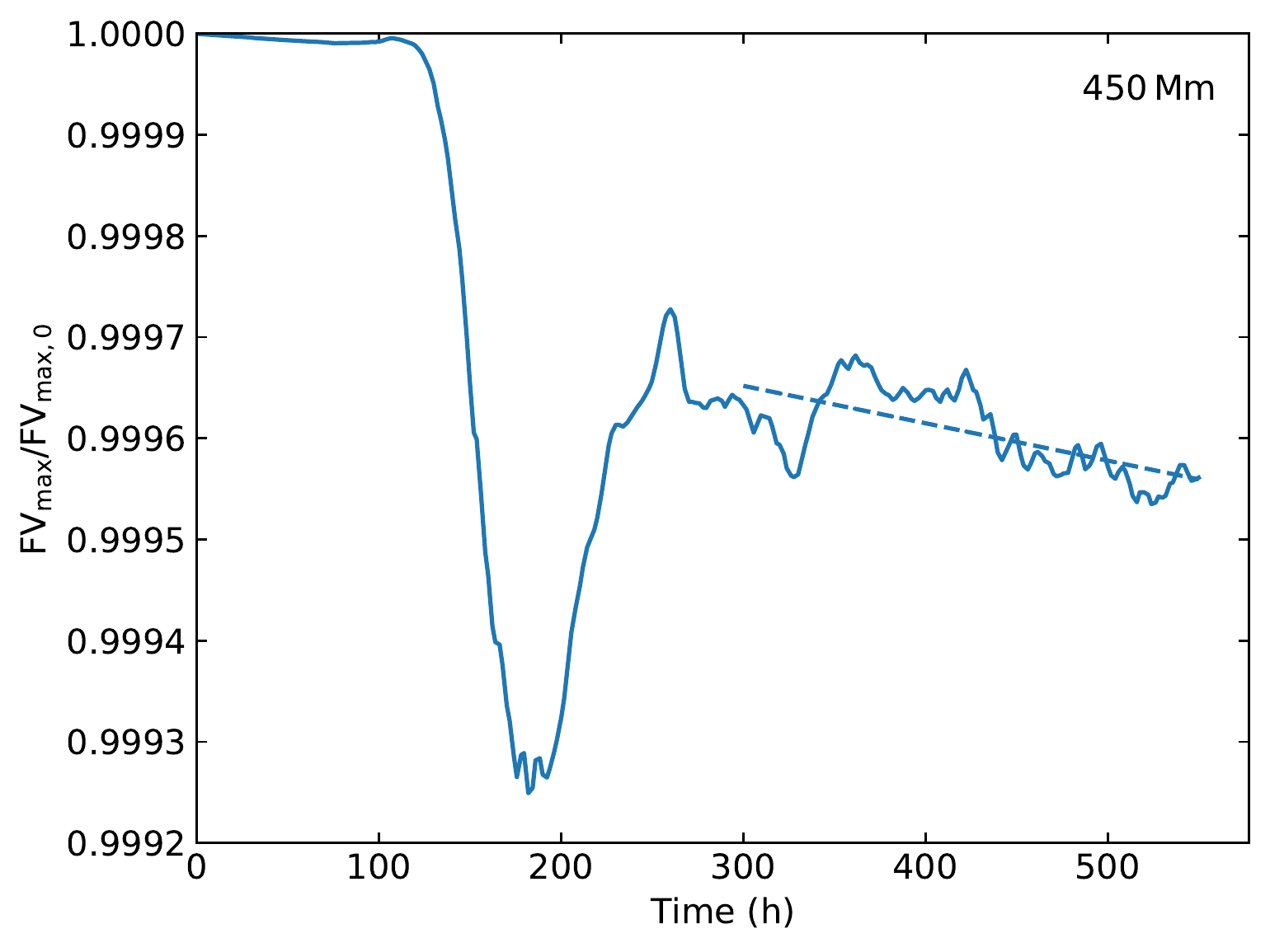}
    \caption{Time evolution of the fractional volume of the peak of the tracer fluid Gaussians inserted at $R=550\,$Mm (top panel) and at $R=450\,$Mm (bottom panel) in run X30. The dashed lines are linear fits for $t \geq 300\,{\rm h}$ and the dotted line shows a lower limit on $d {\rm FV}_{\rm max}/dt$ (see text). Note the different vertical scales for both panels.}
    \label{fig:FVmax}
\end{figure}

Figure~\ref{fig:FVmax} shows the evolution of the height of the two Gaussians closest to the convective boundary (at $R=450$ and 550\,Mm), where IGW mixing is expected to be the strongest (Figure~\ref{fig:Dzahn}). The amplitude of the Gaussians was recovered by fitting the radial FV profiles with a Gaussian. Initially, at $t \lesssim 250\,$h, FV$_{\rm max}$ fluctuates a lot. This is entirely attributable to the initial transient at the beginning of our simulation and for this reason we discard the $t\leq300\,$h portion of the time series from the rest of our analysis. If strong IGW mixing was present, we would expect to observe a decrease of FV$_{\rm max}$ for $t \geq 300\,$h. At $R=550\,{\rm Mm}$, we do not see any evidence for such decline (top panel of Figure~\ref{fig:FVmax}). The Kendall rank correlation coefficient between FV$_{\rm max}$ and $t$ is consistent with the null hypothesis where there is no dependence of FV$_{\rm max}$ on $t$ ($p$-value of 0.057). Similarly, a linear fit to the data gives a positive but statistically insignificant slope, $d {\rm FV}_{\rm max}/dt = (2.6 \pm 1.0) \times 10^{-10}\,{\rm s}^{-1}$ (dashed line in Figure~\ref{fig:FVmax}). From this linear fit, we can extract a lower limit on $d {\rm FV}_{\rm max}/dt$ by taking the 5-sigma lower limit, $d {\rm FV}_{\rm max}/dt > -2.5 \times 10^{-10}\,{\rm s}^{-1}$ (dotted line in Figure~\ref{fig:FVmax}). Using Equation~\eqref{eq:gaussian}, we find that this corresponds to $\log D\,({\rm cm}^2\,{\rm s}^{-1}) < 8.2$. In contrast, a tentative decline of FV$_{\rm max}$ is visible for the 450\,Mm Gaussian thanks to the lower noise level in the FV$_{\rm max}$ time series (compare both panels of Figure~\ref{fig:FVmax}). We find a Kendall rank correlation coefficient of $0.48$, significantly different from 0 with a $p$-value of $10^{-15}$. This time, a linear fit to the data yields $d {\rm FV}_{\rm max}/dt = -(1.03\pm 0.11) \times 10^{-10}\,{\rm s}^{-1}$, which allows a tentative measurement of $\log D\,({\rm cm}^2\,{\rm s}^{-1}) = 7.80 \pm 0.05$. It is possible that this downward trend is only temporary and that a longer simulation would show a stabilization of FV$_{\rm max}$. As for the Gaussians located at smaller radii (250 and 350\,Mm), we find that they do not maintain their Gaussian shapes to a sufficiently high level of accuracy to allow a precise determination of FV$_{\rm max}$. Part of the problem is that the fluctuations of FV at smaller radii are much smaller (as can be inferred from Figure~\ref{fig:FVmax}). This means that FV$_{\rm max}$ must be determined with an increasing precision, making even small departures from perfect Gaussianity problematic. The artifacts introduced by the inner boundary are also a consideration at those small radii.

\begin{figure}
  \includegraphics[width=\columnwidth]{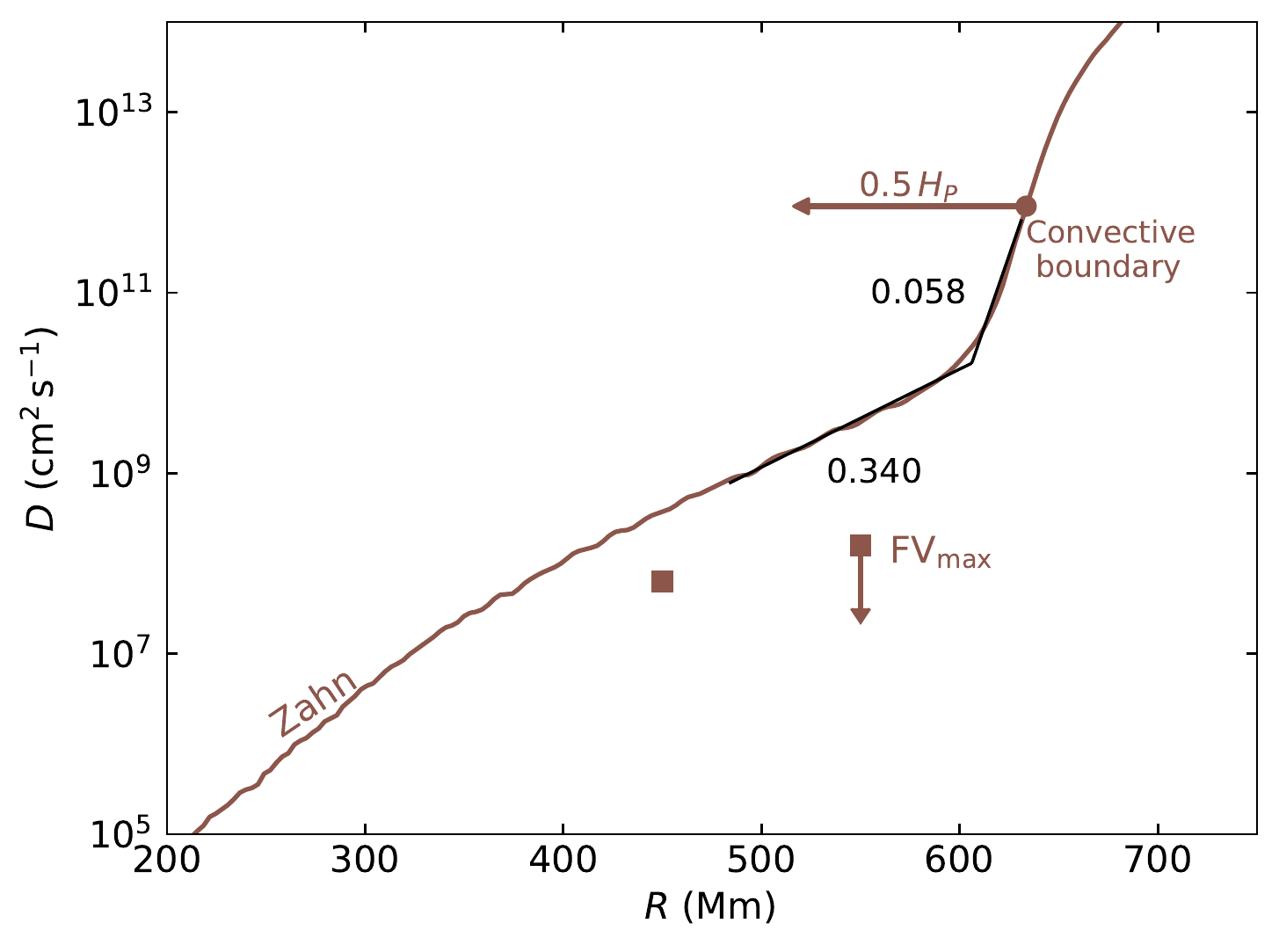}
    \caption{Estimate of the mixing coefficient based on Equation~\eqref{eq:zahn} and the vorticities measured in X30 at dump~700 ($t=552\,$h, solid brown line), and measurements of the diffusion coefficient based on the analysis of the tracer fluid Gaussians (squares). The formal error on the $D$ measurement at 450\,Mm is smaller than the size of the symbol.}
    \label{fig:Dzahn_X30}
\end{figure}

In Figure~\ref{fig:Dzahn_X30}, we compare our estimates of $D$ described in the previous paragraph to $D$ calculated using Equation~\eqref{eq:zahn} and the vorticities measured in X30. The latter $D$ estimates differ from that previously shown in Figure~\ref{fig:Dzahn}, which can be explained by the higher grid resolution of X30, the different boundary conditions, and the slightly larger convective envelope. The main takeaway from Figure~\ref{fig:Dzahn_X30} is that we can rule out IGW mixing on the scale predicted by Zahn's formula with $\eta=0.1$. There is a factor 6 discrepancy between both $D$ estimates at 450\,Mm and a factor $\gtrsim 25$ discrepancy at 550\,Mm. The $\eta=0.1$ value we have assumed so far is only a rough order of magnitude estimate based on existing numerical simulations. A different value of $\eta$ is certainly possible. For instance, \cite{garaud2016} recommend $\eta \sim 0.02$, which would improve the agreement between both estimates in Figure~\ref{fig:Dzahn_X30}. Furthermore, existing $\eta$ determinations are still tentative given that they are based on low Reynolds number numerical simulations \citep{garaud2021}. Taken at face value, our FV-based $D$ determinations also suggest that $\eta$ is not constant through the stable layers: the 450\,Mm Gaussian implies $\eta \sim 0.02$ and the 550\,Mm implies $\eta \lesssim 0.004$. Previous numerical simulations have found a dependence between the turbulent Reynolds number and the value of $\eta$ \citep{prat2016}. This may be related to what we observe here.

In any case, the analysis of the time evolution of the Gaussians presented in this section represents a much more direct assessment of the efficiency of IGW mixing, and the results from this analysis take precedence over those presented in the previous section based on the application of Zahn's formula. Given that we find that $D\,({\rm cm}^2\,{\rm s}^{-1}) \sim 10^8$ at most close to the boundary (remember that the finite grid resolution implies that the measured $D$ at 450\,Mm is formally an upper limit, \citealt{herwig2023b}), our hydrodynamics simulations a priori suggest that IGW mixing is not an important mixing mechanism in the radiative zones of RGB stars and that it cannot provide the missing extra mixing required to explain upper RGB surface compositions. A diffusion coefficient of the order of $10^9\,{\rm cm}^2\,{\rm s}^{-1}$ throughout the upper RGB would be needed, and here we do not even reach that value at the RGB tip, where the luminosity is highest and IGW mixing is expected to be most efficient. That being said, we will see in Section~\ref{sec:envelope} that this conclusion is not definitive and that in a real RGB star IGW mixing may be more efficient.

\section{Effect of the envelope size}
\label{sec:envelope}
\begin{figure}
    \centering
    \includegraphics[width=\columnwidth]{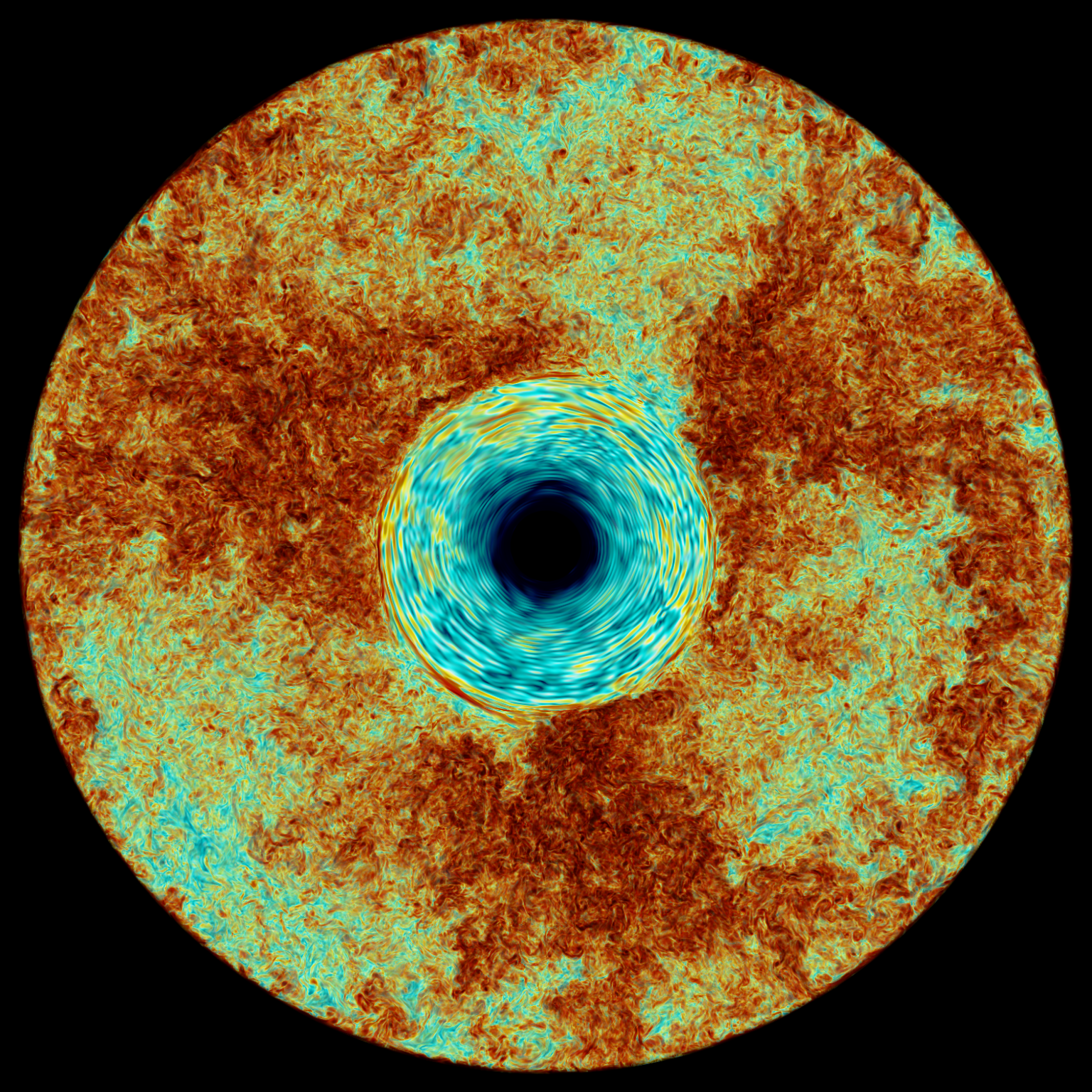}
    \caption{Centre-plane slice rendering of the vorticity magnitude for run X25 at dump~430 ($t=1140\,$h). As in Figure~\ref{fig:bobs2}, $|\nabla \times U|$ increases when going from dark blue to turquoise, yellow, red, and dark red. The inner 120\,Mm was masked to remove artifacts introduced by the inner simulation boundary.}
    \label{fig:X25}
\end{figure}

To document the effect of including a limited portion of the convective envelope in our simulations, we performed an additional run (X25) where we extended our setup to $R_{\rm max}=1800\,$Mm instead of our fiducial 900\,Mm. We performed this run on a 1536$^3$ grid, meaning that it has the same resolution as X24 (performed on a 768$^3$ grid with $R_{\rm max}=900\,$Mm) in the region where both grids overlap. By comparing X24 and X25, we can therefore assess the impact of including a larger envelope while controlling for the grid resolution. Figure~\ref{fig:X25} shows a vorticity magnitude rendering of X25. As expected, larger eddies are able to develop in this extended convective envelope. Throughout the simulation, the flow is dominated by just a few large cells (three are clearly visible in Figure~\ref{fig:X25}), which is to be contrasted with the many small cells that characterize the rest of our simulations (Section~\ref{sec:bobs}). Accordingly, we find that the $U_r$ power spectrum in the convective envelope now exhibits a Kolmogorov $\ell^{-5/3}$ scaling down to $\ell=2$ (Figure~\ref{fig:spectra_envelope}). Extending the envelope even further out would presumably allow a single large dipole mode to develop. However, for a computationally feasible grid size, this would result in a grid resolution that is too poor to properly characterize IGWs in the radiative zone. This conundrum could be resolved in the future by using a nested grid with smaller cells in the central radiative layers and larger cells in the convective envelope. This capability is not yet implemented in \code{PPMstar}. For now, the extended envelope of X25 is enough to document the sensitivity of the IGWs on the size of the convective envelope; we postpone a proper convergence study to future work.

\begin{figure}
    \centering
    \includegraphics[width=\columnwidth]{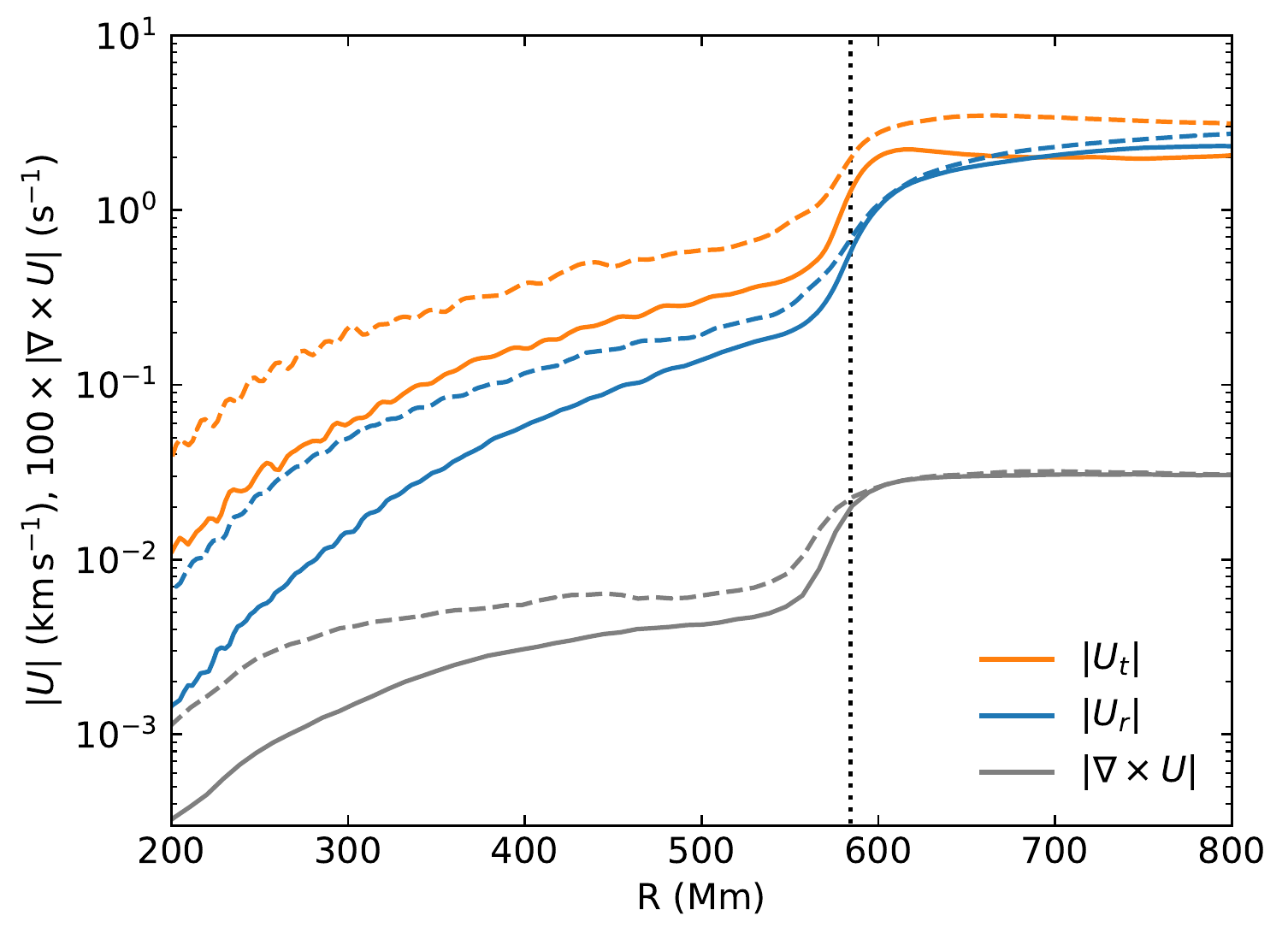}
    \caption{Rms radial velocity (blue), rms tangential velocity (orange) and vorticity magnitude (grey) for X24 (small convective envelope, solid lines) and X25 (large convective envelope, dashed lines) at dump 400. The vertical dotted line indicates the location of the convective boundary.}
    \label{fig:envelope}
\end{figure}

How does the development of larger convective cells affect the properties of the IGW-dominated flow in the radiative zone? Figure~\ref{fig:envelope} shows that the flow is $1.5-2$ times faster in the convection zone with the extended envelope setup. Those faster convective motions in turn excite a more rapid flow below the convective boundary, with $|U_r|$ and $|U_t|$ up to $\simeq 4$ times faster in the radiative zone when a larger envelope is used (see also Figure~\ref{fig:spectra_envelope}). The vorticities are naturally also enhanced by up to a factor $\simeq 3$. Note however that the increase in vorticity is less pronounced in the outermost radiative layers: $|\nabla \times U|$ grows by only $\simeq 50\,\%$ at $0.5\,H_P$ below the convective boundary. This vorticity enhancement would directly impact our Zahn diffusion coefficient estimate (Equation~\ref{eq:zahn}) and increase it by up to one order of magnitude (the radiative diffusivity $K$ and the Brunt--V\"ais\"al\"a frequency $N$ in the stable layers are not affected by the inclusion of a larger envelope). The enhanced vorticity would also conceivably affect our constraints on $D$ based on the analysis of the time evolution of the tracer fluid Gaussians (Section~\ref{sec:gaussians}). Future work should focus on this aspect of the problem. We are forced to conclude that our tentative measurement of $D$ in Section~\ref{sec:gaussians} should be interpreted as a lower limit, as simulations including the full envelope would most likely result in more efficient mixing. Hence, we cannot conclusively rule out the idea that IGW mixing on the upper RGB is an important mixing mechanism and possibly at least part of the solution to the extra mixing problem.

\section{The convective boundary}
\label{sec:boundary}
We now leverage the exceptionally high resolution ($2880^3$) of our X30 simulation to examine the properties of the convective boundary. Because radiative diffusion at the outer boundary has been omitted in X30 (see Section~\ref{sec:gaussians}), heat is removed from the star at the same rate as it is injected. As a result, the Schwarzschild boundary does not migrate as in previous runs, therefore enabling a meaningful study of the boundary region.

\begin{figure}
    \centering
    \includegraphics[width=\columnwidth]{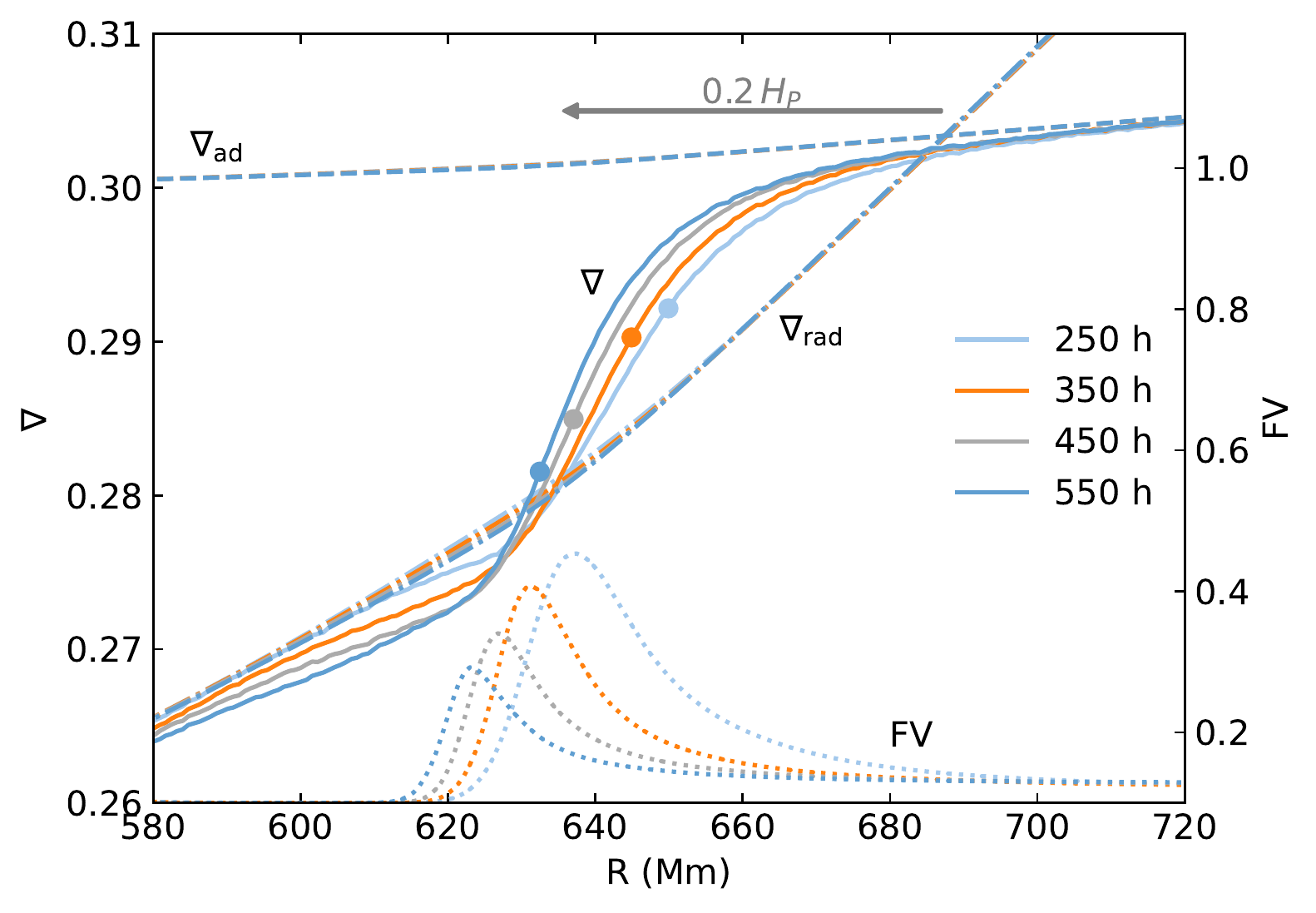}
    \caption{Evolution of the spherically averaged temperature gradient $\nabla$ in run X30 (solid lines). The intersection of the adiabatic ($\nabla_{\rm ad}$, dashed lines) and radiative ($\nabla_{\rm rad}$, dashed-dotted lines) temperature gradients is the Schwarzschild boundary. The convective boundary, defined here as the radius where the maximum $U_t$ gradient is reached, is indicated by a circle. The dotted lines show the evolution of FV (see text). The establishment of a penetration zone beyond the formal Schwarzschild boundary can be observed.}
    \label{fig:penetration}
\end{figure}

Figure~\ref{fig:penetration} tracks the evolution of the spherically averaged temperature gradients and FV profile in the boundary region (for reference, the evolution of the Brunt--V\"ais\"al\"a frequency is also given in Figure~\ref{fig:Npenetration}). In Section~\ref{sec:gaussians}, we analyzed the time evolution of the FV Gaussians located well into the radiative zone, at $R=450\,$Mm and $R=550\,$Mm. X30 also includes an FV Gaussian centered at $R=650\,{\rm Mm}$, very close to the convective boundary. The ingestion of this Gaussian into the convective envelope provides a useful diagnostic for the extent of the fully mixed envelope. There are several things to note in Figure~\ref{fig:penetration}:
\begin{itemize}
\item While the formal Schwarzschild boundary is virtually stationary, the dynamic convective boundary (defined, as previously in this work, as the location of the maximum $U_t$ gradient) migrates inward (circles in Figure~\ref{fig:penetration});
\item The temperature gradient in the region between the Schwarzschild and convective boundaries departs from the $\nabla_{\rm rad}$ value expected in the radiative zone and instead gradually approaches the adiabatic gradient $\nabla_{\rm ad}$;
\item The fully mixed region (i.e., the region where FV is constant) grows past the Schwarzschild boundary.
\end{itemize}
All those properties point to the formation of a convective penetration zone. Formally, a penetration zone is a region where both entropy and composition are mixed by convective motions beyond the Schwarzschild boundary \citep{zahn1991,hurlburt1994,brummell2002,anders2022}. Convective penetration below a convective envelope has been observed in 3D hydrodynamics simulations of a Sun-like star \citep{brun2011} and of a 5\,$M_{\odot}$ star at the end of central He burning \citep{viallet2013}, two examples where the geometry is similar to that of the RGB case. While the limited length of our simulation does not allow the establishment of a fully mixed, stationary penetration zone, it is clear from Figure~\ref{fig:penetration} that such a region is in the process of being formed. The thermal diffusion lengthscale $\sqrt{Kt}$ over the full simulation ($t=552\,$h) is $\simeq 50\,$Mm, which explains why the penetration zone had the time to build up but has not yet reached a steady state.

\begin{figure*}
    \centering
    \includegraphics[width=\columnwidth]{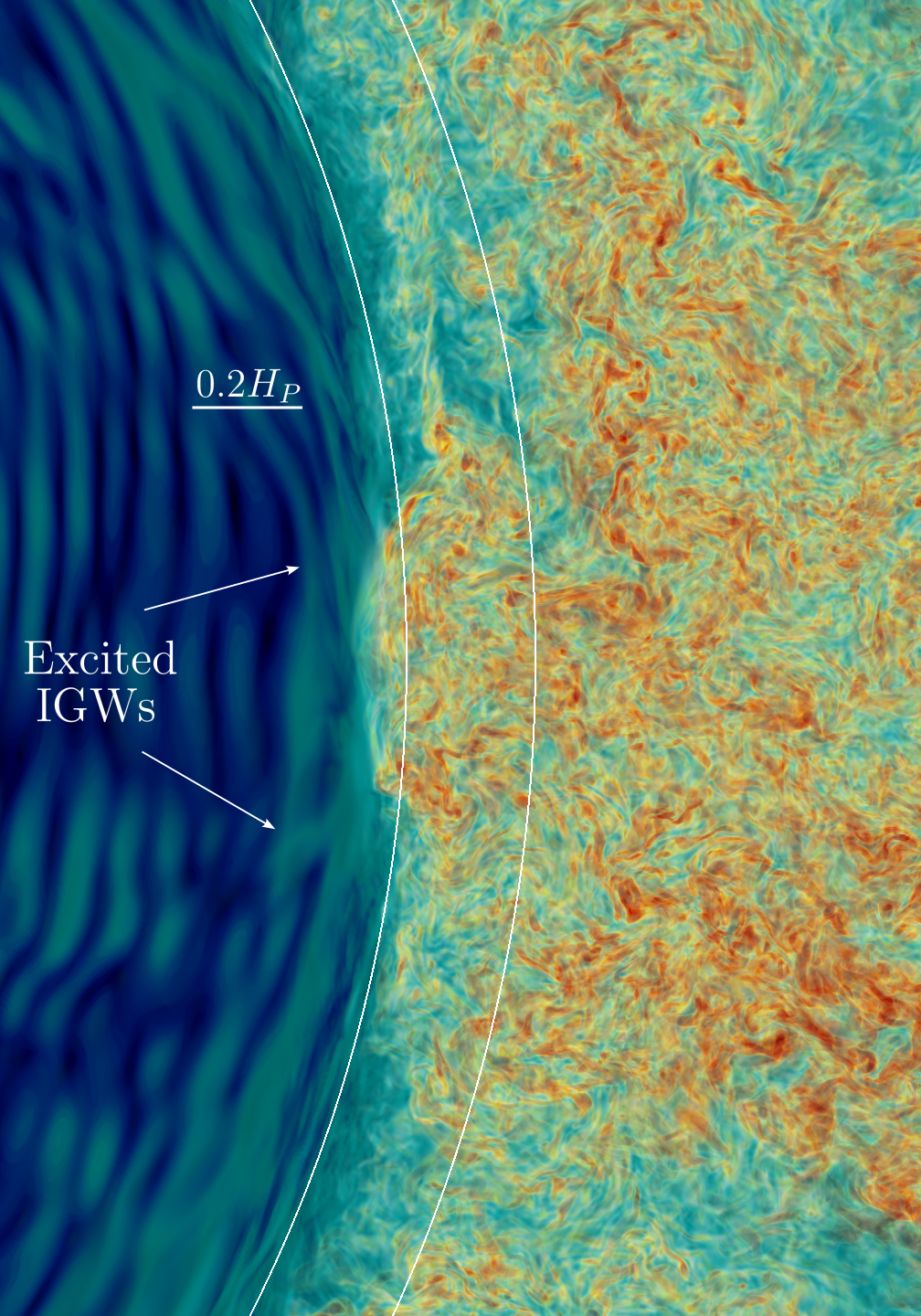}
    \includegraphics[width=\columnwidth]{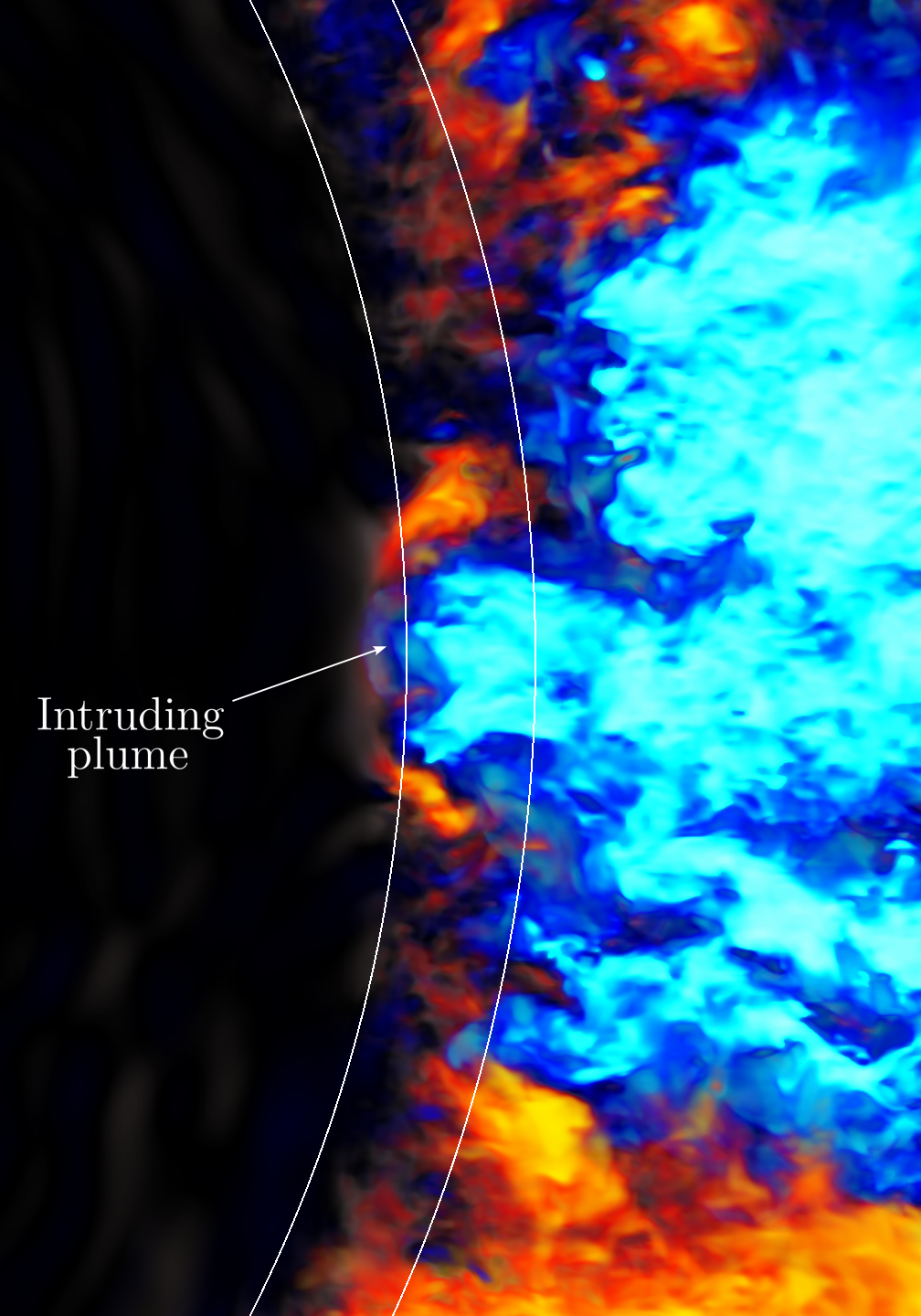}
    \caption{Centre-plane slice rendering in the convective boundary region of the vorticity magnitude (left panel) and radial velocity component (right panel) for run X30 at dump 690 ($t=544\,$h). The colours have the same meanings as in Figures~\ref{fig:bobs} and \ref{fig:bobs2}. In each panel, the inner circular arc marks the location of the spherically averaged convective boundary, defined here as the radius where the maximum $U_t$ gradient is reached. The outer circular arc designates the Schwarzschild boundary. The region between the two circular arcs is the nascent penetration zone. A large plume traversing the penetration zone and intruding into the stable layers excites IGWs (indicated by two arrows). High-resolution movies are available at \url{https://www.ppmstar.org}.}
    \label{fig:plume}
\end{figure*}

To visualize this nascent penetration zone, we show in Figure~\ref{fig:plume} renderings of the vorticity magnitude and radial velocity in the boundary region. The outer white circular arc marks the Schwarzschild boundary. Turbulent motions visible in the vorticity rendering extend well beyond this radius and up to the dynamic convective boundary (identified by the inner white circular arc): this is the convective penetration zone. In fact, this is very similar to the behaviour observed by \cite{anders2022} in their 3D hydrodynamics simulations performed in a simplified plane-parallel geometry (compare the left panel of Figure~\ref{fig:plume} to the left panel of their Figure~1). While the penetration zone is not fully established, it is still useful to compare its extent to existing observational constraints. We can infer from Figures~\ref{fig:penetration} and~\ref{fig:plume} that after 550\,h of simulation time, the penetration zone extends $\sim 0.2\,H_P$ beyond the Schwarzschild boundary. Interestingly, it has been shown that the inclusion of a $\sim 0.25 H_P$ overshooting below the Schwarzschild boundary can eliminate the discrepancy between the observed and predicted location of the RGB bump \citep{cassisi2011,fu2018,khan2018}.

The renderings of Figure~\ref{fig:plume} reveal more than just the formation of a penetration zone. We also see a large plume moving inward (remember that blue colours stand for inward motions in our $U_r$ renderings) and traversing the spherically-averaged dynamic convective boundary to reach the stable, IGW-dominated interior. This points to the presence of an overshoot zone beyond the penetration zone, where the convective motions are too weak to efficiently mix entropy and composition. This is in line with the schematic picture discussed by \cite{zahn1991} and recently illustrated by \citet[Figure~1; see also Figure~14 of \citealt{hotta2017}]{anders2022b}. In the vorticity rendering of Figure~\ref{fig:plume}, we can even see how the intrusion of this plume in the radiative region excites IGWs. On each side of the plume, there are structures that form an angle with respect to the almost circular patterns that otherwise dominate the vorticity rendering of the stable layers. Those structures, indicated by two arrows in Figure~\ref{fig:plume}, are IGWs excited by the intruding plume (this is most clearly seen in the movie). This suggests that penetrative plumes are an important wave excitation mechanism, consistent with our findings of Section~\ref{sec:spectra} based on the IGW power spectra. That being said, it is possible that in a real RGB star, with a much large convective envelope, the dipolar global flow morphology prevents the formation of such plumes.

\begin{figure}
    \centering
    \includegraphics[width=\columnwidth]{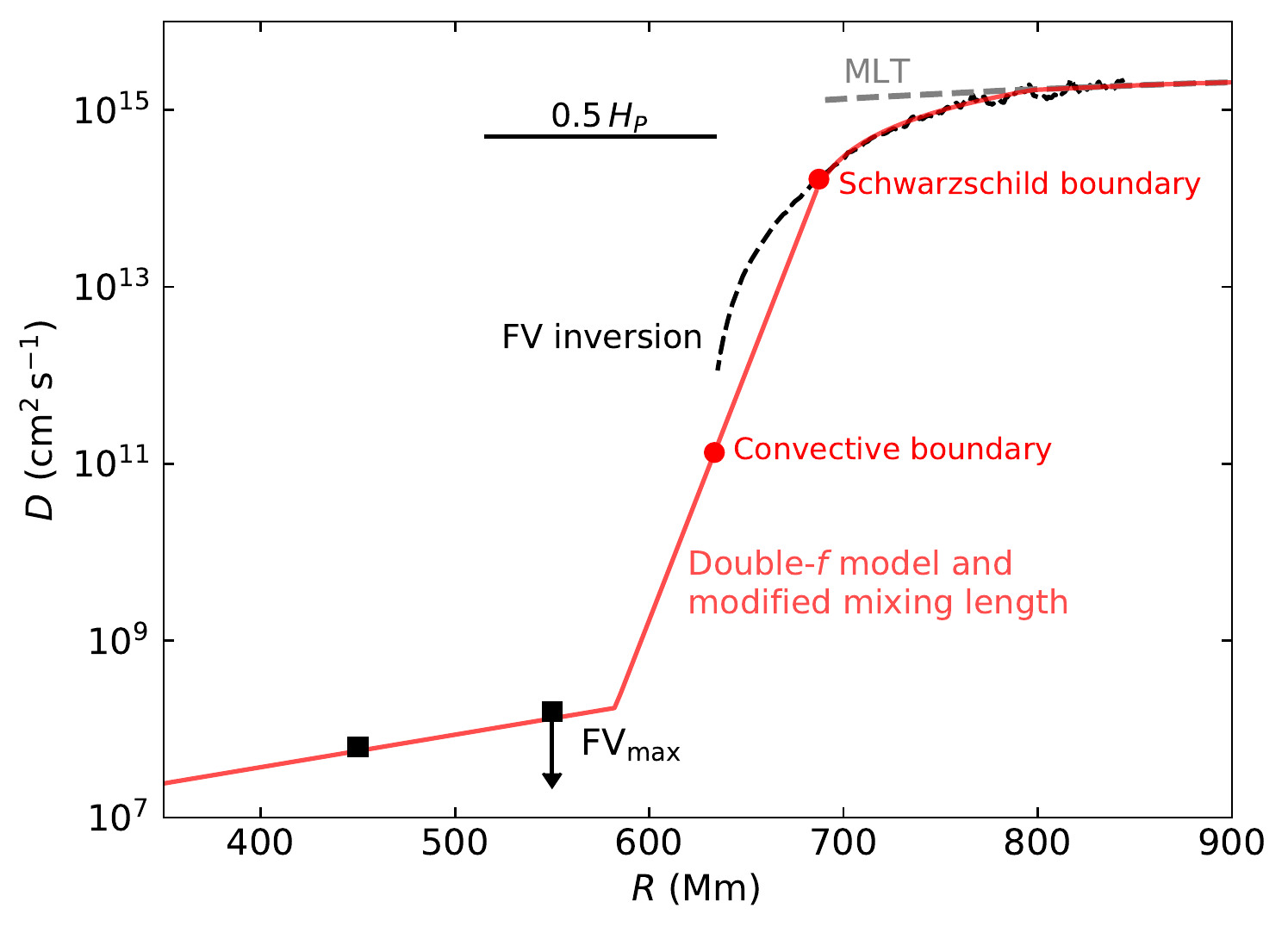}
    \caption{Diffusion coefficient measured in the X30 simulation by inverting the FV profile evolution (black dashed line) and by tracking the FV Gaussian spreading as described in Section~\ref{sec:gaussians} (black squares). The grey dashed line is an MLT estimate of $D$ assuming $\alpha=0.75$. The red line is a simple model that assumes an MLT diffusivity with a variable $\alpha$ (Equation~\ref{eq:alpha}) in the unstable layers and a double-exponential decay below the Schwarzschild boundary ($f_1=0.06$ and $f_2=0.9$, see text for details). Two circles mark the locations of the Schwarzschild and convective boundary (defined as before as the location where the maximum $U_t$ gradient is reached).}
    \label{fig:D_model_X30}
\end{figure}

The evolution of the FV profile in Figure~\ref{fig:penetration} can be used to infer the diffusion coefficient close to the convective boundary. As in \cite{jones2017}, we take the FV radial profiles at different times and invert the 1D diffusion equation to determine the profile $D(R)$ that can reproduce the observed change. The resulting $D(R)$ is shown as a black dashed line in Figure~\ref{fig:D_model_X30}. Our diffusion coefficient inversion technique only works if the gradient of FV is not zero, meaning that we cannot measure $D$ further out in the envelope than what is shown in Figure~\ref{fig:D_model_X30}. Note also that this method cannot be applied in the stable layers, where diffusion is much slower and the approach used in Section~\ref{sec:gaussians} is the best option. In Figure~\ref{fig:D_model_X30}, we show with a grey dashed line the diffusion coefficient predicted by the MLT formula
\begin{equation}
D_{\rm MLT} = \frac{1}{3} U \alpha H_P,
\end{equation}
where we have assumed that $U$ corresponds to the total velocity amplitude in the PPMstar simulation and where we have fixed $\alpha=0.75$ in order to fit the measured diffusivity (black dashed line) in the envelope far from the Schwarzschild boundary. As previously observed in other hydrodynamics simulations, a constant $\alpha$ value leads to an overestimation of $D$ close to the boundary, a problem that can be solved by reducing the mixing length near the boundary \citep{jones2017,herwig2023}. We found that a good prescription for $\alpha$ is given by
\begin{equation}
\alpha = \min(0.75, 1.8 \Delta R + 0.08),
\label{eq:alpha}
\end{equation}
where
\begin{equation}
\Delta R = \frac{R-R_{\rm SB}}{H_P},
\end{equation}
with $R_{\rm SB}$ the radius of the Schwarzschild boundary. This yields an excellent fit to the measured diffusivity in the convection zone (red line in Figure~\ref{fig:D_model_X30} for $R\geq R_{\rm SB}$). The simpler prescription suggested by \citet[Equation~4]{jones2017} and the exponential parametrization of \citet[Equation~9]{herwig2023} cannot reproduce the measured diffusivity to a comparable degree of accuracy.

To extend our diffusivity model below the Schwarzschild boundary, we use the double-$f$ prescription of Equation~\ref{eq:double-exp}. We find that this particular convective boundary mixing model cannot simultaneously match the measured diffusivity in the penetration zone (black dashed line for $R < R_{\rm SB}$ in Figure~\ref{fig:D_model_X30}) and in the stable layers (square symbols as in Figure~\ref{fig:Dzahn_X30}). The best overall match is given by $f_1=0.06$, $f_2=0.9$ and $D_2=1.7\times 10^8\,{\rm cm}^2\,{\rm s}^{-1}$ ($z_2=105\,{\rm Mm} = 0.41 H_{P,0}$), and is displayed as a red line for $R<R_{\rm SB}$ in Figure~\ref{fig:D_model_X30}. This simple diffusivity model, with a modified mixing length in the unstable layers and a double-exponential profile below $R_{\rm SB}$, can be easily implemented in 1D stellar evolution codes, with the caveats that it underestimates mixing in the penetration zone and that it may not apply to the rest of the RGB evolution.

\section{Conclusion}
\label{sec:conclu}
We have presented the first 3D hydrodynamics simulations of IGW excitation and propagation in RGB stars. These simulations clearly show that a rich spectrum of IGWs is generated in the radiative zones of low-mass upper RGB stars (Section~\ref{sec:morpho}). By analysing the time evolution of a tracer fluid, we measured the mixing enabled by IGWs in the radiative interior (Section~\ref{sec:IGWmixing}). In our simulations, we find that IGW mixing is too weak to explain the missing RGB extra mixing, but we cannot rule out that this mixing mechanism is much more efficient in real RGB stars. In fact, our simulations only include a limited portion of the convective envelope, and we have shown how this probably leads to an underestimation of IGW mixing in the radiative zone (Section~\ref{sec:envelope}). This is the most critical aspect of our simulations to be improved in future work. We have also studied the properties of the envelope convective boundary (Section~\ref{sec:boundary}). We found evidence for the establishment of a convective penetration zone and for the excitation of IGWs in the stable layers by plumes that traverse the penetration zone and encroach into the radiative zone. We also provided a simple prescription for the diffusion coefficient in the boundary region.

Promisingly, we also find that the vorticity profiles measured below the convective boundary in our RGB simulations yield support to the idea that IGW mixing may be responsible for the formation of the $^{13}$C pocket in AGB stars. This should be further studied with dedicated AGB hydrodynamics simulations.

\section*{Acknowledgements}
We are grateful to the referee for insighful comments and questions that have improved this manuscript. SB is a Banting Postdoctoral Fellow and a CITA National Fellow, supported by the Natural Sciences and Engineering Research Council of Canada (NSERC). FH acknowledges funding through an NSERC Discovery Grant. PRW acknowledges funding through NSF grants 1814181 and 2032010. FH and PRW have been supported through NSF award PHY-1430152 (JINA Center for the Evolution of the Elements). The simulations for this work were carried out on the NSF Frontera supercomputer operated by the Texas Advanced Computing Center at the University of Texas at Austin and on the Compute Canada Niagara supercomputer operated by SciNet at the University of Toronto. The data analysis was carried on the Astrohub online virtual research environment (\url{https://astrohub.uvic.ca}) developed and operated by the Computational Stellar Astrophysics group (\url{http://csa.phys.uvic.ca}) at the University of Victoria and hosted on the Compute Canada Arbutus Cloud at the University of Victoria.

\section*{Data Availability}
Simulation outputs are available at \url{https://www.ppmstar.org} along with the Python notebooks that have been used to generate the figures presented in this work.

\bibliographystyle{mnras}
\bibliography{references}

\bsp
\label{lastpage}

\appendix
\section{Supplementary figures}
\FloatBarrier

This Appendix contains additional figures that are all referenced in the main text.

\begin{figure}
    \centering
    \includegraphics[width=\columnwidth]{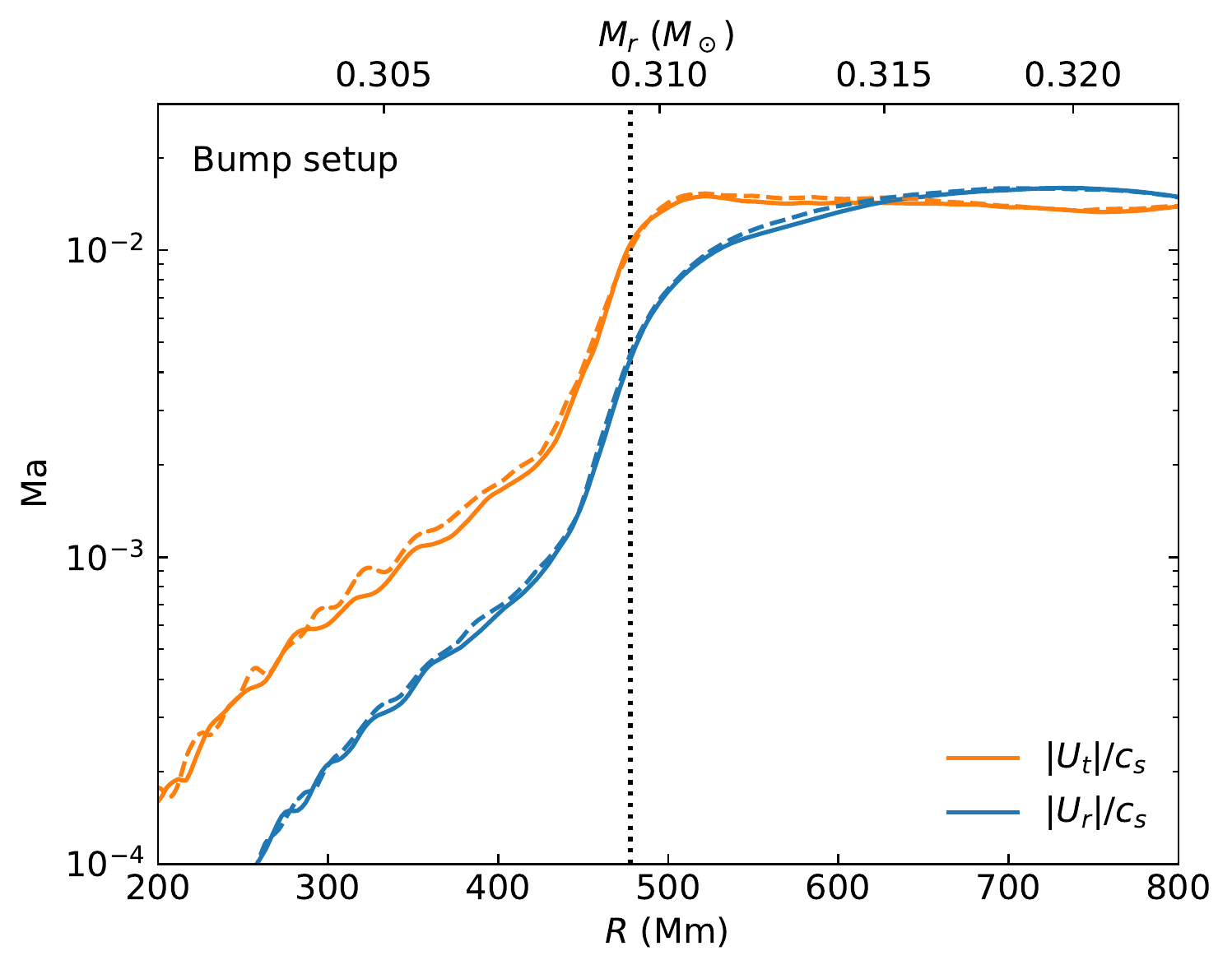}
    \includegraphics[width=\columnwidth]{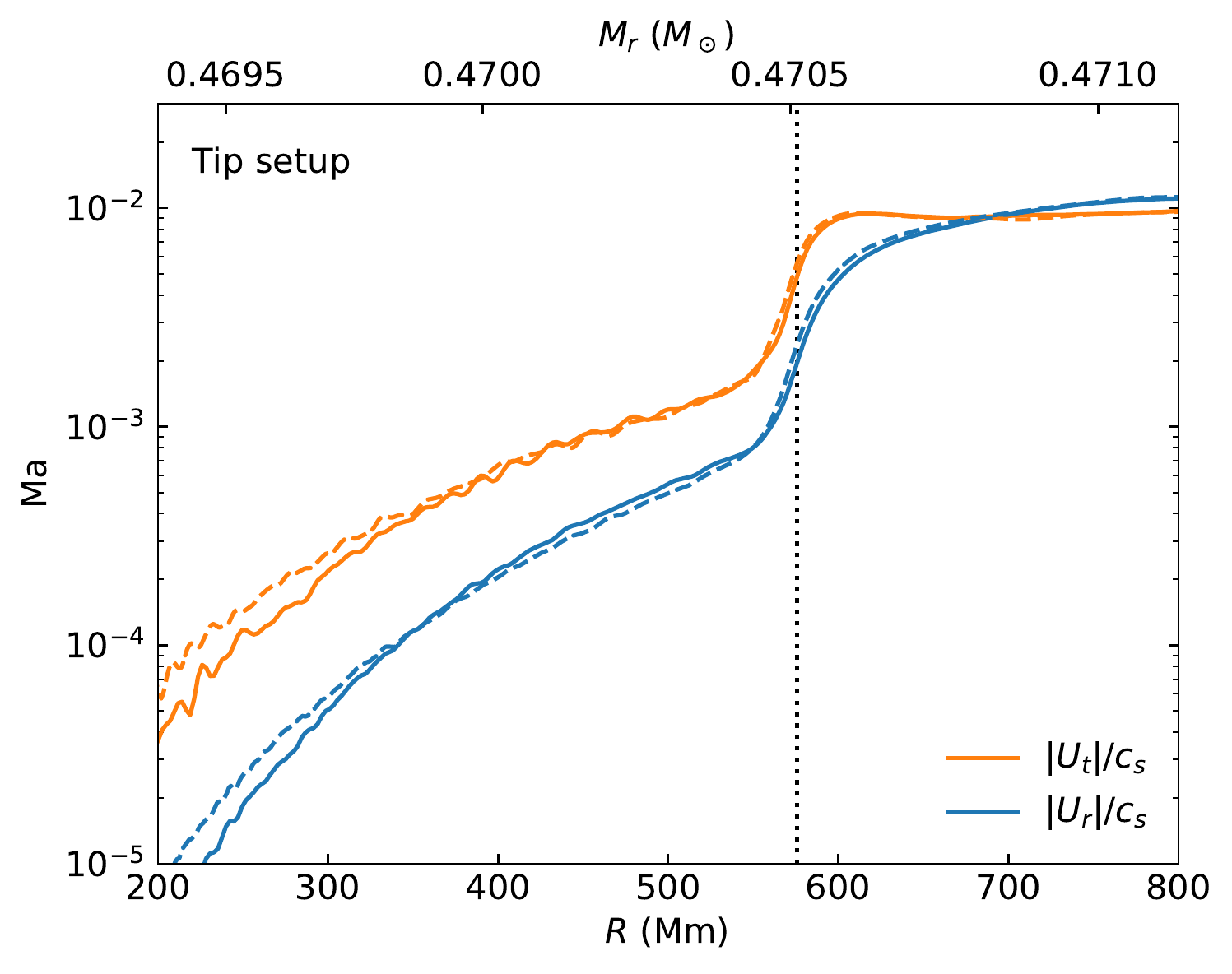}
    \caption{Rms radial and tangential velocity for the bump (top panel, X14 and X22) and tip setups (bottom panel, X24 and X26) at $t=672\,$h (same time as in Figure~\ref{fig:bobs}) in terms of Mach numbers. Profiles calculated from runs on a 768$^3$ grid are shown as solid lines; those from 1536$^3$ grids are shown as dashed lines. The vertical dotted lines indicate the location of the convective boundaries, determined by finding the location of the maximum $U_t$ gradient as in \protect\cite{jones2017}. Note that the X14 and X22 simulations shown on the top panel were driven with $1000 \times$ the nominal luminosity (see Table~\ref{tab:runs})}
    \label{fig:velcomp_mach}
\end{figure}

\begin{figure}
    \centering
    \includegraphics[width=\columnwidth]{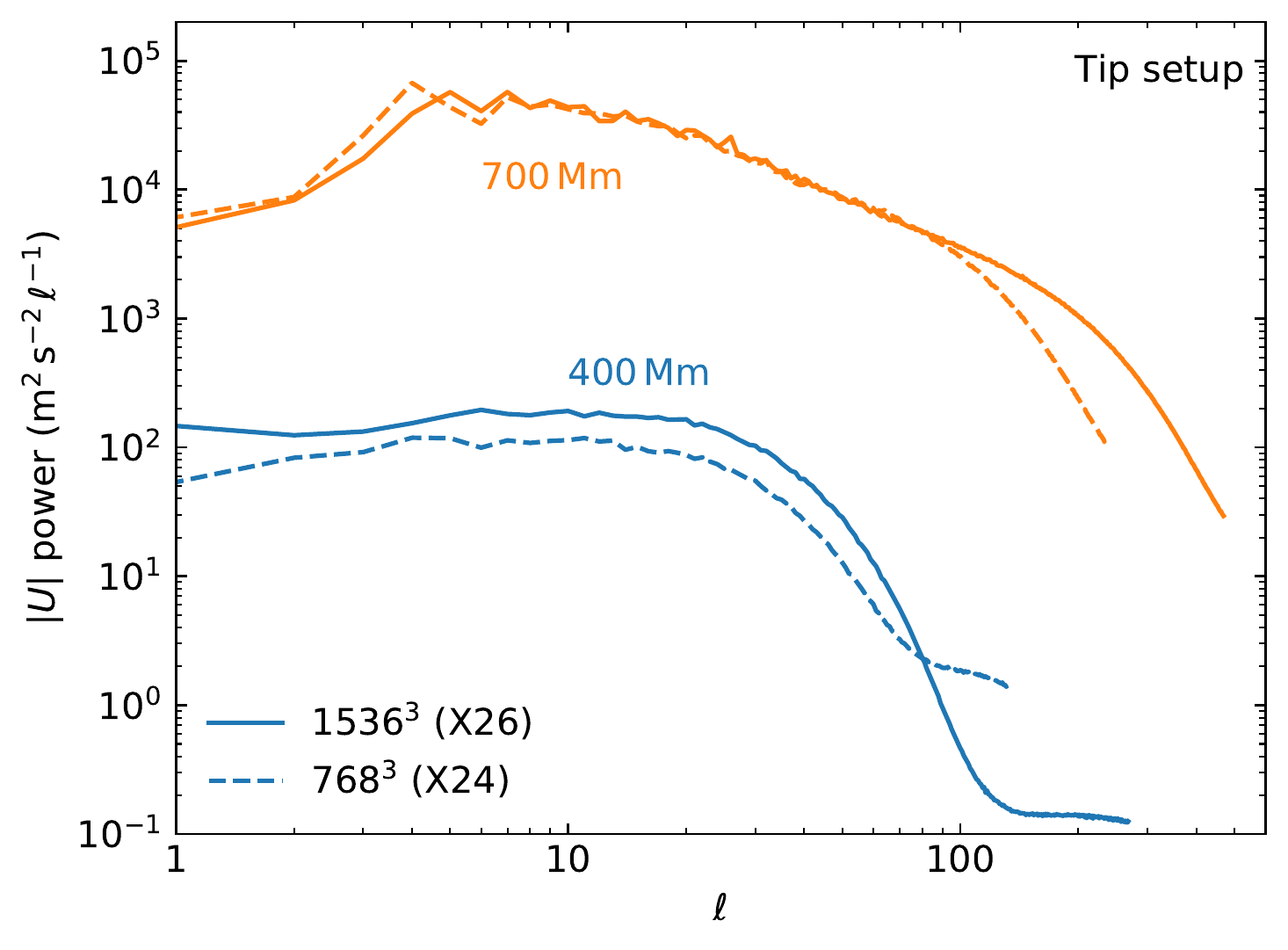}
    \caption{Power spectrum of $|U|$ in the radiative ($R=400\,{\rm Mm}$) and convective ($R=700\,{\rm Mm}$) zones for our RGB tip simulations at different grid resolutions (see legend). The spectra were computed by averaging over dumps 410 to 510.}
    \label{fig:spectra_rescomp}
\end{figure}

\begin{figure}
    \centering
    \includegraphics[width=\columnwidth]{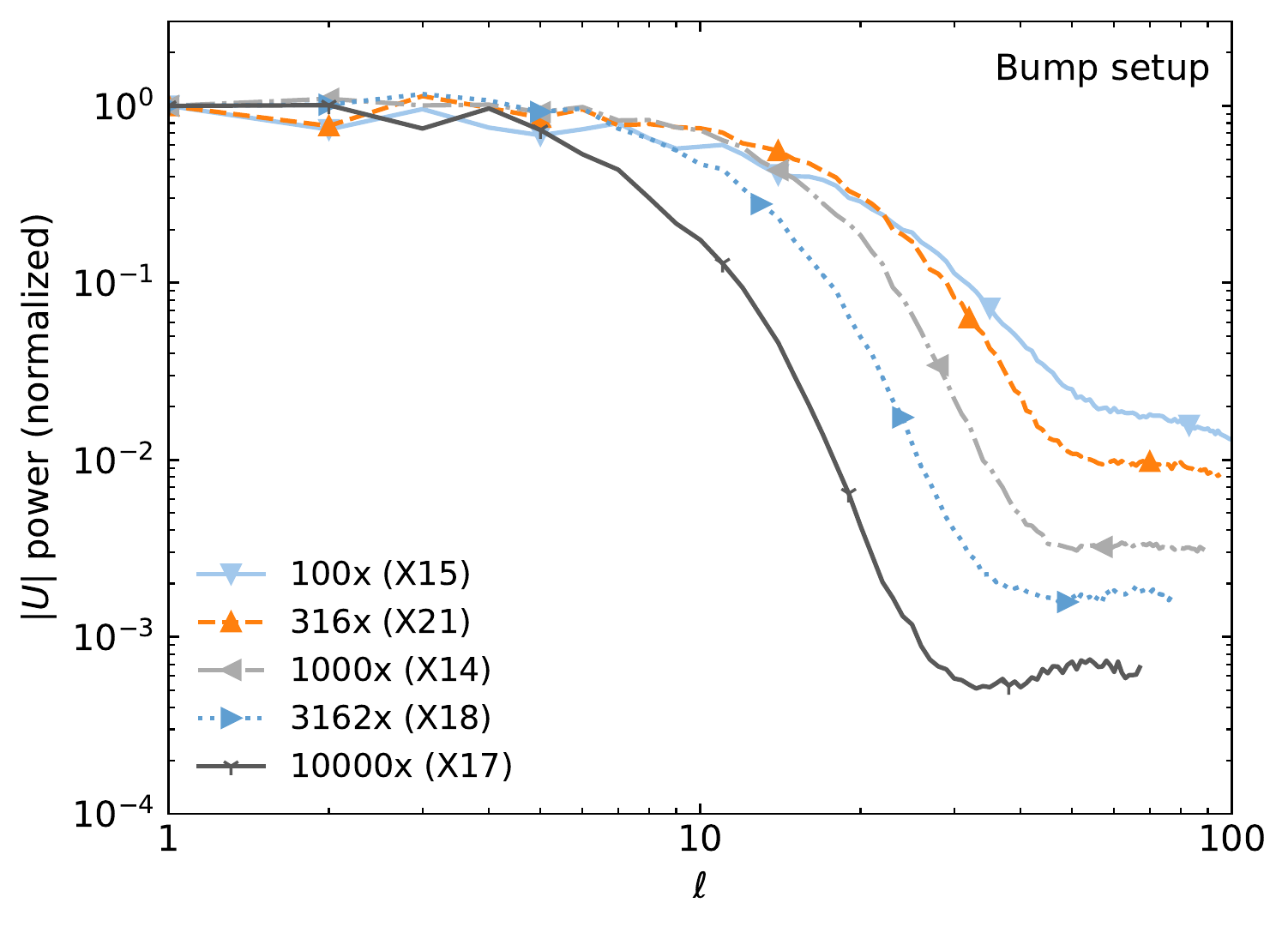}
    \caption{Power spectra of $|U|$ 200\,Mm below the convective boundary for RGB bump simulations with different heating rates (see legend). The same dumps as in Figures~\ref{fig:scaling_U} and~\ref{fig:scaling_vort} are used. To facilitate comparison, all spectra were normalized at $\ell=1$.}
    \label{fig:spectra_heating}
\end{figure}

\begin{figure}
    \centering
    \includegraphics[width=\columnwidth]{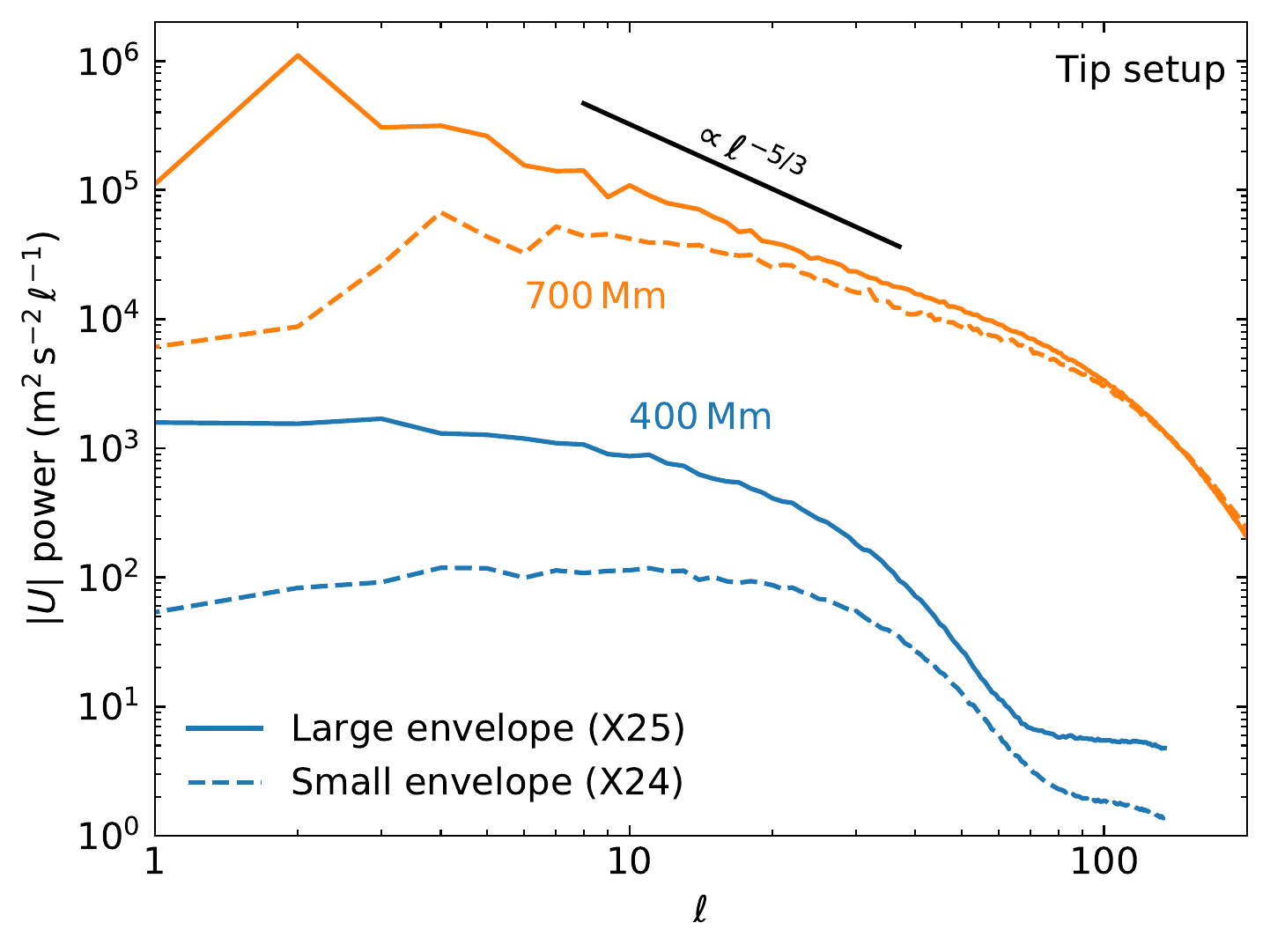}
    \caption{Power spectrum of $|U|$ in the radiative ($R=400\,{\rm Mm}$) and convective ($R=700\,{\rm Mm}$) zones for our RGB tip simulations with different envelope radial extents (see legend). The spectra were computed by averaging over dumps 410 to 510.}
    \label{fig:spectra_envelope}
\end{figure}

\begin{figure}
    \centering
    \includegraphics[width=\columnwidth]{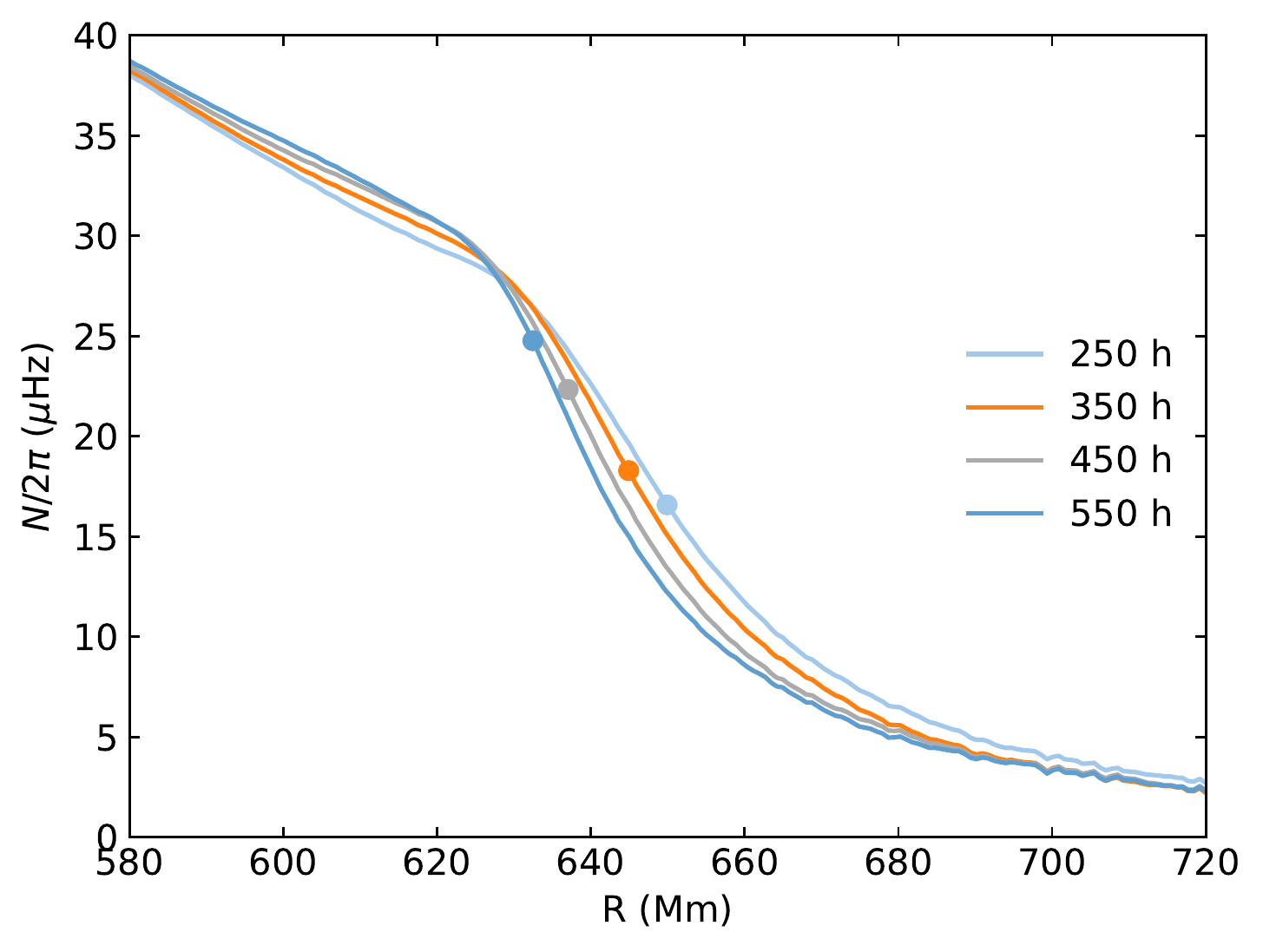}
    \caption{Evolution of the Brunt--V\"ais\"al\"a frequency in run X30 as the penetration zone is established. The convective boundary is indicated by a circle.}
    \label{fig:Npenetration}
\end{figure}

\end{document}